\documentclass[spanish,english]{aims}
\usepackage[T1]{fontenc}
\usepackage[latin9]{inputenc}
\usepackage{amsmath}
  \usepackage{paralist}
\usepackage{amssymb}
\usepackage{esint}

 \usepackage[colorlinks=true]{hyperref}

\def\currentvolume{X}
 \def\currentissue{X}
  \def\currentyear{200X}
   \def\currentmonth{XX}
    \def\ppages{X--XX}

\makeatletter

\newcommand{\noun}[1]{\textsc{#1}}

\newtheorem{thm}{Theorem}[section]

\newtheorem{lem}[thm]{Lemma}
\newtheorem{prop}{Proposition}
\newtheorem*{notation*}{Notation}
\theoremstyle{definition}
\newtheorem{defn}[thm]{Definition}
\newtheorem{note}{Note}
\newtheorem*{note*}{Note}
\newtheorem{rem}{Remark}
\newtheorem{example}{Example}
\newtheorem*{acknowledgement*}{Acknowledgements}

\usepackage{amsthm}
\usepackage{amsfonts}
\usepackage{amscd}
\usepackage{pb-diagram}

\newcommand{\dif}{\textrm{\textbf{d}}}
\newcommand{\dime}[1]{\,\,\text{dim}\,#1}
\newcommand{\Img}[1]{\,\,\text{Im}\,#1}
\newcommand{\bx}{\textbf{x}}
\newcommand{\bv}{\textbf{v}}
\newcommand{\cI}{{\mathcal I}}
\newcommand{\mN}{\mathbb{N}}
\newcommand{\mR}{\mathbb{R}}

\makeatother

\usepackage{babel}
\deactivatetilden

\title[Constraints, Field Theory and EDS]{Dirac constraints in Field theory and
  exterior differential systems}

\author[Santiago Capriotti]{}

\email{caprios2000@gmail.com}


\begin{document}
\maketitle

\centerline{\scshape Santiago Capriotti}
\medskip
{\footnotesize
   \centerline{Instituto Balseiro - U. N. de Cuyo}
   \centerline{Avda. E. Bustillo km. 9,5}
 \centerline{S. C. de Bariloche - Argentina}
} 

\bigskip

 \centerline{(Communicated by Juan-Pablo Ortega)}

\begin{abstract}
The usual treatment of a (first order) classical field theory such
as electromagnetism has a little drawback: It has a primary constraint
submanifold that arise from the fact that the dynamics is governed
by the antisymmetric part of the jet variables. So it is natural to
ask if there exists a formulation of this kind of field theories which
avoids this problem, retaining the versatility of the known
approach. 
The following paper deals with a family of variational problems, namely, the so called non standard variational problems, which intends to capture the data necessary to set up such a formulation for field theories; moreover, we will formulate a multisymplectic
structure for the family of non standard variational problems, and
we will relate this with the (pre)symplectic structure arising on
the space of sections of the bundle of fields. In this setting the Dirac theory of constraints will
be studied, obtaining among other
things a novel characterization of the constraint manifold which arises
in this theory, as generators of an exterior differential system associated to the equations of motion and the chosen slicing. Several examples of application of this formalism are discussed: Two of them motivated from the physical point of view, that is, electromagnetism and Poisson sigma models, and two examples of mathematical application. In the case of electromagnetism, it is shown that this formulation avoids the problems arising in the usual approach.
\end{abstract}

\section{Introduction}

In this work, the term \emph{field theory} will refer to a particular kind of variational problems on the sections of some bundle on space-time (for definitions, see section \ref{Section:4}). In the usual approach, the Dirac constraints in a field theory are found using the Gotay-Nester-Hinds algorithm on an infinite dimensional presymplectic manifold associated to the underlying variational problem. One of the purposes of this article is to show another way to build up these constraints, namely, by using a geometrical representation of the corresponding Euler-Lagrange equations into the exterior algebra of a bundle. The tools used for this task were taken from the theory of exterior differential systems (EDS from now on, for references see appendix \ref{Section:2}). In this way the Cauchy-Kowalevsky existence theorem can be introduced into the realm of field theory; on the contrary, the usual setting has to do with functional analytic conditions, hiding these existence conditions into the (infinite dimensional) manifold structure.
\\
In more precise terms, the variational problems we are concerned with are
initially characterized by a double fibration\[
\Lambda\rightarrow\Lambda_{1}\rightarrow M^{n}\]
together with an exterior differential system $\cI\subset\Omega^{\bullet}\left(\Lambda\right)$
and an $n$-form $\lambda$. These kind of problems will be called \emph{non standard variational
problems} throughout the work. The idea for this formulation can be traced back to \cite{PGriffithsVariations}, for the case in which the base manifold has dimension 
$n=1$; another work on the subject, in the same vein, can be found in \cite{HsuVariations}. It is worth remarking that both references provide examples from geometry (not only mechanics) where this scheme can be applied. The basic idea is that, for every section $\sigma$ of the bundle $\Lambda_{1}$,
we can build a section $\textsf{\textbf{pr}}\sigma$ for the bundle
$\Lambda$ by using these data (see details below); the non standard
problem consists of finding those sections $\sigma$ of the bundle $\Lambda_{1}$
which are extremals of the functional\[
S_{\lambda}\left[\sigma\right]:=\int_{M}\left(\textsf{\textbf{pr}}\sigma\right)^{*}\lambda.\]
 This setting includes what can be called \emph{standard variational problems}, like the usual Lagragian mechanics and the lagrangian viewpoint of first order field theories as in \cite{Gotay:1997eg}: For example, in the latter case the bundle $\Lambda_1$ corresponds to the bundle of fields, $\Lambda$ is the $1$-jet space of sections for this bundle, and the prolongation of a section $\sigma$ is given by its $1$-jet $j^1\sigma$.
\\
Our aim is twofold:
\begin{enumerate}
\item In first place, to set up an infinite dimensional presymplectic manifold with a hamiltonian for every non standard variational problem. This is done by appealing to a \emph{bivariant Lepagean equivalent} variational problem associated to the non standard problem; it provides us with a multisymplectic-like structure, and introducing an slicing of the bundle where this structure lives, the presymplectic manifold can be defined.
\item Secondly, we want to prove that the Dirac constraints obtained from the successful application of Gotay-Nester algorithm to the data found in the previous item can be calculated as the generators of some EDS closely related to the (Lepagean equivalent) variational problem. We must remark here that these constraints generalize in some sense the Dirac constraints to the non standard setting: The usual Dirac constraints can be found through this procedure by using the canonical Lepage equivalent (in the sense of definition \ref{DefCanonicalLepageEquivalent} below) for a first order field theory on the $1$-jet bundle associated to the bundle of fields, as shown in example \ref{ExampleFirstoOrderDirac}.
\end{enumerate}

The present article is structured as follows: In section \ref{Section:4} the non standard variational problem is defined, and some relevant examples are presented. In particular, it is shown that several dynamical problems of mathematical physics can
be formulated as non standard problems; as a particular example, we discuss in this setting the electromagnetism. The Euler-Lagrange equations of the non standard problem are found here, and the ideas of \emph{Lepage equivalent} and \emph{canonical Lepage equivalent} system associated to a non standard variational problem are introduced. 
This gives us a kind of multisymplectic structure for every non standard variational problem. It is important to note that no use of Legendre transform is made in the construction of this multisymplectic space; the main disadvantage of this approach is that the equivalence between the original equations of motion and the equations of motion in the multisymplectic space must be done in each case separately (this is an issue related to the so called \emph{bivariance} of the chosen Lepage equivalent problem, see below). By adding a
compatible slicing of the space-time, a Hamiltonian
version of the equations governing the extremals is developed in section \ref{sec:Introduction-to-Initial}: it is described a procedure that associates a presymplectic manifold (infinite dimensional if $n>1$) and a function on it to every (Lepagean equivalent of a) non standard problem. Moreover it is proved here that the solutions of the dynamical system determined by these data are extremals of the underlying variational problem. Section \ref{Section:7} contains the main result of the article, which deals with the description of the constraint submanifold arising from the Gotay-Nester
algorithm \cite{GotayNester} in terms of an exterior differential system associated to the
data of the theory. The section \ref{Section:Examples} contains four examples
where these considerations are applied. The first two constitues the examples with
physical content, namely, electromagnetism and the Poisson sigma model, and
the last two explores other aspects of the method: the calculation of
integrability conditions for a system of PDEs via Gotay-Nester algorithm, and a toy model where this algorithm fails to reach a succesful termination. The relevance of the last example is that its features could be related to the singular behaviour of some field theories (cf. remark \ref{RemarkFieldTheory}). We show also how the EDS perspective help us in the interpretation of the results. A note of caution about these examples is in order here: It will be assumed that there exists a well-defined differential structure on the sets of sections which we work with, so several of the manipulations in this context are of formal nature. 
\\
It is worth noting that the idea giving rise
to the definition of the EDS associated to the constraints is an extension of the one used in \cite{HartleyTucker} for the treatment of classical mechanics
from the EDS viewpoint. Similar results, but from the viewpoint of
Janet-Riquier theory, instead of Cartan's EDSs, and oriented to the
construction of an algorithm, can be found in \cite{Seiler:1995bm,Seiler:1995ne,Gerdt:2003ty,Gerdt:2006ui}. The introductory material about variational calculus and exterior differential systems was deferred to Appendices \ref{Section:1} and \ref{Section:2}.

\begin{notation*}
Given a bundle $F\rightarrow B$, $\Gamma\left(F\right)$
denotes the set of differentiable sections of this bundle. The subbundle
$VF\subset TF$ denotes the set of vertical vectors. The symbol $\Omega^{\bullet}\left(X\right)$
indicates the set of forms of any order on the manifold $X$, and
correspondingly $\Omega^{k}\left(X\right)$ is the set of $k$-forms
on $X$. Although in some cases this fact was proved (see proposition
\ref{pro:Existe-extension} below), it is assumed that some vector fields
defined on certain subsets of the bundles which we work with admits
an extension to a neighborhood of these subsets. We do not discuss
the convergence of the integrals used elsewhere; it is supposed that the hypothesis ensuring
its convergence are satisfied. The notation $\tau_{X}$ (resp. $\bar{\tau}_{X}$)
used in \cite{A-M} in order to denote the canonical projection of
the tangent bundle (resp. cotangent bundle) of $X$ is adopted. If
$i:Y\hookrightarrow X$ is the canonical injection of some submanifold $Y$ of $X$, the symbol $T_YX$ will denotes
the pullback bundle $i^*\left(TX\right)$. Given
a set of forms $S\subset\Omega^{\bullet}\left(X\right)$ we denote
by $\left\langle S\right\rangle _{\text{alg}}\subset\Omega^{\bullet}\left(X\right)$
the algebraic ideal generated by these forms, and by $\left\langle S\right\rangle _{\text{diff}}\subset\Omega^{\bullet}\left(X\right)$
the minimal algebraic ideal containing $S$ and is closed with respect
to the exterior derivative operator. If $S\subset V$ is a set of
vectors in the vector space $V$, we indicate by $\left\langle S\right\rangle $
the subspace spanned by the vectors in $S$.
\end{notation*}

\section{Formal structure for variational problems}\label{Section:4}

The essential data in both the formulation of the usual field theories
\cite{Gotay:1997eg,Gotay:2004ib,Blee} and classical mechanics \cite{A-M,HartleyTucker}
are the following: A double fibration $\Lambda\stackrel{p}{\longrightarrow}\Lambda_{1}\stackrel{\pi_{1}}{\longrightarrow}M$
(with composition $\pi:=\pi_{1}\circ p$), an exterior differential
system $\mathcal{I}\subset\Omega^{\bullet}\left(\Lambda\right)$
and a semibasic $n$-form $\lambda\in\Omega^{n}\left(\Lambda\right)$,
where $n=\dime M$. In terms of these objects, the prolongation of
a section $\sigma\in\Gamma\left(\Lambda_{1}\right)$ can be defined as a section $\textsf{\textbf{pr}}\sigma\in\Gamma\left(\Lambda\right)$
such that

\begin{enumerate}
\item Its graph is an integral submanifold for $\cI$, that is, $\textsf{\textbf{pr}}\sigma^{*}\left(\cI\right)=0$,
and
\item The section $\textsf{\textbf{pr}}\sigma$ covers $\sigma$, that
is, the following diagram is commutative

\[
\begin{diagram}
\node[2]{\Lambda}\arrow{s,r}{p}\\
\node{M}\arrow{ne,t}{\textsf{\textbf{pr}}\sigma}\arrow{e,b}{\sigma}\node{\Lambda_1}
\end{diagram}
\]

\end{enumerate}
Hence we introduce the action as a real-valued function on the set
of sections of $\Lambda_{1}$
\[
S\left[\sigma\right]:=\int_{M}\left(\textsf{\textbf{pr}}\sigma\right)^{*}\left(\lambda\right).\]
\begin{defn}[Non standard variational problem]
In the previous setting, the \emph{non standard variational problem }consists in finding sections $\sigma\in\Gamma\left(\Lambda_{1}\right)$
which are extremals of the action $S$.
\end{defn}

The usual point of view in geometric mechanics is to take the contact
structure on jet manifolds as the basic prolongation structure for the
underlying variational problem; in this view it is a piece in the
formalism. The non standard approach consists in the weakening of this
assumption, regarding the prolongation structure as an additional data of the
problem, related to some aspects of its underlying geometry. The following examples may help us to clarify this assertion. Moreover, they are in order to illustrate the ubiquity of this notion; subsequent sections in the article will intend to show its usefulness.

\begin{example}[Classical Mechanics]
Taking $M:=I\subset\mR$ a real interval, $Q$ a manifold, $\Lambda_{1}:=Q\times I$
and $\Lambda:=TQ\times I$, with $\pi_{1}\left(q,t\right)=t$ and
$p\left(q,\dot{q};t\right)=\left(q,t\right)$, choosing as the prolongation
EDS the differential ideal generated by the forms $\theta^{i}:=\dif q^{i}-\dot{q}^{i}\dif t$,
and with $\lambda:=L\dif t$, we obtain the usual variational problem
of the classical mechanics. In this case if we have that $\sigma:t\mapsto\left(q\left(t\right),t\right)$,
then \[
\textsf{\textbf{pr}}\sigma:t\mapsto\left(q\left(t\right),\frac{\mbox{d}q}{\mbox{d}t}\left(t\right);t\right)\]
 is the formula defining the prolongation of sections. 
\end{example}

\begin{example}[First order field theories]
The following example show how first order field theories fits in our scheme, that is, how one might formulate a first order field theory as a non standard problem. So let us suppose that the fields are sections of certain bundle $F\stackrel{\pi}{\longrightarrow}M$ on space-time. The double fibration is in this case
\[
\Lambda:=J^1\left(\pi\right)\stackrel{\pi_{1,0}}{\longrightarrow}\Lambda:=J^0\left(\pi\right)=F\rightarrow M,
\]
where $J^k\left(\pi\right)$ denotes the $k$-jet manifold of  $F$.
By taking on $J^1\left(\pi\right)$ the contact EDS as prolongation structure, it can be shown that $\mathbf{pr}\sigma=j^1\sigma$. If $L\in C^\infty\left(J^1\left(\pi\right)\right)$, the $n$-form $\lambda:=L\dif x^1\wedge\cdots\wedge\dif x^n$ satisfies
\[
\left(\mathbf{pr}\sigma\right)^*\lambda=\left(L\circ j^1\sigma\right)\dif x^1\wedge\cdots\wedge\dif x^n;
\]
the variational problem consisting in finding the extremals to
\[
\sigma\mapsto\int_M\left(L\circ j^1\sigma\right)\dif x^1\wedge\cdots\wedge\dif x^n
\]
is the usual Hamilton's principle for the first order field theory with lagrangian density $L$.
\end{example}

\begin{example}[Electromagnetism]
Let us consider now the first example on non standard variational
problem, that is, the electromagnetic field on a spacetime $M$; previous work on the subject can be found in \cite{delphenich-2005-14,Horava:1990ba}.
Let us take\begin{eqnarray*}
\Lambda: & = & \bigwedge^{2}\left(T^{*}M\right)\oplus T^{*}M\\
\Lambda_{1}: & = & T^{*}M\end{eqnarray*}
$\pi_{1}:=\bar{\tau}_{M}$ and for $p$ the projection $p_{2}:\bigwedge^{2}\left(T^{*}M\right)\oplus T^{*}M\rightarrow T^{*}M$
onto the second summand. An arbitrary element of $\Lambda$ is given
by a pair $\left(F,A\right)$ where $F\in\bigwedge^{2}\left(T_{m}^{*}M\right)$
and $A\in T_{m}^{*}M$ for some $m\in M$. For every $k\in\mN$ there
exists a canonical $k$-form on $\bigwedge^{k}\left(T^{*}M\right)$
defined by\[
\Theta_{k}|_{\alpha}:=\alpha\circ\bar{\tau}_{M*},\qquad\forall\alpha\in\bigwedge^{k}\left(T^{*}M\right);\]
as prolongation EDS $\cI$ we take the differential ideal generated
by the form\[
\Gamma:=\dif\Theta_{1}-\Theta_{2}.\]
So a section $m\mapsto\left(F\left(m\right),A\left(m\right)\right)$
of $\Lambda$ is an integral section for $\cI$ iff $F=\dif A$. 
\end{example}

\begin{example}[EDS as a non standard problem]\label{ExampleEDSasNonStandard}
This is the first example of a non standard problem of mathematical nature: Every EDS $\cI$ on a bundle $F\rightarrow M$ can be considered as a non standard problem, by using the double fibration
\[
F\rightarrow M\stackrel{\text{id}}{\longrightarrow}M,
\]
with prolongation structure $\cI$ and lagrangian $n$-form $\lambda=0$. Therefore the intermediate bundle has just one section, namely $\sigma_0=\text{id}_M$, and the possible prolongations of it are the integral sections of $\cI$. Later on we will use this remark in order to employ the Gotay-Nester algorithm in searching the integrability conditions of a system of PDEs (see example in section \ref{ExampleOfProlongation}).
\end{example}

\begin{example}[Tetrad gravity]
Let $M$ be a pseudoriemannian $4$-manifold. Let us introduce a formulation of the variational problem for tetrad gravity on $M$ in these terms. The fields in this case are the components $e_\mu\in\Omega^1\left(M\right),\mu=0,\cdots,3$ of a tetrad on the spacetime $M$. The natural ``velocities'' for such an object are not its prolongation into $J^1\left(\bar{\tau}_M\right)$; instead, it is more geometric to take as velocities for a tedrad the associated connection forms $\omega_{\mu\nu}\in\Omega^1\left(M\right)$. Thus it is possible to formulate tetrad gravity as a non standard problem on the double fibration
\[
F_{\mathcal{O}}^*\left(M\right)\oplus\left(T^*M\right)^{\oplus4}\rightarrow F_{\mathcal{O}}^*\left(M\right)\rightarrow M
\]
where $\bar{p}_M:F^*_{\mathcal{O}}\left(M\right)\rightarrow M$ stands for the coframe bundle on $M$, with prolongation structure generated by the forms
\[
\Theta^\alpha:=\dif E^\alpha-\Omega^\alpha_\nu\wedge E^\nu,\nu=0,\cdots,3.
\]
Here with capital letters we denote the corresponding canonical forms. The lagrangian is the scalar curvature, so
\[
\lambda:=\star\left(E^\alpha\wedge E^\beta\right)\wedge\left(\dif \Omega_{\alpha\beta}+\Omega_\alpha^\nu\wedge\Omega_{\nu\beta}\right).
\]
The prolongation means that it is a torsionless connection.
\end{example}

The purpose of the forthcoming sections is to show that in the non standard setting it is possible to build a reasonable multisymplectic space and even to deal with its Dirac constraints, althought no use of Legendre transformation will be made in the process. But first let us show how the Euler-Lagrange equations for a non standard variational problem can be built.

\subsection{Euler-Lagrange equations for non standard variational problems}

In order to perform variations on the class of functionals relevant
to our study, we need to adapt the proposition \ref{pro:The-variational-derivative}
to the case in which the section involved in the integrand is the
prolongation of another one by using some prolongation structure.
To this end, let us introduce the following definition.

\begin{defn}
An (infinitesimal) \emph{allowed variation} for a non standard variational
problem with prolongation structure given by the EDS $\cI$ is an
element $V\in\Gamma\left(\sigma^{*}\left(V\Lambda\right)\right),\sigma\in\Gamma\left(\cI\right)$
admitting an extension $\hat{V}\in\Gamma\left(V\Lambda\right)\subset\mathfrak{X}\left(\Lambda\right)$
which is an infinitesimal symmetry for $\cI$, that is\[
\sigma^{*}\left(\mathcal{L}_{\hat{V}}\cI\right)=0.\]

\end{defn}
It is assumed, unless explicitly stated, that whenever a manifold
has a boundary, the allowed variations annihilates on this boundary.
Then we have the following proposition \cite{GotayCartan,PGriffithsVariations}.

\begin{prop}
\label{pro:The-Euler-Lagrange}The Euler-Lagrange equations associated
to the functional $S\left[\sigma\right]=\int_{M}\left(\textsf{\textbf{pr}}\sigma\right)^{*}\lambda$
defined on $\Gamma\left(\Lambda_{1}\right)$ are given by\[
\left(\textsf{\textbf{pr}}\sigma\right)^{*}\left(V\lrcorner\dif\lambda\right)\equiv0\mod\text{ exact }n\text{-forms on }\Lambda,\]
where $V\in\Gamma\left(\left(\textsf{\textbf{pr}}\sigma\right)^{*}\left(V\Lambda\right)\right)$
is an arbitrary allowed variation.
\end{prop}
\begin{example}[Electromagnetism - Cont.]
We return to the electromagnetic field in order to find the Euler-Lagrange
equations for the lagrangian $\lambda:=\Theta_{2}\wedge\ast\Theta_{2}$;
here\[
\ast\Theta_{2}|_{\alpha}:=\left(\ast\alpha\right)\circ\bar{\tau}_{M*}\]
by using the Hodge star associated to the spacetime metric. Now, as in
every vector bundle, the vertical space can be identified at each
point with the fibre through that point, so any vertical vector field
on $\Lambda$ can be denoted as an application $\left(F,A\right)\in\Lambda\mapsto\left(\delta F,\delta A\right)\in\pi^{-1}\left(\pi\left(F,A\right)\right)\simeq V_{\left(F,A\right)}\Lambda$.
This identification has the following interesting property: For the
Lie derivative the following holds\[
\left(\mathcal{L}_{\delta\alpha}\Theta_{k}\right)|_{\alpha}=\delta\alpha\circ\bar{\tau}_{M*},\qquad\forall\alpha\in\bigwedge^{k}\left(T^{*}M\right).\]
So if $V:=\left(\delta F,\delta A\right)$ is an allowed variation with extension $\hat{V}:=\left(\hat{\delta F},\hat{\delta A}\right)$
to a vertical vector field on $\Lambda$, this extension must be such
that\[
\mathcal{L}_{\left(\hat{\delta F},\hat{\delta A}\right)}\left(\dif\Theta_{1}-\Theta_{2}\right)=\mu\left(\dif\Theta_{1}-\Theta_{2}\right)\]
for some function $\mu\in C^{\infty}\left(\Lambda\right)$. By restricting
to the integral submanifold $\mbox{Im}\left(\mbox{pr}\sigma\right),\sigma\in\Gamma\left(\cI\right)$
we obtain the following differential condition\[
\dif\delta A-\delta F=0.\]
Then the Euler-Lagrange equations are expressed as\[
2\dif\delta A\wedge\ast F\equiv0\qquad\Leftrightarrow\qquad\delta A\wedge\dif\left(\ast F\right)\equiv0\qquad\Leftrightarrow\qquad\dif\left(\ast F\right)=0\]
to which the prolongation condition must be added\[
F=\dif A.\]

\end{example}

\subsection{(Multi)hamiltonian formalism through Lepagean equivalent problems}

We want to construct a hamiltonian version for the non standard variational
problem. The usual approach \cite{Gotay:1997eg} seems useless here, because of the following facts:
\begin{itemize}
\item The covariant multimomentum space is a bundle in some sense dual to the velocity space, which is a jet space. 
\item The dynamics in the multimomentum space is defined through the Legendre transform, and it is not easy to generalize to a non standard problem this notion.
\end{itemize}
The trick to circumvect the difficulties is to mimic the passage from Hamilton's principle to Hamilton-Pontryaguin principle. This is done by including the generators of the prolongation structure in the lagrangian density by means of a kind of Lagrange multipliers. This procedure will be formalized below, where the hamiltonian version is defined by associating a first order variational
problem to the non standard variational problem, whose extremals are in one to one correspondence with its extremals. This is called \emph{canonical bivariant Lepage equivalent problem}.

\subsubsection{Lepagean equivalent problems}
Here we will follow closely the exposition of the subject in the article \cite{GotayCartan}. Before going into details, let us introduce a bit of terminology: If $\Lambda\stackrel{\pi}{\longrightarrow}M$ is a bundle, $\lambda\in\Omega^n\left(\Lambda\right)$ ($n=\text{dim}\,M$) and $\cI$ is an EDS on $\Lambda$, the symbol $\left(\Lambda\stackrel{\pi}{\longrightarrow}M,\cI,\lambda\right)$ indicates the variational problem consisting in extremize the action
\[
S\left[\sigma\right]=\int_M\sigma^*\left(\lambda\right)
\]
with $\sigma\in\Gamma\left(\Lambda\right)$ restricted to the set of integral sections of $\cI$. Furthermore, $\mathcal{E}\left(\lambda\right)$ will denote the set of extremals for $\left(\Lambda\stackrel{\pi}{\longrightarrow}M,\cI,\lambda\right)$.

The idea is to eliminate in some way the constraints imposed by the elements of $\mathcal{I}$; intuitively, it is expected that the number of unknown increase when this is done. The following concept captures these ingredients formally.

\begin{defn}[Lepage equivalent variational problem]
A \emph{Lepagean equivalent} of a variational problem $\left(\Lambda\stackrel{\pi}{\longrightarrow}M,\cI,\lambda\right)$ is another variational problem
\[
\left(\tilde{\Lambda}\stackrel{\rho}{\longrightarrow}M,\left\{0\right\},\tilde{\lambda}\right)
\]
together with a surjective submersion $\nu:\tilde{\Lambda}\rightarrow\Lambda$ such that
\begin{itemize}
\item $\rho=\pi\circ\nu$, and
\item if $\gamma\in\Gamma\left(\tilde{\Lambda}\right)$ is such that $\nu\circ\gamma$ is an integral section of $\cI$, then
\[
\gamma^*\tilde{\lambda}=\left(\nu\circ\gamma\right)^*\lambda.
\]
\end{itemize}
\end{defn}
There exists a canonical way to build up a Lepage equivalent problem associated to a given variational problem $\left(\Lambda\stackrel{\pi}{\longrightarrow}M,\cI,\lambda\right)$, the so called canonical Lepage equivalent problem. Let $\cI$ be differentially generated by the sections of a graded subbundle $I\subset\bigwedge^\bullet\left(T^*\Lambda\right)$ (this is a ``constant rank'' hypothesis, ensuring the existence of a bundle in the construction, see below). Define $\cI^{\text{alg}}$ as the algebraic ideal in $\Omega^\bullet\left(\Lambda\right)$ generated by $\Gamma\left(I\right)$, and
\[
\left(\cI^{\text{alg}}\right)^l:=\cI^{\text{alg}}\cap\Omega^l\left(\Lambda\right).
\]
For $\lambda\in\Omega^n\left(\Lambda\right)$, define the affine subbundle $W^\lambda\subset\bigwedge^n\left(T^*\Lambda\right)$ whose fiber above $p\in\Lambda$ is
\[
\left.W^\lambda\right|_p:=\left\{\left.\lambda\right|_p+\left.\beta\right|_p:\beta\in\left(\cI^{\text{alg}}\right)^n\right\}.
\]
\begin{defn}[Canonical Lepage equivalent problem]\label{DefCanonicalLepageEquivalent}
In the previous setting, it is the triple $\left(W^\lambda\stackrel{\rho}{\longrightarrow}M,\left\{0\right\},\tilde{\Theta}\right)$, where $\nu$ is the canonical projection $\bar{\tau}^n_\Lambda:\bigwedge^n\left(T^*\Lambda\right)\rightarrow\Lambda$ restricted to $W^\lambda$, $\rho:=\pi\circ\nu$ and $\tilde{\Theta}$ is the pullback of the canonical $n$-form
\[
\left.\Theta_n\right|_\alpha:=\alpha\circ\left(\bar{\tau}^n_\Lambda\right)_*
\]
to $W^\lambda$. The form $\tilde\Theta$ will be called \emph{Cartan form} of the variational problem.
\end{defn}
It is worth remarking here that the terminology adopted here correspond to \cite{GotayCartan}, which is slightly different from the terminology of the classical theory, as exposed in e.g. \cite{Krupka:1987}; it is fully explained in the former work how can be related the notions developed in each case.
\newline
Returning to our main concern, it can be proved that the canonical Lepage equivalent is a Lepagean equivalent problem of $\left(\Lambda\stackrel{\pi}{\longrightarrow}M,\cI,\lambda\right)$. Now, the extremals of some variational problem has, in general, nothing to do with the extremals of its Lepagean equivalent problem, so it is necessary to introduce the following definition.
\begin{defn}[Covariant and contravariant Lepage equivalent problems]
We say that a Lepagean equivalent problem $\left(W^\lambda\stackrel{\rho}{\longrightarrow}M,\left\{0\right\},\tilde{\Theta}\right)$ for the variational problem $\left(\Lambda\stackrel{\pi}{\longrightarrow}M,\cI,\lambda\right)$ is \emph{covariant} if $\nu\circ\gamma\in\mathcal{E}\left(\lambda\right)$ for all $\gamma\in\mathcal{E}\left(\tilde{\Theta}\right)$; on the contrary, it is called \emph{contravariant} if every $\sigma\in\mathcal{E}\left(\lambda\right)$ is the projection of some extremal in $\mathcal{E}\left(\tilde{\Theta}\right)$ through $\nu$. A Lepagean equivalent problem is \emph{bivariant} if and only if it is both covariant and contravariant.
\end{defn}
There exists a fundamental relation between the extremals of a variational problem and the extremals of its canonical Lepage equivalent.
\begin{thm}
The canonical Lepage equivalent is covariant.
\end{thm}
For a proof, see \cite{GotayCartan}. The contravariant nature of a Lepage
equivalent problem is more subtle to deal with. Next we will describe another construction for the Lepage equivalent problem, by reducing the number of additional variables that we put in order to get the prolongation EDS into the variational equations; although in this way there exist more chances to find a non contravariant Lepage equivalent problem, we will see that in many cases it leads to well-behaved Lepage equivalent problems.

\subsubsection{Another canonical Lepage equivalent problem}
The canonical Lepage problem has to deal with general EDSs; the examples we
will try to manage has some nice features that allow us to simplify this scheme. Namely, define the subbundle
\[
Z_1\left(\Lambda\right):=\left\{\alpha\in\bigwedge^\bullet\left(T^*\Lambda\right):v_1\lrcorner v_2\lrcorner\alpha=0\quad\forall v_1,v_2\,\left(\Lambda\rightarrow M\right)\text{-vertical vectors}\right\}
\]
and suppose further that the generators of $\cI$ are in $Z_1\left(\Lambda\right)$; as before, we have a subbundle $\left(\cI^{\text{alg}}\right)^n\subset\bigwedge^n\left(T^*\Lambda\right)$. Let $K^n\subset I^n$ be the subbundle whose fibers are the subspaces $\left.K^n\right|_\gamma\subset\left.I^n\right|_\gamma$ of forms of (algebraic) degree $1$ in the generators; set $\tilde{K}^n:=K^n\cap Z_1\left(\Lambda\right)$. Depending on $\cI$, it can leads to an affine subbundle $\tilde{W}^\lambda$ in $\bigwedge^\bullet\left(T^*\Lambda\right)$ with fibers
\[
\left.\tilde{W}^\lambda\right|_\gamma:=\left\{\left.\lambda\right|_\gamma+\psi:\psi\in\left.\tilde{K}^n\right|_\gamma\right\}.
\]
As before, the map $\nu:\tilde{W}^\lambda\rightarrow\Lambda$ is the restriction of the canonical projection $\bar{\tau}^\bullet_\Lambda:\bigwedge^\bullet\left(T^*\Lambda\right)\rightarrow\Lambda$; moreover, it can be shown that it is a covariant Lepagean equivalent problem also. The reasons for this choice will become apparent later: The variational problems associated to forms in $Z_1\left(\Lambda\right)$ will have a well defined description in terms of dynamics on a (infinite dimensional) presymplectic manifold (cf. prop. \ref{PropUnoaUno}).
\paragraph*{\textit{A local version of the alternate canonical Lepage equivalent.}}
Let us now define the pullback bundle $\pi^*\left(\bigwedge^\bullet\left(T^*M\right)\right)$ on $\Lambda$; then we have the commutative diagram
\[
\begin{diagram}
\node{\pi^*\left(\bigwedge^\bullet\left(T^*M\right)\right)}\arrow{s,l}{\pi^*\left(\bar{\tau}_M^\bullet\right)}\arrow{e,t}{f}\node{\bigwedge^\bullet\left(T^*M\right)}\arrow{s,r}{\bar{\tau}_M^\bullet}\\
\node{\Lambda}\arrow{e,b}{\pi}\node{M}
\end{diagram}
\]
Let $Z_0\left(\Lambda\right)\subset\bigwedge^\bullet\left(T^*\Lambda\right)$
be the subbundle of semibasic forms on $\Lambda$, and $Z_0^k\left(\Lambda\right):=Z_0\left(\Lambda\right)\cap\bigwedge^k\left(T^*\Lambda\right)$ for all $k\in\mathbb{N}$. The following lemma can be found in \cite{saunders89:_geomet_jet_bundl} (cf. lemma $3.3.5$ there).
\begin{lem}
There exists a natural identification between the bundles $\pi^*\left(\bigwedge^\bullet\left(T^*M\right)\right)$ and $Z_0\left(\Lambda\right)$.
\end{lem}
This identification provides us the horizontal map $f$, which has the following nice property.
\begin{prop}\label{PropPropsF}
The submersive bundle map $f:Z_0\left(\Lambda\right)\rightarrow\bigwedge^\bullet\left(T^*M\right)$ is graded (of degree $0$), and verifies that
\begin{equation}\label{DefPropertySubmersion}
\left(\left.f\right|Z_0^k\left(\Lambda\right)\right)^*\Theta^M_k=\left.\Theta^\Lambda_k\right|Z_0^k\left(\Lambda\right),\qquad\forall k\in\mathbb{N}
\end{equation}
where $\Theta^\Lambda_k\in\Omega^k\left(\bigwedge^k\left(T^*\Lambda\right)\right)$ and $\Theta_k^M\in\Omega^k\left(\bigwedge^k\left(T^*M\right)\right)$ are the canonical $k$-forms on the corresponding spaces.
\end{prop}
Further simplifications can be performed whenever the EDS admits a set of
global generators 
\[
\left.I\right|_\gamma=\mathbb{R}\left\langle\left.\alpha_{1}^{1}\right|_\gamma,\cdots,\left.\alpha_{k_{1}}^{1}\right|_\gamma,\left.\alpha_{1}^{2}\right|_\gamma,\cdots,\left.\alpha_{k_{2}}^{2}\right|_\gamma,\cdots,\left.\alpha_{1}^{p}\right|_\gamma,\cdots,\left.\alpha_{k_{p}}^{p}\right|_\gamma\right\rangle,\qquad\forall\gamma\in\Lambda,
\]
where $\left\langle \alpha_{1}^{j},\cdots,\alpha_{k_{j}}^{j}\right\rangle
_{\text{diff}}=\cI^{\left(j\right)}=:\cI\cap\Omega^{j}\left(\Lambda\right)$. It is supposed here that $\alpha^j_i\in Z_1\left(\Lambda\right)$ for all $j=1,\cdots,p,i=1,\cdots,k_j$, although neither of them is in $Z_0\left(\Lambda\right)$. 
Let us define
$\tilde{\Lambda}:=\Lambda\times_M\bigoplus_{l=1}^p\left(\bigwedge^{m_l}T^*M\right)^{\oplus k_l}$ where\footnote{Without loss of generality, it will be assumed that $p\leq\text{dim}M=n$, because we are dealing with integral sections, a subset of integral manifolds of dimension $n$ for $\mathcal{I}$.} $m_l:=n-l$; the
previous lemma leads us to the map
$F_\lambda:\tilde{W}^\lambda\rightarrow\tilde\Lambda$ given by
\begin{multline*}
F_\lambda:\left.\lambda\right|_\gamma+\sum_{l=1}^p\sum_{i=1}^{k_l}\left.\alpha_i^l\right|_\gamma\wedge\beta_{m_l}^i\mapsto\cr
\mapsto\left(\gamma,f\left(\beta_{m_1}^1\right),\cdots,f\left(\beta_{m_1}^{k_1}\right),\cdots,f\left(\beta_{m_p}^1\right),\cdots,f\left(\beta_{m_p}^{k_p}\right)\right).
\end{multline*}
Let us introduce the definitions
\[
\mathcal{Z}_{m_l}:=Z_0\left(\Lambda\right)\cap\bigwedge^{m_l}\left(T^*\Lambda\right),\qquad\mathcal{Z}_{m_l}^{k_l}:=\underbrace{\mathcal{Z}_{m_l}\times\cdots\times\mathcal{Z}_{m_l}}_{k_l\text{ times}}
\]
for $l=1,\cdots,p$. The map $F_\lambda$ will be well-defined provided that between the generators there are no ``zyzygies of degree $n$'', namely, that the unique solution for the system
\[
\sum_{l=1}^p\sum_{i=1}^{k_l}\left.\alpha_i^l\right|_\gamma\wedge\gamma_{m_l}^i=0
\]
in $\mathcal{Z}^{k_1}_{m_1}\times\cdots\times\mathcal{Z}^{k_p}_{m_p}$ is $\gamma_{m_l}^i=0$. Let us suppose that it is indeed the case; on $\tilde{\Lambda}$ we will define the $n$-form
\[
\tilde{\lambda}:=\sum_{l=1}^{p}\left(\sum_{j=1}^{k_{l}}\alpha_{j}^{l}\wedge\beta^{j}_{m_l}\right)-\lambda.
\]
In this setting the proposition \ref{PropPropsF} implies the following result, that allow us to use $\tilde\Lambda$
as canonical Lepage equivalent whenever our prolongation EDS $\cI$ admits
global generators.
\begin{lem}
Let us suppose that $I$ admits a set of global generators with the properties detailed above; then the map
$F_\lambda:\tilde{W}^\lambda\rightarrow\tilde{\Lambda}$ is a diffeomorphism,
and $F_\lambda^*\tilde\lambda=\tilde\Theta$.
\end{lem}
 These considerations can be applied to build a local version for the canonical Lepage equivalent variational problem, because it is always possible to find an open set where the considered EDS is described by a set of forms as above, so this procedure provides a local description (i.e. on the given open set) for the
modified canonical Lepage equivalent.
\newline
As we said before, the main difference between our constructions and the construction of the canonical Lepage equivalent problem is that in our case the affine bundle has lower rank. This means that there are less variables to take care of; the main drawback is that such a system has lower chances to be contravariant. Nevertheless, it will be shown below that it works very well in many important circunstances.

\subsubsection{Canonical Lepage equivalent of a non standard problem}
We will apply these considerations to our problem. The important thing to note is that, if a variational problem has a covariant and contravariant Lepagean equivalent problem, then the latter can be considered as a kind of Hamilton-Pontryaguin's principle for the given variational problem; in fact, it is shown below that the canonical Lepagean equivalent problem associated to the variational problem underlying the Hamilton's principle gives rise to the classical Hamilton-Pontryaguin's principle (see example \ref{HamiltonPontryaguin}). This will be our starting point for assigning a multisymplectic space to the variational problem we are dealing with. The bivariance ensures us that every extremal has been taken into account in the new setting. Otherwise, namely, for non contravariant canonical Lepage equivalent problems, some extremals for the original variational problem could be lost in the process.

So let
us suppose that we have the non standard problem defined by the following
data\[
\begin{cases}
\Lambda\rightarrow\Lambda_{1}\rightarrow M,\\
\mathcal{I}\subset\Omega^{\bullet}\left(\Lambda\right),\\
S\left[\sigma\right]:=\int_{M}\left(\textsf{\textbf{pr}}\sigma\right)^{*}\left(\lambda\right).\end{cases}\]
If $\cI$ in this non standard problem has the required regularity (i.e. the ``constant rank'' hypothesis), then the variational problem $\left(\Lambda\rightarrow M,\cI,\lambda\right)$ will have a canonical Lepage equivalent problem $\left(\tilde{W}^\lambda\rightarrow M,\left\{0\right\},\tilde{\Theta}\right)$, and we can apply the scheme described above. In order
to carry out this task locally, let us suppose as above that the fibers of the bundle $I$ on an open set $U\subset\Lambda$ can be written as
\[
\left.I\right|_\gamma=\mathbb{R}\left\langle\left.\alpha_{1}^{1}\right|_\gamma,\cdots,\left.\alpha_{k_{1}}^{1}\right|_\gamma,\left.\alpha_{1}^{2}\right|_\gamma,\cdots,\left.\alpha_{k_{2}}^{2}\right|_\gamma,\cdots,\left.\alpha_{1}^{p}\right|_\gamma,\cdots,\left.\alpha_{k_{p}}^{p}\right|_\gamma\right\rangle,\quad\gamma\in U
\]
so that the prolongation
structure $\cI$ is (differentially) generated by\[
\cI=\left\langle \alpha_{i}^{j}:1\leq i\leq p,1\leq j\leq k_i\right\rangle _{\text{diff}}\]
on $U$, where $\left\langle \alpha_{1}^{j},\cdots,\alpha_{k_{j}}^{j}\right\rangle _{\text{diff}}=\cI^{\left(j\right)}=:\cI\cap\Omega^{j}\left(U\right)$ and $\alpha_i^j\in Z_1\left(\Lambda\right)$ for all $i,j$.
Then if $U_0:=\pi\left(U\right)\subset M$, we can define for each $1\leq l\leq p$ the numbers $m_{l}:=\mbox{dim}\left(M\right)-l$
and the $n$-form on $\tilde{\Lambda}_U:=U\times_{U_0}\bigoplus_{l=1}^p\left[\bigwedge^{m_l}\left(T^*U_0\right)\right]^{\oplus k_l}$ will reads\[
\tilde{\lambda}:=\sum_{l=1}^{p}\left(\sum_{j=1}^{k_{l}}\alpha_{j}^{l}\wedge\beta^{j}_{m_l}\right)-\lambda\]
where $\left(\beta^{1}_{m_{1}},\cdots,\beta^{k_{1}}_{m_{1}},\cdots,\beta^{1}_{m_{p}},\cdots,\beta^{k_{p}}_{m_{p}}\right)$
denotes sections of $\bigwedge^{m_l}\left(T^*U_0\right)$; the subscript in these sections thus indicates their degree (in the exterior algebra sense). It will be shown in the
examples below that, in many important cases, the Euler-Lagrange
equations associated to the non standard problem defined by the data\[
\begin{cases}
\tilde{\Lambda}_U\stackrel{\mbox{id}}{\longrightarrow}\tilde{\Lambda}_U\rightarrow M,\\
0\subset\Omega^{\bullet}\left(\tilde\Lambda_U\right),\\
S\left[\sigma\right]:=\int_{M}\left(\textsf{\textbf{pr}}\sigma\right)^{*}\tilde{\lambda},\end{cases}\]
has a family of solutions which is isomorphic to the family of solutions
of the previous system; this means that in these cases the canonical
Lepage equivalent problem is also contravariant. By proposition \ref{pro:The-Euler-Lagrange},
the Euler-Lagrange eqs for an stationary section $\sigma\in\Gamma\left(\tilde{\Lambda}_U\right)$
are\begin{equation}
\sigma^{*}\left(V\lrcorner\dif\tilde{\lambda}\right)=0,\qquad\forall V\in\Gamma\left(V\tilde{\Lambda}_U\right)\label{eq:HamEqElectro1}\end{equation}
because there are no conditions for admisibility of variations; these
sections are then integral sections for the EDS\begin{equation}
\cI:=\left\langle V\lrcorner\dif\tilde{\lambda}:V\in\Gamma\left(V\tilde{\Lambda}_U\right)\right\rangle _{\text{diff}}.\label{eq:EDSHam}\end{equation}
We call this EDS the (local version of) \noun{Hamilton-Cartan EDS}.

\subsubsection{Some examples}
We will illustrate with some examples how the standard approach fits into the
scheme developed above, and moreover we will use it in variational problems
with non standard description.
\begin{example}\label{HamiltonPontryaguin}
Let us start with the standard variational problem of mechanics. This
scheme can be found in \cite{PGriffithsVariations,HartleyTucker}. Our Lepage equivalent problem
will have as underlying bundle
\[
\bigwedge^\bullet\left(T^*\left(\mathbb{R}\times
    TQ\right)\right)=T^*\left(\mathbb{R}\times
  TQ\right)=\mathbb{R}\times\mathbb{R}\times T^*TQ.
\]
By fixing $U'\subset Q$ a coordinate neighborhood, we have that on $U:=\mathbb{R}\times TU'$ the EDS $\cI$ is generated by the collection of forms $\theta^i:=\dif q^i-\dot{q}^i\dif t$; 
therefore the bundle $I\subset T^*\left(\mathbb{R}\times TQ\right)$ will have the fibers
\[
\left.I\right|_{\left(t,q,\dot{q}\right)}=\left\{\sum_{j=1}^na_j\theta^j:a_i\in\mathbb{R}^n\text{ and }i=1,\cdots,n\right\}.
\]
Thus $I^1=K^1$ (the elements in $I$ are of first degree in the generators); moreover
\[
T^*\left(\mathbb{R}\times TQ\right)\subset Z_1\left(\mathbb{R}\times TQ\right)
\]
implying $\tilde{K}^1=K^1\cap Z_1\left(\mathbb{R}\times TQ\right)=I^1$.
In terms of our local description we obtain that $\left.\tilde{W}^\lambda\right|U=U\times\mathbb{R}^n$, where
\[
\left.\tilde{W}^\lambda\right|_{\left(t,q,\dot{q}\right)}=\left\{-L\dif
  t+\sum_{j=1}^na_j\theta^j:a_i\in\mathbb{R},i=1,\cdots,n\right\}\qquad\forall\left(t,q^i,\dot{q}^i\right)\in U.
\]
The $a$'s in the previous formula have an important meaning concerning the
manifold $\tilde\Lambda:=\mathbb{R}\times\left(TQ\oplus
  T^*Q\right)$ with the $1$-form
$\tilde\lambda\in\Omega^1\left(\tilde\Lambda\right)$ defined through
$\left.\tilde\lambda\right|_{\left(t,v,\alpha\right)}:=-\left[L\left(\alpha\right)-\alpha\left(v\right)\right]\dif
t +\left.\Theta^Q_1\right|_\alpha$ (where $\Theta^Q_1$ is the canonical $1$-form on $T^*Q$), as the following
lemma shows.
\begin{lem}
The map $P_L:\tilde{W}_\lambda\rightarrow\tilde\Lambda$ given locally as
\[
P_L\left(-L\dif t+\sum_{i=1}^na_i\theta^i\right)=\left(t,\sum_{i=1}^n\dot{q}^i\frac{\partial}{\partial
    q^i}\oplus\sum_{i=1}^na_i\dif q^i\right)
\]
is a diffeomorphism, and $P_L^*\tilde\lambda=\tilde{\Omega}$.

\end{lem}
We can work then on the simpler space $\tilde\Lambda=\mR\times\left(TQ\oplus
  T^{*}Q\right)$, where locally\[
\tilde{\lambda}=\sum_{i=1}^np_{i}\left(\dif q^{i}-\dot{q}^{i}\dif t\right)-L\dif t.\]
The canonical Lepage equivalent problem consist in finding sections $\sigma$ of $\mathbb{R}\times\left(TQ\oplus T^*Q\right)$ which extremize
\[
S\left[\sigma\right]=\int_{\mathbb{R}}\sigma^*\left(\sum_{i=1}^np_{i}\left(\dif q^{i}-\dot{q}^{i}\dif t\right)-L\dif t\right);
\]
the equations for an stationary section $\sigma:t\mapsto\left(q\left(t\right),\dot{q}\left(t\right),p\left(t\right)\right)$
are then\[
\begin{cases}
\dif q^{i}-\dot{q}^{i}\dif t=0\\
\dif p_{i}-L_{qi}\dif t=0\\
\left(p_{i}-L_{\dot{q}^
i}\right)\dif t=0.\end{cases}\]
We identify here the Hamilton-Pontryaguin variational principle.
\end{example}

\begin{example}\label{MultiSympFieldTheory}
The standard first order field theories deals with sections $\sigma$ of a bundle $\pi:F\rightarrow M$ with typical fiber $V$; a lagrangian density $L\in C^\infty\left(J^1\left(\pi\right)\right)$ allow us to define the action
\begin{align*}
S\left[\sigma\right]&:=\int_M\left(L\circ\mathbf{pr}\sigma\right)\dif x^0\wedge\cdots\wedge\dif x^{n-1}\cr
&=\int_M\left(L\circ j^1\sigma\right)\dif x^0\wedge\cdots\wedge\dif x^{n-1}
\end{align*}
according to considerations previously made. Let $\mathcal{U}\subset F$ be an adapted coordinate chart with coordinates $\left(x^k,u^\beta\right)$; then $\cI$ on $\hat{\mathcal{U}}:=\left.J^1\left(\pi\right)\right|\mathcal{U}$ has the generators
\[
\left\{\theta^\alpha:=\dif u^\alpha-u^\alpha_k\dif x^k:\alpha=1,\cdots,\text{dim}V\right\}\subset\Omega^1\left(\hat{\mathcal{U}}\right).
\]
where the Einstein's summation convention was asummed. 
So in this case we will have
\[
\left.\tilde{W}^\lambda\right|\hat{\mathcal{U}}=\left\{\lambda+\tilde{\beta}^{n-1}_{\alpha}\wedge\theta^{\alpha}:\tilde{\beta}^{n-1}_{\alpha}\in Z_0^{n-1}\left(\hat{\mathcal{U}}\right)\right\};
\]
by using the $F_\lambda$ map it can be written
\[
\left.\tilde{W}^\lambda\right|\hat{\mathcal{U}}\simeq\hat{\mathcal{U}}\oplus\left[\bigwedge^{n-1}\left(T^*\mathcal{U}_0\right)\right]^{\oplus\text{dim}V},
\]
(where $\mathcal{U}_0:=\pi\left(U\right)$) and this is the local description that we will adopt.
Therefore, we can consider the Cartan
form defined on $\tilde{\Lambda}:=\hat{\mathcal{U}}\oplus\mathcal{Z}$
with local coordinates $\left(x^{k};u^{\alpha},u_{k}^{\alpha},m_{\alpha}\right)$; here $\mathcal{Z}:=\left[\bigwedge^{n-1}\left(T^{*}\mathcal{U}_0\right)\right]^{\oplus\text{dim}V}$
denotes the set of $\text{dim}V$ forms of degree $n-1$ on $M$. So
we can write\[
\tilde{\lambda}:=m_{\alpha}\wedge\left(\dif u^{\alpha}-u_{k}^{\alpha}\dif x^{k}\right)-L\dif x^{0}\wedge\cdots\wedge\dif x^{n-1}\]
for the Cartan form. If the variations are $\left(0;\delta u^{\alpha},\delta u_{k}^{\alpha},\delta m_{\alpha}\right)$
then the Hamilton-Cartan equations reads\begin{multline*}
\left(0;\delta u^{\alpha},\delta u_{k}^{\alpha},\delta m_{\alpha}\right)\lrcorner\dif\tilde{\lambda}=0\Rightarrow\\
\Rightarrow\left(\frac{\partial L}{\partial u^{\alpha}}\delta u^{\alpha}+\frac{\partial L}{\partial u_{k}^{\alpha}}\delta u_{k}^{\alpha}\right)\dif x^{0}\wedge\cdots\wedge\dif x^{n-1}+\delta m_{\alpha}\wedge\left(\dif u^{\alpha}-u_{k}^{\alpha}\dif x^{k}\right)-\\
-\delta u^{\alpha}\dif m_{\alpha}- m_{\alpha}\delta u_{k}^{\alpha}\dif x^{k}=0.\end{multline*}
Now we can define the collection of $(n-1)$-forms $\left(\Gamma_{\alpha}\right)$ on $M$ via \[
\frac{\partial L}{\partial u_{k}^{\alpha}}\dif x^{0}\wedge\cdots\wedge\dif x^{n-1}=\Gamma_{\alpha}\wedge\dif x^{k}\qquad\forall\alpha,k,\]
from which we obtain that the solutions \[
x^{k}\mapsto\left(x^{k};u^{\alpha}\left(x\right),u_{k}^{\alpha}\left(x\right),m_{\alpha}\left(x\right)\right)\]
 must satisfies\[
\begin{cases}
\Gamma_{\alpha}-m_{\alpha}=0\\
\dif u^{\alpha}-u_{k}^{\alpha}\dif x^{k}=0\\
\dif m_{\alpha}-\left(\frac{\partial L}{\partial u^{\alpha}}\right)\dif x^{0}\wedge\cdots\wedge\dif x^{n-1}=0.\end{cases}\]
This system projects onto the local version of Euler-Lagrange equations for first order
field theory. The space of (multi)momenta is locally $\hat{\mathcal{U}}\oplus\mathcal{Z}$.
\end{example}

\begin{example}[Electromagnetism - Cont.]
The first true hamiltonian non standard theory which we want to deal with
is electromagnetism; in this case we have that $\tilde{\Lambda}:=\bigwedge^{2}\left(T^{*}M\right)\oplus\Lambda$
and $\tilde{\lambda}=p_{1}^{*}\left(\ast\Theta_{2}\right)\wedge\left(\dif p_{3}^{*}\Theta_{1}-p_{2}^{*}\Theta_{2}\right)+\lambda$,
being $p_{i}$ the projection in the $i$-th summand of $\tilde{\Lambda}$ and $\lambda$ the pullback of the corresponding form on $\Lambda$.
As before, let us suppose that we indicate by $\left(P,F,A\right)\in\tilde{\Lambda}$
an element of this (multi)phase space; then a section $\sigma:x\mapsto\left(P\left(x\right),F\left(x\right),A\left(x\right)\right)$
will be a solution for the eqs \ref{eq:HamEqElectro1} iff\[
\begin{cases}
F=\dif A\\
P=\ast F\\
\dif P=0.\end{cases}\]
Thus the space of momenta coincides
with the space of velocities.
\end{example}

\begin{example}[Poisson sigma models]\label{PoissonSigmaExample}
The Poisson sigma model \cite{Cattaneo:2001bp,Bojowald:2003pz,Bonechi:2003hd,Calvo07}
can be analysed from this point of view. This is an interesting system, because its usual description (cf. the references above) is non standard, in the sense that it deals with variations without a prolongation structure. Concretely, for a surface
$\Sigma$ and a Poisson manifold $\left(M,\pi\right)$ we have the
fibration\[
\tilde\Lambda:=\bigwedge^2\left(T^{*}\left(M\times\Sigma\right)\right)\rightarrow\Sigma\times M\rightarrow\Sigma.\]
Then the open set $\mathcal{S}:=Z_1^2\left(\Sigma\times M\right)-Z_0^2\left(\Sigma\times M\right)$ is a subbundle; let us call $\nu:\mathcal{S}\rightarrow\Sigma$ the restriction of the above projection. The non standard variational problem equivalent to the Poisson sigma model is the triple $\left(\mathcal{S}\stackrel{\nu}{\longrightarrow}\Sigma,0,\tilde{\lambda}\right)$, where in the local coordinates $\left(\xi^\alpha,x^\mu,\eta_{\beta\nu}\right)$ associated to a coordinate system on $M\times\Sigma$ via
\[
\alpha\in\left.\mathcal{S}\right|_{\left(x,\xi\right)}\mapsto-\eta_{\beta\nu}\dif\xi^\beta\wedge\dif x^\nu,
\]
the $2$-form $\tilde{\lambda}$ reads
\[
\left.\tilde{\lambda}\right|_{\left(\xi,x,\eta\right)}:=-\eta_{\beta\nu}\dif x^\nu\wedge\dif\xi^\beta+\frac{1}{2}\pi^{\mu\nu}\eta_{\alpha\mu}\eta_{\beta\nu}\dif\xi^{\alpha}\wedge\dif\xi^{\beta}.
\]
The Poisson structure on $M$ defines a bundle map $\Pi:T^*M\rightarrow TM$ and therefore induces another bundle map (which we denote with the same symbol) $\Pi:T^*\left(M\times\Sigma\right)\rightarrow TM\times T^*\Sigma$; then the form $\tilde{\lambda}$ can be written in global terms as
\[
\left.\tilde\lambda\right|_\eta=\left.\Theta_2\right|_\eta+\frac{1}{2}\left<\Pi\left(\eta\right)\stackrel{\wedge}{,}\eta\right>
\]
where $\left<\cdot,\cdot\right>$ is the pairing of an element of $TM\times T^*\Sigma$ with an element of $T^*\left(M\times\Sigma\right)\simeq T^*M\times T^*\Sigma$ and $\Theta_2$ is the canonical $2$-form (restricted to $\mathcal{S}$).
The local generators of the Hamilton-Cartan EDS are in this case\[
\begin{cases}
\dif x^{\mu}-\pi^{\mu\nu}\eta_{\alpha\nu}\dif\xi^{\alpha},\\
\dif\eta_{\alpha\sigma}\wedge\dif\xi^\alpha-\frac{1}{2}\left(\partial_{\sigma}\pi^{\mu\nu}\right)\eta_{\alpha\mu}\eta_{\beta\nu}\dif\xi^{\beta}\wedge\dif\xi^\alpha.\end{cases}\]
Thus these systems are naturally formulated in a multisymplectic space.
\end{example}

\section{Slicings and presymplectic structures\label{sec:Introduction-to-Initial}}

Given a Lepagean variational problem (that is, a non standard problem with trivial prolongation structure) and a decomposition of the underlying bundle in a family of hypersurfaces (i.e., the constant time slices), the set of sections of a fixed slice can be dressed up with a presymplectic structure and a hamiltonian, under convenient regularity hypothesis; these data yields through Gotay-Nester algorithm to a set of constraints on the sections under consideration. The procedure is explained with some extent in the following paragraphs; additionally, it is settled the relationship between the extremals of the variational problem and the solutions of the Hamilton eqs associated to the hamiltonian and the presymplectic structure on the space of sections of the fixed slice. Finally, it is shown how the usual Dirac theory of constraints for a field theory can be derived in this setting.

\subsection{Preliminaries}

The purpose now is to set the framework allowing us to relate the solutions of the Hamilton eqs on $n-1$-sections
with regular and ordinary $n-1$-integral elements of some associated
EDS. The initial data is a bundle $F\stackrel{\pi}{\longrightarrow}B$
and a $n$-form $\lambda$ ($n=\mbox{dim}B$) on $F$. Then the Euler-Lagrange
eqs for the action\[
S_{\lambda}\left[\sigma\right]:=\int_{B}\sigma^{*}\lambda\]
characterizes the sections $\sigma:B\rightarrow F$ which extremizes
it. These sections are integral for the EDS (we assume that $\partial B=0$)
\begin{equation}\label{EDSconK}
\cI:=\left\langle Z\lrcorner\dif\lambda:Z\in\Gamma\left(VF\right)\right\rangle _{\text{diff}}.
\end{equation}
We would like to introduce a presymplectic manifold with a hamiltonian, whose solution curves are in one-to one correspondence with the integral sections of $\cI$. In this vein it is necessary to introduce the concept of \emph{slicings}.

\subsection{Slicings}

The first thing that we need \cite{Gotay:2004ib} is a \emph{compatible slicing} $\left(s_{F},s_{B}\right)$
of the bundle of fields: It consist of a pair of diffeomorphisms\begin{eqnarray*}
s_{F}:\mR\times K & \rightarrow & F\\
s_{B}:\mR\times\Sigma & \rightarrow & B\end{eqnarray*}
making commutative the diagram\[
\begin{diagram}
\node{\mR\times K}\arrow{s,l}{p_2}\arrow{e,t}{s_{F}}\node{F}\arrow{s,r}{p_2}\\
\node{\mR\times\Sigma}\arrow{e,b}{s_{B}}\node{B}
\end{diagram}
\]
\begin{note}
Although not strictly necessary, we will consider that the factor $\Sigma$ in the decomposition of the base manifold is compact. This makes sense to definitions like \ref{eq:DefHam} and \ref{eq:DefFormSimp} below, and turns the $2$-form automatically closed.
\end{note}
Let us define $F_{\tau}:=s_{F}\left(\left\{ \tau\right\} \times K\right)$
and $\Sigma_{\tau}:=s_{B}\left(\left\{ \tau\right\} \times\Sigma\right)$
for each $\tau\in\mR$. A useful consequence of this property is that
the space of sections $\Gamma\left(F\right)$ admits a trivialization;
in fact, defining for each $\tau\in\mR$ the embedding\[
i_{\tau}:\Sigma\rightarrow B:x\mapsto s_{B}\left(\tau,x\right)\]
we have that the map\begin{equation}
s_{\Gamma\left(F\right)}:\mR\times\Gamma\left(K\right)\rightarrow\Gamma\left(F\right):\left(\tau,\sigma_{K}\right)\mapsto s_{F}\circ\sigma_{K}\circ\left(i_{\tau}\right)^{-1}\label{eq:SlicingSections}\end{equation}
makes it works (it is well-defined because of the compatibility of the slicing, meaning
in particular that $\mbox{Im}\left(\left.s_{F}\right|\Sigma_{\tau}\right)\subset F_{\tau}.$)

\subsection{(Pre)symplectic structure on the space $\Gamma\left(K\right)$}

For each $\tau\in\mR$ we have the diffeomorphism $i_{\tau}:K\rightarrow F_{\tau}:x\mapsto s_{F}\left(\tau,x\right)$
so we can define \cite{Gawedzki:1990jc,CrnkovicWitten2661} the following $\tau$-family of forms on $K$\begin{eqnarray*}
\Omega_{\tau}: & = & i_{\tau}^{*}\left(\left.\dif\lambda\right|F_{\tau}\right)\in\Omega^{n+1}\left(K\right)\\
\mathcal{H}_{\tau}: & = & i_{\tau}^{*}\left(\left.\partial_{0}\lrcorner\lambda\right|F_{\tau}\right)\in\Omega^{n-1}\left(K\right)\end{eqnarray*}
where the symbol $\partial_{0}$ is used in order to denote the canonical vector
field associated to the $\tau$-direction in both spaces. These forms
can be used to define the following data on $\Gamma\left(K\right)$
for each $\tau\in\mR$: The $2$-form $\omega_{\tau}\in\Omega^{2}\left(\Gamma\left(K\right)\right)$
and the function $H_{\tau}\in C^{\infty}\left(\Gamma\left(K\right)\right)$,
defined through\begin{eqnarray*}
\omega_{\tau}|_{\gamma}\left(V_{1},V_{2}\right): & = & \int_{\Sigma}\gamma^{*}\left(\Omega_{\tau}\left(\hat{V}_{1},\hat{V}_{2}\right)\right)\\
H_{\tau}\left(\gamma\right): & = & \int_{\Sigma}\gamma^{*}\left(\mathcal{H}_{\tau}\right).\end{eqnarray*}
Here $\gamma\in\Gamma\left(K\right)$ and $V_{1},V_{2}\in\Gamma\left(\gamma^{*}\left(VF\right)\right)\equiv T_{\gamma}\Gamma\left(K\right)$;
as before, $\hat{V}\in\Gamma\left(VK\right)$ denotes an extension
to a vector field on $K$ for the element $V\in\Gamma\left(\gamma^{*}\left(VK\right)\right).$
With this data we can define a Hamiltonian system on $\Gamma\left(K\right)$.

\subsection{Forms on $\Gamma\left(K\right)$ and forms on $K$}

We want to establish a useful result concerning forms on a space
of sections; that is, by taking on a bundle $F\stackrel{\pi}{\longrightarrow}B$
an $n+k$-form $\phi\in\Omega^{n+k}\left(F\right),k\in\mathbb{Z},n=\mbox{dim}B,$
we can define the following $k$-form $\Phi$ on $\Gamma\left(F\right)$
via\begin{equation}\label{DefPhiForm}
\Phi|_{\sigma}\left(V_{1},\cdots,V_{k}\right):=\int_{B}\sigma^{*}\left(\hat{V}_{1}\lrcorner\cdots\hat{V}_{k}\lrcorner\phi\right).\end{equation}

\begin{defn}
For each $k\in\mathbb{Z}$ define the bundle $Z^{n+k}_{k}\left(F\right)\stackrel{\bar{\tau}_{F}^{n+k}}{\longrightarrow}F$
such that the fiber on $e\in F$ is given by\[
\left.Z^{n+k}_{k}\left(F\right)\right|_{e}:=\left\{ \alpha\in\bigwedge^{n+k}\left(T_{e}^{*}F\right):V_{1}\lrcorner\cdots V_{k+1}\lrcorner\alpha=0\quad\forall V_{1},\cdots,V_{k+1}\in V_{e}F\right\}. \]This set can be called the \emph{space of $k$-semibasic $n+k$-forms.} We will say that a nonzero form $\alpha$ belongs to $Z_k^\bullet$ \emph{properly} iff $\alpha\in Z_k^\bullet$ but $\alpha\notin Z_{k-1}^\bullet\left(F\right)$.

\end{defn}
Then we can prove the next result.

\begin{prop}
\label{pro:phiPhiMono}Let us suppose that $\phi\in\Gamma\left(Z^{n+l}_{k}\left(F\right)\right)$ properly ($l\leq k$), and let $\Phi$ be as in Eq. \ref{DefPhiForm}.
Then $\Phi=0$ iff $\phi=0$.
\end{prop}
\begin{note}
The previous proposition can be rephrased as follows: The map $\phi\mapsto\Phi$
is injective when restricted to $\Gamma\left(Z^{n+l}_{k}\left(F\right)\right)$.
\end{note}
\begin{proof}
It is clear that $\phi=0$ implies $\Phi=0$. For the converse implication, $\Phi=0$ implies that $\int_B\sigma^*\left(\hat{V}_1\lrcorner\cdots\hat{V}_l\lrcorner\phi\right)=0$ for all $\sigma\in\Gamma\left(F\right)$ and $V_1,\cdots,V_l$ vertical vectors defined on $\text{Im}\,\sigma$; therefore  $\hat{V}_1\lrcorner\cdots\hat{V}_l\lrcorner\phi=0$ for all collection of $l$ vertical vectors, and it means that $\phi\in Z_{l-1}\left(F\right)$. Therefore $\phi=0$.
\end{proof}
For each $\tau\in\mR$ we can define the following EDS on $K$\begin{eqnarray}
\cI_{\tau}: & = & i_{\tau}^{*}\left(\left.\cI\right|F_{\tau}\right),\label{eq:IsubTau}\end{eqnarray}
a fundamental object in the subsequent sections, because it is closely
related to the Dirac constraints of the theory.
\\
Finally we state a very important result, because it relates the solutions (if any) of the presymplectic dynamical system with the extremals of the underlying variational problem.

\begin{prop}
\label{pro:HamEqs-and-EDS}Let $\lambda$ be a $n$-form on $F$ such that $\lambda\in Z_1^n\left(F\right)$ properly. Then for each $\tau\in\mR$, a pair $\left(\sigma_{\tau},X_{\tau}\right)\in T_{\sigma_{\tau}}\Gamma\left(K\right)$
satisfies the Hamilton equations for sections\begin{equation}
\left.\left(X_{\tau}\lrcorner\omega_{\tau}\right)\right|_{\sigma_{\tau}}=\left.\dif H_{\tau}\right|_{\sigma_{\tau}}\label{eq:HamEqonSections}\end{equation}
iff it satifies the following equation\begin{equation}
\sigma_{\tau}^{*}\left(V\lrcorner\left(X_{\tau}\lrcorner\Omega_{\tau}-\dif\mathcal{H}_{\tau}\right)\right)=0,\qquad\forall V\in\Gamma\left(\sigma_{\tau}^{*}\left(VK\right)\right).\label{eq:HamEqInterm}\end{equation}

\end{prop}
\begin{proof}\label{PropUnoaUno}
The key is to realize that the $n$-form $X_{\tau}\lrcorner\Omega_{\tau}-\dif\mathcal{H}_{\tau}$
belongs to $
Z^n_{2}\left(F\right).$
On the other side it was shown in proposition \ref{pro:phiPhiMono}
that there are no non zero $n$-forms $\beta$ in $Z^n_2\left(F\right)$ such that the map $V\mapsto\int_{\Sigma}\sigma^*\left(V\lrcorner\beta\right)$
is zero.
\end{proof}

The meaning of this proposition must be clarified: It says that the pair $\left(\sigma_\tau,X_\tau\right)$ is a solution of the Hamilton equations on $\Gamma\left(K\right)$ defined by $\omega_\tau$ and $H_\tau$ if and only if $V_\tau:=\partial_0+X_\tau$ belongs to the polar space of the $n-1$-integral element $\sigma_{\tau*}\left(T\Sigma\right)$ for the Hamilton-Cartan EDS in this context, Eq. \ref{EDSconK}. Then if there exists a regular flag for this EDS with last terms $0\subset\cdots\subset\sigma_{\tau*}\left(T\Sigma\right)\subset\left<\sigma_{\tau*}\left(T\Sigma\right),V_\tau\right>$, the $n$-integral manifold passing through it will generate solutions for the Hamilton equations, and conversely. The following sections will use this remarkable relation.

\subsection{Presymplectic structures on a multimomentum space}

Let us now apply the setting of the previous section to our multisymplectic space. In particular, let us suppose that there exists a compatible slicing of the bundle $\tilde\Lambda\rightarrow M$: So besides of the spacetime decomposition $M=\mR\times\Sigma$ (as above, it will be assumed that $\Sigma$ is a compact manifold),
we have a decomposition $\tilde{\Lambda}=\mR\times\tilde{L}$ such
that $\tilde{\Lambda}_{\tau}=\left\{ \tau\right\} \times\tilde{L}$
for every $\tau\in\mR$; thus $p:\tilde{L}\rightarrow\Sigma$ is a fibration
that makes commutative the following diagram \[
\begin{diagram}
\node{\mR\times\tilde{L}\simeq\tilde{\Lambda}}\arrow{s,l}{p_2}\arrow{e,t}{\text{id}\times p}\node{\mR\times\Sigma\simeq M}\arrow{s,r}{p_2}\\
\node{\tilde{L}}\arrow{e,b}{p}\node{\Sigma}
\end{diagram}
\]In such a case (cf. the previous section) we can define
a presymplectic structure on the space of sections $\Gamma\left(\tilde{L}\right)$
together with a function on it such that the solutions of the classical
mechanics system defined by these data induce solutions for the EDS
generated by \ref{eq:EDSHam}. That is, if we define\[
i_{\tau}:\tilde{L}\rightarrow\tilde{\Lambda}_{\tau}:l\mapsto\left(\tau,l\right)\]
 we can use these structures to define the forms
\begin{align}
\mathcal{H}_{\tau}: & =i_{\tau}^{*}\left[\left(\partial_{0}\lrcorner\tilde{\lambda}\right)|\tilde{\Lambda}_{\tau}\right]\label{eq:DefHam}\\
\Omega_{\tau}: & =i_{\tau}^{*}\left[\left(\dif\tilde{\lambda}\right)|\tilde{\Lambda}_{\tau}\right];\label{eq:DefFormSimp}
\end{align}
so, we have $H_{\tau}\in C^{\infty}\left(\Gamma\left(\tilde{L}\right)\right)$
and $\omega_{\tau}\in\Omega^{2}\left(\Gamma\left(\tilde{L}\right)\right)$
(which preserves a $\tau$-dependence) defined by\begin{align}
H_{\tau}\left(\sigma\right): & =\int_{\Sigma}\sigma^{*}\left(\mathcal{H}_{\tau}\right),\label{HamiltonianoSobreSecciones}\\
\omega_{\tau}|_{\sigma}\left(\tilde{X}_{\tau},\tilde{Y}_{\tau}\right): & =\int_{\Sigma}\sigma^{*}\left(\Omega_{\tau}\left(\tilde{X}_{\tau},\tilde{Y}_{\tau}\right)\right),\label{FormaSimplecticaSobreSecciones}\end{align}
for every pair $\tilde{X}_{\tau},\tilde{Y}_{\tau}$ of vertical vector
fields that cover $\sigma$, i.e., elements of $T_{\sigma}\Gamma\left(\tilde{L}\right)$.
In this setting the proposition \ref{pro:HamEqs-and-EDS} is true, and we have to our disposal (under adequate regularity conditions, i.e., if the foliation induced by the slicing is regular in the sense of definition \ref{RegularityForFoliationsDefn} below) the identification between the dynamical equations on the space of sections and the Hamilton-Cartan EDS.
\\
Now it will be necessary to introduce further simplifications; in particular, we take
$\mathcal{L}_{\partial_{0}}\tilde\lambda=0$. Using the slicing introduced on $\tilde{\Lambda}$
and the Hamilton-Cartan EDS, given by \ref{eq:EDSHam}, we can define
a parameterized EDS $\cI_{\tau}$ according to the Eq. \ref{eq:IsubTau}.
Under this simplification it is true the following result. 

\begin{lem}
If $\mathcal{L}_{\partial_{0}}\tilde{\lambda}=0$, then $H_{\tau}=H_{\tau'}=:H$
and $\omega_{\tau}=\omega_{\tau'}=:\omega$ for all $\tau,\tau'\in\mR$.
Moreover, the EDS $\cI_{\tau}$ does not depends on $\tau$. 
\end{lem}
In particular, this lemma allow us to choose a slice to work with.
\\
Finally we want to show how the usual theory of constraints in a field theory can be obtained from this scheme. In the following example we set up the relation between the presymplectic structures defined above and the symplectic structures defined in the usual approach to (first order) field theories.

\begin{example}[First order field theories - Cont]\label{ExampleFirstoOrderDirac}
Our aim is to calculate the dynamical system associated to the multisymplectic structure for field theory found in example \ref{MultiSympFieldTheory}; locally we have that $\tilde\lambda=m_\alpha\wedge\left(\dif u^\alpha-u^\alpha_k\dif x^k\right)-L\dif x^0\wedge\cdots\wedge\dif x^{n-1}$, so
\[
\dif\tilde\lambda=\dif m_\alpha\wedge\left(\dif u^\alpha-u^\alpha_k\dif x^k\right)-\dif L\wedge\dif x^0\wedge\cdots\wedge\dif x^{n-1}-\left(-1\right)^{n-1}m_\alpha\wedge\dif u^\alpha_k\wedge\dif x^k,
\]
and taking into account that $m_\alpha\in\Omega^{n-1}\left(M\right)$, we will have that
\[
\left.\left(u^\alpha_k\dif m_\alpha\wedge\dif x^k\right)\right|_{N_0}=\left.\left(m_\alpha\wedge\dif u^\alpha_k\wedge\dif x^k\right)\right|_{N_0}=0
\]
where $N_0:=\left\{P\in\hat{\mathcal{U}}\oplus\mathcal{Z}:x^0\left(P\right)=0\right\}$. Then if $\check{N}_0:=\left\{m\in M:x^0\left(m\right)=0\right\}$, by using Eq. \ref{FormaSimplecticaSobreSecciones}, the presymplectic form on $\Gamma\left(N_0\right)$ will be
\[
\left.\omega^x\right|_\sigma\left(X,Y\right)=\int_{\check{N}_0}\left(\delta m^0_\alpha\delta v^\alpha-\delta n^0_\alpha\delta u^{\alpha}\right)\dif x^1\wedge\cdots\wedge\dif x^{n-1}
\]
and $X:=\left(0;\delta u^{\alpha},\delta u^{\alpha}_{k},\delta m_\alpha\right),Y:=\left(0;\delta v^{\alpha},\delta v^{\alpha}_{k},\delta n_\alpha\right)$ indicates a pair of tangent vectors at $\sigma\in\Gamma\left(N_0\right)$. By introducing the coordinates
\[
m_\alpha:=m_\alpha^k\dif_kx,
\]
with the symbol $\dif_kx$ defined through
\[
\dif_kx:=\partial_{x^k}\lrcorner\left(\dif x^0\wedge\cdots\wedge\dif x^{n-1}\right)=\left(-1\right)^k\dif x^0\wedge\cdots\wedge\widehat{\dif x^k}\wedge\cdots\wedge\dif x^{n-1},
\]
then we will see that for the restriction of the $n-1$-form
\[
\partial_{x^0}\lrcorner\tilde\lambda=\left(\partial_{x^0}\lrcorner m_\alpha\right)\wedge\left(\dif u^\alpha-u^\alpha_k\dif x^k\right)-L\dif x^1\wedge\cdots\wedge\dif x^{n-1}-\left(-1\right)^{n-1}u^\alpha_0m_\alpha
\]
to $N_0$, it can be proved that
\begin{multline*}
\left.\partial_{x^0}\lrcorner\tilde\lambda\right|_{N_0}=\cr
=\left.\left(\partial_{x^0}\lrcorner m_\alpha\right)\right|_{N_0}\wedge\left(\dif u^\alpha-u^\alpha_A\dif x^A\right)-\left[L-\left(-1\right)^{n}u^\alpha_0m^0_\alpha\right]\dif x^1\wedge\cdots\wedge\dif x^{n-1},
\end{multline*}
where the capital letter indices takes values in the range $1,\cdots,n-1$. Finally
\begin{align*}
\left(\partial_{x^0}\lrcorner m_\alpha\right)\wedge\dif x^B&=m_{\alpha}^A\left(\partial_{x^0}\lrcorner\dif_Ax\right)\wedge\dif x^B\cr
&=\left(-1\right)^{n-1}m_\alpha^B\dif x^1\wedge\cdots\wedge\dif x^{n-1}
\end{align*}
and from Eq. \ref{HamiltonianoSobreSecciones}, the hamiltonian function $H^x\in C^\infty\left(\Gamma\left(N_0\right)\right)$ will becomes
\begin{multline*}
H^x\left(\sigma\right):=\int_{\check{N}_0}\sigma^*\left(\left.\partial_{x^0}\lrcorner\tilde\lambda\right|_{N_0}\right)\cr
=\int_{\check{N}_0}\left\{\left(-1\right)^{n-1}m_\alpha^B\left[\left(\partial_Bu^\alpha\right)-u_B^\alpha\right]-L-\left(-1\right)^{n-1}u_0^\alpha m_\alpha^0\right\}\dif x^1\wedge\cdots\wedge\dif x^{n-1}.
\end{multline*}
We are now ready to make contact with the usual theory of constraints. In order to do that, it is necessary to calculate the primary constraints associated to the dynamical data found above. This can be achieved by realizing that $\text{ker}\,\omega^x=\left\{\left(0;0,\delta
    u^\alpha_k,\delta m^A_\alpha\right)\right\}$, so the primary constraints for the dynamical system will be
\begin{align*}
0&=\dif H^x\left(\delta u^\alpha_k\right)\cr
&=\int_{\check{N}_0}\left[\left(-1\right)^nm_\alpha^k-\frac{\partial
    L}{\partial u_k^\alpha}\right]\delta u^\alpha_k\dif x^1\wedge\cdots\wedge\dif x^{n-1},\cr
0&=\dif H^x\left(\delta m_\alpha^A\right)\cr
&=\int_{\check{N}_0}\left(-1\right)^{n-1}\left[\left(\partial_Bu^\alpha\right)-u_B^\alpha\right]\delta m_\alpha^B\dif x^1\wedge\cdots\wedge\dif x^{n-1}.
\end{align*}
Let us now define the set of primary constraints
\begin{multline*}
C_1:=\Bigg\{\left(-1\right)^nm_\alpha^k-\frac{\partial
    L}{\partial u_k^\alpha},\left(\partial_Bu^\alpha\right)-u_B^\alpha:\cr
:\alpha=1,\cdots\text{dim}V,k=0,\cdots,n-1,B=1,\cdots,n-1\Bigg\};
\end{multline*}
then $M_1\subset\Gamma\left(N_0\right)$, the zero locus of $C_1$, is a presymplectic manifold with hamiltonian, both structures being defined through restriction. Let $\mathcal{N}_0$ be the trivial bundle $\check{N}_0\times\mathbb{R}^m\times\mathbb{R}^{m\left(n-1\right)}\times\mathbb{R}^m\times\mathbb{R}^m$ on $\check{N}_0$ with coordinates $\left(x;\phi^\alpha,\phi^\alpha_B,\phi^\alpha_0,\pi_\alpha\right)$; then the map $\Pi:N_0\rightarrow\mathcal{N}_0$ defined through
\[
\Pi:
\begin{cases}
\phi^\alpha=u^\alpha,\cr
\phi_B^\alpha=u_B^\alpha,\cr
\phi_0^\alpha=u_0^\alpha,\cr
\pi_\alpha=\left(-1\right)^nm_\alpha^0
\end{cases}
\]
has the following properties.
\begin{lem}
Let $L_1\in C^\infty\left(\mathcal{N}_0\right)$ be the unique map such that $L_1\circ\Pi=L$. The set $\tilde{M}_1:=\Pi\circ\left(M_1\right)$ is composed by the sections of $\mathcal{N}_0$ given by
\[
\sigma:x\mapsto\left(x;\phi^\alpha\left(x\right),\phi_B^\alpha\left(x\right),\phi^\alpha_0\left(x\right),\pi_\alpha\left(x\right)\right)
\]
such that
\[
\pi_\alpha=\frac{\partial L_1}{\partial\phi^\alpha_0},\quad\phi_B^\alpha=\frac{\partial\phi^\alpha}{\partial x^B}.
\]
If $\tilde{M}_1$ is a manifold, the the presymplectic form defined through
\[
\omega_1=\int_{\check{N}_0}\dif\pi_\alpha\wedge\dif\phi^\alpha\wedge\dif x^1\wedge\cdots\wedge\dif x^{n-1}
\]
will verify that
$\left(\Pi\circ\right)^*\omega_1=\left.\omega^x\right|_{M_1}$; also the hamiltonian function
\[
H_1\left(\sigma\right)=\int_{\check{N}_0}\sigma^*\left(\phi_0^\alpha\pi_\alpha-L_1\right)\dif x^1\wedge\cdots\wedge\dif x^{n-1}
\]
will satisfy $\left(\Pi\circ\right)^*H_1=\left.H^x\right|_{M_1}$. Due to the injectivity of
$\left(\Pi\circ\right)^*$, the structures are uniquely determined by these requeriments.
\end{lem}
Then the dynamical problem reduces to solve
\[
X_{H_1}\lrcorner\omega_1=\dif H_1
\]
on each
$\sigma:x\mapsto\left(x;\phi^\alpha\left(x\right),\phi^\alpha_B\left(x\right),\phi^\alpha_0\left(x\right),\pi_\alpha\left(x\right)\right)$
such that
\[
\pi_\alpha=\frac{\partial L_1}{\partial\phi^\alpha_0},\quad\phi_B^\alpha=\frac{\partial\phi^\alpha}{\partial x^B},
\]
and this is the basic scheme in the usual theory of constraints for a (first order) field theory; therefore, the primary constraints in the standard approach are related through $\Pi$ with the primary constraints found in the non standard viewpoint.
\end{example}

\section{Constraints and EDS}\label{Section:7}

The aim in this section is to prove the main result of the article: The fact that, under certain regularity hypothesis (cf. definition \ref{RegularityForFoliationsDefn} below), it is possible to describe the set of constraints arising from the application of the Gotay-Nester algorithm on the dynamical data \ref{HamiltonianoSobreSecciones}, \ref{FormaSimplecticaSobreSecciones} as a set of generators of the EDS induced on the chosen slice by the (prolongation of the) Hamilton-Cartan EDS.

\subsection{Useful characterization for constraints}

As we said above, we want to relate the Dirac constraints arising from a field theory (in the non standard setting) with a EDS associated to its eqs of motion. In order to achieve this, it is necessary to adopt a meaningful picture for these constraints. So we use the following description for the constraints obtained through
the Gotay-Nester algorithm \cite{Gotay:2004ib}.

\begin{defn}
Given a presymplectic manifold $\left(N,\omega\right)$ with a hamiltonian
$H\in C^{\infty}\left(N\right)$, the \emph{constraint submanifold}
$C\subset N$ is the maximal submanifold of $N$ according to property\begin{equation}
\dif H\left(TC^{\perp}\right)=0\label{eq:StabilityConditionforSections}\end{equation}
where $TC^{\perp}$ denotes the (perhaps singular) subbundle of $TN$
composed by the symplectic complements of the fibres of the subbundle
$TC$. A submanifold in $M$ satisfying property \ref{eq:StabilityConditionforSections} is said to be \emph{invariant} respect to the dynamics defined by $\left(\omega,H\right)$.
\end{defn}
Because we need to know the tangent spaces for submanifolds $\Gamma\left(\mathcal{K}\right)\subset\Gamma\left(\tilde{L}\right)$ composed of integral sections for
some EDS $\mathcal{K}$, we prove the next lemma.

\begin{lem}\label{lem:TangentSectionsCharacterization}
If the set $\Gamma\left(\mathcal{K}\right),$ with $\mathcal{K}\subset\Omega^{\bullet}\left(\tilde{L}\right)$
an EDS, is a true submanifold in $\Gamma\left(\tilde{L}\right)$,
then its tangent space in $\sigma\in\Gamma\left(\mathcal{K}\right)$
is the vector space\[
T_{\sigma}\Gamma\left(\mathcal{K}\right)=\left\{ X\in\Gamma\left(\sigma^{*}\left(V\tilde{L}\right)\right):\sigma^{*}\left(\mathcal{L}_{\hat{X}}\mathcal{K}\right)=0\right\} ,\]
where the hat denotes some extension to $\mathfrak{X}\left(\tilde{L}\right)$
of an element of $\Gamma\left(\sigma^{*}\left(V\tilde{L}\right)\right)$.
\end{lem}
\begin{proof}
Under our hypothesis, we just need to derive along a curve to obtain
the lemma.
\end{proof}

\subsection{Admissible sections and involution}
We start with a (non standard) variational problem on a bundle $\Lambda\rightarrow M$, where $\text{dim}M=n$; by assuming that there exists a (bivariant) Lepage-equivalent problem, we pass to a non restricted variational problem on a new bundle $\tilde{\Lambda}\rightarrow M$. The variational equations defines here an EDS $\cI\subset\Omega^{\bullet}\left(\tilde{\Lambda}\right)$; the integral sections for this EDS are the classical solutions to our field theory. So we have the EDS \[
\left(\cI,\Omega:=\dif x^{1}\wedge\cdots\wedge\dif x^{n}\right)\]
 generated by the Hamilton-Cartan equations, i.e.
\begin{equation}\label{eq:EqMotionCharacterizationEDS}
\cI:=\left\langle V\lrcorner\dif\tilde{\lambda}:V\in\Gamma\left(V\tilde{\Lambda}\right)\right\rangle _{\text{diff}},
\end{equation}
and by using the Cartan-Kuranishi theorem we found a bundle $\tilde{\Lambda}'\rightarrow M$,
a submersion $\Pi:\tilde{\Lambda}'\rightarrow\tilde{\Lambda}$ and
an EDS $\left(\cI',\Omega'\right)\subset\Omega^{\bullet}\left(\tilde{\Lambda}'\right)$
which is $n$-involutive and verifies that\[
\Pi_{*}\left(V_{n}\left(\cI',\Omega'\right)\right)=V_{n}\left(\cI,\Omega\right).\]

\begin{rem}It is important to pointing out the following facts.
\begin{itemize}
\item The original EDS does not work if there exists $n-1$-integral manifolds
on each $\tau$-slice whose possible thickening directions rests in
the same slice. The way to avoid this annoying fact is to select
in $V_{n}\left(\cI\right)$ only those integral elements satisfying
the independence condition; this subset of sections can be described
as $n$-integral submanifolds for an involutive EDS $\cI'$, obtained
from $\cI$ by performing enough prolongations.
\item It is assumed that there are no $0$-forms to take care of in $\cI$ and
  $\cI'$; if it occurs, one must to include them into the bundles
  $\tilde\Lambda$ and $\tilde{\Lambda}'$, that is, redefine these sets by taking
  into account that the new $0$-forms annihilates on them. The corresponding
  EDSs are the pullback of the original ones.
\item We know that $\tilde{\Lambda}\rightarrow M$ admits a compatible slicing; we will suppose further that the
same is true for $\tilde{\Lambda}'\rightarrow M$ with $\tilde{\Lambda}'\simeq\mR\times\tilde{L}'$. This diffeomorphism will be denoted as $s_{\tilde{\Lambda}'}$.
Moreover, it will be assumed the existence of a submersion $\Pi_{0}:\tilde{L}'\rightarrow\tilde{L}$
such that the following diagram is commutative\[
\begin{diagram}
\node{\tilde{\Lambda}'}\arrow{s,l}{p_2}\arrow{e,t}{\Pi}
\node{\tilde{\Lambda}}\arrow{s,r}{p_2}\\
\node{\tilde{L}'}\arrow{e,b}{\Pi_0}\node{\tilde{L}}
\end{diagram}
\]
\item For each $\tau\in\mathbb{R}$ it is tempting to define the EDS
\[
\cI_\tau':=i_\tau^*\left(\left.\cI'\right|\tilde{\Lambda}'_\tau\right)\subset\Omega^{\bullet}\left(\tilde{L}'\right)
\]
where $\tilde{\Lambda}'_\tau:=i_\tau\left(\{\tau\}\times\tilde{L}'\right)$ and
\[
i_\tau:\tilde{L}'\hookrightarrow\tilde{\Lambda}':l\mapsto s_{\tilde{\Lambda}'}\left(\tau,l\right),
\]
claiming that\begin{equation}\label{TheNonTrueShit}
\Gamma_{n-1}\left(\cI'_{\tau},\omega'\right)=i_{\tau}^{*}\left(\Gamma_{n}\left(\cI',\Omega'\right)\right)\end{equation}
where $\omega':=\partial_0\lrcorner\Omega'$.
It is obvious that $\Gamma_{n-1}\left(\cI'_{\tau},\omega'\right)\supset i_{\tau}^{*}\left(\Gamma_{n}\left(\cI',\Omega'\right)\right)$. The $n$-involutivity of $\cI'$ ensures that every $n$-integral element is tangent to some solution; however, the opposite inclusion $\Gamma_{n-1}\left(\cI'_{\tau},\omega'\right)\subset i_{\tau}^{*}\left(\Gamma_{n}\left(\cI',\Omega'\right)\right)$ is achieved if every $n-1$-integral element of $\cI'_\tau$ can be thickened out (in a regular fashion) to a $n$-integral element for $\cI'$. The requeriment of involutivity is not enough; instead, we need the regularity of every $n-1$-integral element of $\cI'_\tau$.
\end{itemize}
\end{rem}


\begin{note}[System where \ref{TheNonTrueShit} is not verified]
There exists an almost canonical counterexample to this claim: The EDS describing orthogonal coordinates in three dimensions. See example 3.2 and theorem 3.3 in \cite{BCG}.
\end{note}

Due to the previous discussion, we need to introduce a new definition.

\begin{defn}\label{RegularityForFoliationsDefn}
Let $\cI$ be an involutive EDS on $M$ and let us suppose that $M\simeq\mathbb{R}\times N$ for some manifold $N$; as above, for each $\tau\in\mathbb{R}$ let us define the following EDS on $N$
\[
\cI_\tau:=i_\tau^*\left(\left.\cI\right|\{\tau\}\times N\right).
\]
We say that the induced foliation is \emph{regular with respect to the EDS $\cI$} if every $n-1$-integral element of $\cI_\tau,\tau\in\mathbb{R},$ is regular. A slicing inducing a regular folitation will be called \emph{regular}.
\end{defn}

The regularity notion for a foliation ensures that one can extend any $n-1$-integral manifold for the EDS contained in a leaf to a $n$-integral manifold; for suitable independence conditions the required involutivity ensures that the extension is not included in that leaf. So we can state the following result.

\begin{prop}\label{Prop:RealTrueShit}
If $\cI$ is an involutive EDS on $P$, and $s_P:P\stackrel{\sim}{\longrightarrow}Q\times\mathbb{R}$ is a regular slicing for $\cI$, then
\[
\Gamma_{n-1}\left(\cI_{\tau},\omega\right)=i_{\tau}^{*}\left(\Gamma_{n}\left(\cI,\Omega\right)\right),
\]
where $\omega:=\partial_0\lrcorner\Omega$ and $\partial_0$ is the vector field pointing in the $\mathbb{R}$-direction of the slicing.
\end{prop}
\begin{proof}
The hard inclusion is
\[
\Gamma_{n-1}\left(\cI_{\tau},\omega\right)\subset i_{\tau}^{*}\left(\Gamma_{n}\left(\cI,\Omega\right)\right).
\]
Because of the regularity, it follows from the Cartan-K\"{a}hler theorem that any $n-1$-integral section for $\cI_\tau$ can be extended to a $n$-integral section for $\cI$.
\end{proof}
It remains to show some revelant properties of the solutions of an EDS. In
order to do that, it is important to establish the following fact.

\begin{lem}\label{ProofPropertySols}
Let $\cI$ be an EDS on the manifold $M$, $N\subset M$ an integral
submanifold and $X\in\Gamma\left(T_NM\right)$ such that
\[
\left.\left(X\lrcorner\alpha\right)\right|N=0
\]
for all $\alpha\in\cI$. If $\hat{X}\in\mathfrak{X}\left(M\right)$ is any
extension for $X$, then
\[
\left.\left(\mathcal{L}_{\hat{X}}\cI\right)\right|N=0.
\]
\end{lem}
\begin{proof}
If $i:N\hookrightarrow M$ is the canonical injection, we have that
\[
i^*\left(\hat{X}\lrcorner\alpha\right)=i^*\left(X\lrcorner\alpha\right),
\]
so
\begin{eqnarray*}
\left.\left(\mathcal{L}_{\hat{X}}\alpha\right)\right|N&=&i^*\left(\hat{X}\lrcorner
\dif\alpha+\dif\left(\hat{X}\lrcorner\alpha\right)\right)\cr
&=&i^*\left(X\lrcorner
\dif\alpha\right)+\dif i^*\left(X\lrcorner\alpha\right)\cr
&=&0
\end{eqnarray*}
for all $\alpha\in\cI$.
\end{proof}
By using the definition \ref{eq:IsubTau} we obtain an EDS $\left(\cI_{\tau}',\omega':=\partial_{0}\lrcorner\Omega'\right)$
on $\tilde{L}'$. As we will see in the next sections, this exterior system
gives another description for the Dirac constraints, namely, as (algebraic)
generators of an EDS.

\subsection{The constraint submanifold as an space of sections}

\subsubsection{An outline}
Here we describe the essential elements involved in proving that the constraint submanifold
coincides with the set of maps $\Pi_{0}\circ\Gamma_{n-1}\left(\cI'_{\tau},\omega'\right)$.
Because this proof uses the characterization of $C$ as the maximal
invariant submanifold of the set of section of the bundle $\tilde{L}$,
it is necessary to prove that

\begin{enumerate}
\item $\Pi_{0}\circ\Gamma_{n-1}\left(\cI'_{\tau},\omega'\right)$ is invariant,
and
\item $C\subset\Pi_{0}\circ\Gamma_{n-1}\left(\cI'_{\tau},\omega'\right)$.
\end{enumerate}
It seems that the first condition is consequence of the fact that
the original PDEs \ref{eq:HamEqInterm} are satisfied by the elements
of this set. The second condition can be verified by building a $n$-section
from a $n-1$-section in $C$ via the formula \[
\sigma:\left(\tau,x\right)\mapsto\left(\tau,\sigma_{\tau}\left(x\right)\right)\]
realizing that it is an integral section for $\left(\cI',\Omega'\right)$,
and then restricting
to a $\tau$-slice.
It was previously stated that a submanifold $Q\subset P$ in our phase space $\Gamma\left(\tilde{L}\right)$ is \emph{invariant} iff 
\[
\dif H\left(TQ^{\perp}\right)=0.
\]
According to a result of Marsden \emph{et al} \cite{Gotay:2004ib}, the invariance property is equivalent to the existence of tangent solutions to the Hamilton eqs restricted to $Q$; that is, for every $q\in Q$ the system
\[
\left.\left(X_H\lrcorner\omega\right)\right|_q=\left.\dif H\right|_q
\]
has a solution $\left.X_H\right|_q$ belonging to $T_qQ$. In the first part of the proof the construction is performed in order to ensure that the submanifold defined by the EDS admits tangent solutions to the Hamilton eqs, implying the invariance. In the second part the strategy is to prove that our submanifold is maximal with respect to the property of invariance; now the characterization of invariance is used in the opposite direction, that is, it is initially supposed that certain submanifold is invariant, and by using the characterization, arriving to the conclusion that the Hamilton eqs. admits tangent solutions to this submanifold. Finally it can be related to the manifold defined by the EDS because of the interpretation of the Hamilton eqs as generators of the EDS.
\newline
After this warm-up, we are then ready to formulate the theorem with its proof.

\subsubsection{The main theorem and its proof}

There are several elements to be included in the final proof of the main theorem of the paper. In the previous paragraphs it was argued the necessity of having a submersion $\Pi:\tilde{\Lambda}'\rightarrow\tilde{\Lambda}$ between bundles fitting in the diagram
\[
\begin{diagram}
\node{\tilde{\Lambda}'}\arrow[2]{e,t}{\Pi}\arrow{se,b}{}\node[2]{\tilde{\Lambda}}\arrow{sw,b}{}\\
\node[2]{M}
\end{diagram}
\]
Next it is assumed that there exists a compatible (with respect to the both bundle structures) slicing of $M$ which is well-behaved with the maps in this diagram. That is, there exists a map $\Pi_0:\tilde{L}'\rightarrow\tilde{L}$ making commutative the following
\[
\begin{diagram}
\node{\tilde{\Lambda}'}\arrow{e,t}{s_{\tilde{\Lambda}'}}\arrow{s,l}{\Pi}\node{\tilde{L}'\times\mathbb{R}}\arrow{s,r}{\Pi_0\times\text{id}}\\
\node{\tilde{\Lambda}}\arrow{e,b}{s_{\tilde{\Lambda}}}\arrow{s,l}{}\node{\tilde{L}\times\mathbb{R}}\arrow{s,r}{}\\
\node{M}\arrow{e,b}{s_M}\node{\Sigma\times\mathbb{R}}
\end{diagram}
\]
The slicing on $\tilde{\Lambda}'$ allows us to define the EDS
\[
\cI_\tau':=i_\tau^*\left(\left.\cI\right|\tilde{L}'\times\{\tau\}\right),
\]
which is nothing but the original involutive EDS $\cI'$ restricted to the leaf $\tilde{L}'\times\{\tau\}$. In order to ensures that $V_{n-1}\left(\cI_\tau'\right)\subset V_{n-1}^r\left(\cI'\right)$, it is required furthermore that the slicing be regular with respect to the EDS $\cI'$. Now we define
\[
\Gamma\left(\cI_{\tau}'\right):=\Pi_{0}\circ\Gamma_{n-1}\left(\cI'_{\tau},\omega'\right),
\]
and the main theorem can be stated as follows.
\begin{thm}\label{TeoremaDiracEDS}
Let us suppose that the EDS $\cI$ defined through \ref{eq:EqMotionCharacterizationEDS} is $\tau$-invariant, namely
\[
\mathcal{L}_{\partial_0}\cI\subset\cI.
\]
Then the constraint submanifold $C$ in the presymplectic space of sections is equal to $\Gamma\left(\cI_{\tau}'\right)$.
\end{thm}
\begin{proof}
The constraint submanifold $C$ can be characterized as the maximal invariant submanifold of the set of sections of $\tilde{L}$ \cite{Gotay:2004ib}, where invariance for $J\subset\Gamma\left(\tilde{L}\right)$ means that
\[
\dif H\left(TJ^\perp\right)=0,
\]
being $H$ the Hamiltonian function for our system and the orthogonal complement is taken with respect to the presymplectic structure in the space of sections. It can be proved (prop. 6.9 in \cite{Gotay:2004ib}) that the invariance of a submanifold $J$ in a presymplectic manifold is equivalent to the fact that through any point of $J$ there exists solutions to the Hamilton equations tangent to this submanifold. So let us take $\sigma_\tau\in\Gamma\left(\cI_\tau'\right)$; according to the very definition of this space, there exists a unique $\sigma_\tau'\in\Gamma_{n-1}\left(\cI_\tau',\omega'\right)$ such that
\[
\sigma_\tau=\Pi_0\circ\sigma_\tau'.
\]
Because of the regularity of the submanifold $\Img\left(\sigma_\tau'\right)\subset\tilde{\Lambda}'$, by Cartan-K\"{a}hler theorem there exists a (non necessarily unique!) integral section $\sigma'\in\Gamma_n\left(\cI',\Omega'\right)$ such that $\left.\sigma'\right|\left(\tilde{L}'\times\{\tau\}\right)=\sigma_\tau'$. If $\partial_0$ denotes alternatively the vector fields defined through
\begin{eqnarray*}
\left.\partial_0\right|_{\left(\lambda,\tau\right)}:&=&\left.\left(s_{\tilde{\Lambda}'}\right)^{-1}_*\right|_{\left(\lambda,\tau\right)}\left(0,\left.\frac{\text{d}}{\text{d}\tau}\right|_\tau\right)\qquad\text{on }\tilde{\Lambda}'\cr
\left.\partial_0\right|_{\left(m,\tau\right)}:&=&\left.\left(s_{M}\right)^{-1}_*\right|_{\left(m,\tau\right)}\left(0,\left.\frac{\text{d}}{\text{d}\tau}\right|_\tau\right)\qquad\text{on }M
\end{eqnarray*}
then we can define the section $X_\tau'\in\Gamma\left(\left(\sigma_\tau'\right)^*\left(V\tilde{L}'\right)\right)=T_{\sigma_\tau'}\left(\Gamma\left(\tilde{L}'\right)\right)$ via
\begin{equation}\label{eq:EvolutionEDSDefinition}
\left(\sigma'\right)_*\left(\partial_0\right)=\partial_0+i_{\tau*}\left(X_\tau'\right).
\end{equation}
Thus the expression $X_\tau:=\Pi_{0*}\circ X_\tau'$ is a well defined section of the pullback bundle $\left(\sigma_\tau\right)^*\left(V\tilde{L}\right)$, because the prolongation procedure and the proposition \ref{Prop:RealTrueShit} ensures us that the map $\Pi_0\circ\left(\cdot\right)$ between $n-1$-sections of the relevant bundles is one-to-one. Furthermore, by projecting both sides of \ref{eq:EvolutionEDSDefinition} along $\Pi_*\circ\left(\cdot\right)$ we obtain that
\begin{equation}
\sigma_*\left(\partial_0\right)=\partial_0+i_{\tau*}\left(X_\tau\right)
\end{equation}
where $\sigma:=\Pi\circ\sigma'$ is a $n$-integral section for $\cI$ on $\tilde{\Lambda}$, due to prolongation property. In particular, $\left.\sigma_*\right|_{\left(m,\tau\right)}\left(T_mM\oplus\partial_0\right)\subset T\tilde{\Lambda}$ is a $n$-integral element for $\cI$ containing the subspace $\left.\sigma_{\tau*}\right|_m\left(T_mM\right)$, which is $n-1$-integral for $\cI_\tau$. Then $\sigma_*\left(\partial_0\right)\in H\left(\left.\sigma_{\tau*}\right|_m\left(T_mM\right)\right)$, so we must have that
\[
\left.\left(\sigma_*\left(\partial_0\right)\lrcorner\alpha\right)\right|\left(\left.\sigma_{\tau*}\right|_m\left(T_mM\right)\right)=0,\qquad\forall\alpha\in\cI.
\]
By using eq. \ref{eq:EqMotionCharacterizationEDS}, we obtain that $X_\tau$ satisfies
\[
\sigma_{\tau}^{*}\left(V\lrcorner\left(X_{\tau}\lrcorner\Omega-\dif\mathcal{H}\right)\right)=0,\qquad\forall V\in\Gamma\left(\sigma_{\tau}^{*}\left(V\tilde{L}\right)\right),
\]
meaning through proposition \ref{pro:HamEqs-and-EDS} that $X_\tau$ is solution of the Hamilton eqs. associated to $H$. It remains to show that this solution is tangent to $\Gamma\left(\cI_\tau'\right)$; according to lemma \ref{lem:TangentSectionsCharacterization}, it is equivalent to show that
\[
\sigma_{\tau}^{*}\left(\mathcal{L}_{\hat{X}_{\tau}}\cI_{\tau}\right)=0
\]
for some extension $\hat{X}_\tau\in\mathfrak{X}\left(\tilde{L}\right)$ of
$X_\tau$. Because of lemma \ref{ProofPropertySols} and taking into account
that $\sigma_\tau$ is a $n-1$-dimensional integral submanifold of $\cI$, we
conclude that
\begin{equation}\label{EqFullEDSInvariance}
\sigma_{\tau}^{*}\left(\mathcal{L}_{\hat{X}}\cI\right)=0,
\end{equation}
where $\hat{X}:=\partial_0+i_\tau\left(\hat{X}_\tau\right)$, and
$\hat{X}_\tau\in\mathfrak{X}\left(\tilde{L}\right)$ is some extension to
$X_\tau$. The $\tau$-invariance
\[
\mathcal{L}_{\partial_0}\cI\in\cI
\]
implies that\footnote{Abuse of language: The symbol
  $i_{\tau*}\left(\hat{X}_\tau\right)$ in the following equation represents the
  vector field $\left(l,s\right)\mapsto i_{s*}\left(\hat{X}_{s}\left(m\right)\right)$ on $\tilde{\Lambda}\simeq\tilde{L}\times\mathbb{R}$, where the
  number $\tau$ refers to the second factor in this decomposition.}
\begin{equation}\label{RestByTauInvariance}
\sigma_{\tau}^{*}\left(\mathcal{L}_{i_{\tau*}\left(\hat{X}_\tau\right)}\cI\right)=0,
\end{equation}
and furthermore, because $X_\tau\left(m\right)\in V\tilde{L}$ for all $m\in M$, for each $\alpha\in\cI$ we will have that
\begin{eqnarray*}
\sigma_\tau^*\left(i_{\tau*}\left(X_\tau\right)\lrcorner\alpha\right)&=&\sigma_\tau^*\left(i_{\tau*}\left(X_\tau\right)\lrcorner\left(\left.\alpha\right|\tilde{L}\times\{\tau\}\right)\right)\cr
&=&\sigma_\tau^*\left(X_\tau\lrcorner\left(i_{\tau}^*\left.\alpha\right|\tilde{L}\times\{\tau\}\right)\right).
\end{eqnarray*}
Then the Cartan's Magic Formula allows us
to rewrite the expression \ref{RestByTauInvariance} as
\[
\sigma_{\tau}^{*}\left(\mathcal{L}_{\hat{X}_\tau}\cI_\tau\right)=0,
\]
which is the desired tangency condition. Therefore we have shown that
$\Gamma\left(\cI'_\tau\right)$ is an invariant submanifold of
$\Gamma\left(\tilde{L}\right)$.
\newline
Our next task is to show that $\Gamma\left(\cI'_\tau\right)$ is maximal among
all the invariant submanifolds of the space of sections
$\Gamma\left(\tilde{L}\right)$. So let us take
$Q\subset\Gamma\left(\tilde{L}\right)$ an invariant submanifold. Then for every $\sigma_{0}\in Q$
there exists $X_{0}\in\Gamma\left(\sigma_{0}^{*}\left(V\tilde{L}\right)\right)$
such that
\begin{itemize}
\item $X_{0}\in T_{\sigma_{0}}Q$, and
\item $\sigma_{0}^{*}\left(V\lrcorner\left(X_{0}\lrcorner\Omega-\dif\mathcal{H}\right)\right)=0$
for all $V\in\Gamma\left(\sigma_{0}^{*}\left(V\tilde{L}\right)\right)$.
\end{itemize}
Assuming that any vector field $Z\in\mathfrak{X}\left(Q\right)$ assigning
to every section $\sigma_{0}$ a vector $X_{0}$ with these characteristics
has integral curves\footnote{We hide behind this assumption some issues
  concerning hard analysis.} $\tau\mapsto\sigma_{\tau}$ passing through $\sigma_{0}\in Q$
when $\tau=0$, we are able to build the $n$-section $\sigma:\left(\tau,x\right)\mapsto\left(\tau,\sigma_{\tau}\left(x\right)\right)$
of $\tilde{\Lambda}$. This section is $n$-integral for $\cI$ because
of the Hamilton equations.
In fact, both conditions above implies that
$\sigma_{\tau*}\left(T_x\Sigma\right)\oplus\left<\left.\partial_0\right|_\tau\right>$
($\tau$ belonging to some neighborhood of $0$) annihilates
the set of forms
\[
\mathcal{S}:=\left\{V\lrcorner\dif\tilde{\lambda}:V\in\Gamma\left(V\tilde{\Lambda}\right)\right\};
\]
because $\mathcal{S}\subset\cI$, we have that
$\left<\mathcal{S}\right>_{\text{diff}}=\cI$ (by the definition of $\cI$, see
Hamilton-Cartan EDS, equation \ref{eq:EDSHam}) and thus
$V_n\left(\left<\mathcal{S}\right>_{\text{diff}}\right)=V_n\left(\cI\right)$.
Then by using the prolongation properties we can find an integral section
$\sigma'$ which is $n$-integral for $\cI'$ and $\Pi\circ\sigma'=\sigma$;
moreover, \begin{eqnarray*}
i_{0}\circ\sigma_{0} & = & \Pi_{0}\circ\left(\left.\sigma'\right|\tilde{\Lambda}_{0}'\right)\end{eqnarray*}
and $\left.\sigma\right|\tilde{\Lambda}_{0}'\in\Gamma_{n-1}\left(\cI_{0}'\right)$,
and then $\sigma_{0}\in\Pi_{0}\circ\Gamma_{n-1}\left(\cI_{0}'\right)=\Gamma\left(\cI_0'\right)$.
\end{proof}

In the following section we will use the relation settled by this theorem between Dirac constraints and the EDS $\cI_\tau$ in some interesting examples.




\section{Examples}\label{Section:Examples}
This section contains examples where the techniques developed in the work will be applied. The first two of them deals with variational problems related to field theories; the last two shows some applications with more mathematical taste.

\subsection{Techniques in the EDS treatment}
Before to start with, it is necessary to mention some specific issues related to the verification of involutivity for a system with independence condition. Namely, it consists of two parts:
\begin{enumerate}
\item The verification that the reduced characters are equal to the original
characters.
\item The verification of the equality hypothesis of the second part of
the Cartan's test.
\end{enumerate}
If any of these fails, we use some of the tools discussed in the appendix \ref{Section:2} in order to find an EDS with the same integral manifolds
(in some dimension) which make them hold. In order to work with these
conditions, let us suppose that $E=\left\langle \bv_{1},\cdots,\bv_{n}\right\rangle \subset T_{x}X$
is an $n$-integral element of an EDS $\left(\cI,\omega_{1}\wedge\cdots\wedge\omega_{n}\right)$
on a manifold $X$ and we take the flag induced by the subspaces\begin{eqnarray*}
E_{0}: & = & 0,\\
E_{k}: & = & \left\langle \bv_{1},\cdots,\bv_{k}\right\rangle ,\qquad k=1,\cdots,n-1;\end{eqnarray*}
for each $k$, the linear system \ref{eq:Linear-System-for-polar}
gives us the space $H\left(E_{k}\right)$ as the kernel of a linear
operator $M\left(\bv_{1},\cdots,\bv_{k}\right):T_{x}X\rightarrow T_{x}^{*}X$.
Then we have that the Cartan characters%
\footnote{These are not the true Cartan characters as they are defined in the
literature, although they are closely related.%
} are\[
c_{k}=\text{rank }M\left(\bv_{1},\cdots,\bv_{k}\right).\]
For the calculation of the reduced Cartan characters, we must write
the operators $M\left(\bv_{1},\cdots,\bv_{k}\right)$ in terms of
a basis $\left\{ \omega_{1},\cdots,\omega_{n},\cdots\right\} \subset T^{*}X$
and its dual: The reduced Cartan characters $\tilde{c}_{k}$ are the
ranks of the matrices obtained through this procedure, after deleting
the columns corresponding to the elements $\omega_{1},\cdots,\omega_{n}$
of the chosen basis.

\subsection{Electromagnetism}
We apply our results in the case of the classical theory of electromagnetic
field; as we know, the EDS associated to their Hamilton-Cartan equations
is defined on $\tilde{\Lambda}=\bigwedge^{2}\left(T^{*}M\right)\oplus\bigwedge^{2}\left(T^{*}M\right)\oplus T^{*}M$
with coordinates $\left(P,F,A\right)$ by\[
\cI:=\left\langle F-\dif A,P-\ast F,\dif P,\dif F,\dif P-\dif\left(\ast F\right)\right\rangle _{\text{alg}}\]
with the independence condition $\Omega:=\left(p_{3}\circ\bar{\tau}_{M}\right)^{*}\left(\omega_{M}\right)\not=0$,
where $\omega_{M}\in\Omega^{4}\left(M\right)$ is some volume form
defined on the space-time (by supposing $M$ orientable). For this system the reduced characters
$\tilde{c}_{2}$ and $\tilde{c}_{3}$ are less than the corresponding
Cartan characters; this is because of the appeareance of the form
$P-\ast F$ in the system, which reduces to a collection of functions
on $\tilde{\Lambda}$ when it is evaluated on integral elements satisfying
the independence condition. In order to circumvect this, we will deal
with the associated EDS on $\Lambda:=\bigwedge^{2}\left(T^{*}M\right)\oplus T^{*}M$
defined through\[
\mathcal{J}:=\left\langle F-\dif A,\dif\left(\ast F\right),\dif F\right\rangle _{\text{alg}};\]
it is the EDS induced by pullback of our original one to the submanifold
described by the equation $P=\ast F$. Anyway, by using the identification
of the vertical directions with the fibres in our bundle, we use the
following vectors\[
V^{\mu}:=\left(\partial_{\mu};F^{\mu},A^{\mu}\right)\]
as the basis inducing the flag\[
E_{0}:=\left\{ 0\right\} \subset E_{1}:=\left\langle V^{1}\right\rangle \subset E_{2}:=\left\langle V^{1},V^{2}\right\rangle \subset E_{3}:=\left\langle V^{1},V^{2},V^{3}\right\rangle \subset E\]
in $E:=\left\langle V^{0},\cdots,V^{3}\right\rangle \subset T_{\left(F,A\right)}\Lambda$
whose Cartan characters are being calculated. Concretely
\begin{itemize}
\item Because $\mathcal{J}^{\left(1\right)}:=\mathcal{J}\cap\Omega^{1}\left(\Lambda\right)=\left\{ 0\right\} $
we obtain that\begin{eqnarray*}
H\left(E_{0}\right) & = & \left\{ v\in T_{\left(F,A\right)}\Lambda:v\lrcorner\phi=0\text{ for all }\phi\in\mathcal{J}^{\left(1\right)}\right\} \\
 & = & T_{\left(F,A\right)}\Lambda\end{eqnarray*}
and so $c_{0}=\text{codim }H\left(E_{0}\right)=0$.
\item In this case\[
H\left(E_{1}\right)=\left\{ v\in T_{\left(F,A\right)}\Lambda:v\lrcorner\left(w\lrcorner\phi\right)=0\text{ for all }\phi\in\mathcal{J}^{\left(2\right)}\text{ and }w\in E_{1}\right\} ,\]
so $v\in H\left(E_{1}\right)$ if and only if\[
v\lrcorner\left(\partial_{1}\lrcorner F-A^{1}-\partial_{1}\lrcorner\dif A\right)=0;\]
then $c_{1}=\text{rank }M\left(V^{1}\right)=1$.
\item Now\begin{eqnarray*}
H\left(E_{2}\right) & = & \Big\{v\in T_{\left(F,A\right)}\Lambda:v\lrcorner\left(w_{1}\lrcorner\phi\right)=0\text{ and }v\lrcorner\left(w_{2}\lrcorner w_{3}\lrcorner\psi\right)=0\\
 &  & \qquad\qquad\text{ for all }\phi\in\mathcal{J}^{\left(2\right)},\psi\in\mathcal{J}^{\left(3\right)}\text{ and }w_{1},w_{2},w_{3}\in E_{2}\Big\}\end{eqnarray*}
and then $v\in H\left(E_{2}\right)$ iff it is solution for the system\begin{eqnarray*}
v\lrcorner\left(\partial_{1}\lrcorner F-A^{1}-\partial_{1}\lrcorner\dif A\right) & = & 0\\
v\lrcorner\left(\partial_{2}\lrcorner F-A^{2}-\partial_{2}\lrcorner\dif A\right) & = & 0\\
v\lrcorner\left(\partial_{1}\lrcorner\left(\ast F^{2}\right)-\partial_{2}\lrcorner\left(\ast F^{1}\right)+\partial_{1}\lrcorner\partial_{2}\lrcorner\dif\left(\ast F\right)\right) & = & 0\\
v\lrcorner\left(\partial_{1}\lrcorner F^{2}-\partial_{2}\lrcorner F^{1}+\partial_{1}\lrcorner\partial_{2}\lrcorner\dif F\right) & = & 0;\end{eqnarray*}
this means that $c_{2}=4$.
\item Accordingly, the equations determining $H\left(E_{3}\right)$ will
be\begin{eqnarray*}
v\lrcorner\left(\partial_{1}\lrcorner F-A^{1}-\partial_{1}\lrcorner\dif A\right) & = & 0\\
v\lrcorner\left(\partial_{2}\lrcorner F-A^{2}-\partial_{2}\lrcorner\dif A\right) & = & 0\\
v\lrcorner\left(\partial_{3}\lrcorner F-A^{3}-\partial_{3}\lrcorner\dif A\right) & = & 0\\
v\lrcorner\left(\partial_{1}\lrcorner\left(\ast F^{2}\right)-\partial_{2}\lrcorner\left(\ast F^{1}\right)+\partial_{1}\lrcorner\partial_{2}\lrcorner\dif\left(\ast F\right)\right) & = & 0\\
v\lrcorner\left(\partial_{1}\lrcorner\left(\ast F^{3}\right)-\partial_{3}\lrcorner\left(\ast F^{1}\right)+\partial_{1}\lrcorner\partial_{3}\lrcorner\dif\left(\ast F\right)\right) & = & 0\\
v\lrcorner\left(\partial_{2}\lrcorner\left(\ast F^{3}\right)-\partial_{3}\lrcorner\left(\ast F^{2}\right)+\partial_{2}\lrcorner\partial_{3}\lrcorner\dif\left(\ast F\right)\right) & = & 0\\
v\lrcorner\left(\partial_{1}\lrcorner F^{2}-\partial_{2}\lrcorner F^{1}+\partial_{1}\lrcorner\partial_{2}\lrcorner\dif F\right) & = & 0\\
v\lrcorner\left(\partial_{1}\lrcorner F^{3}-\partial_{3}\lrcorner F^{1}+\partial_{1}\lrcorner\partial_{3}\lrcorner\dif F\right) & = & 0\\
v\lrcorner\left(\partial_{2}\lrcorner F^{3}-\partial_{3}\lrcorner F^{2}+\partial_{2}\lrcorner\partial_{3}\lrcorner\dif F\right) & = & 0;\end{eqnarray*}
the associated Cartan character being\[
c_{3}=9\]
because of the Hodge operator.
\end{itemize}
Let us now calculate the codimension of $V_{4}\left(\mathcal{J}\right)\subset G_{4}\left(T\Lambda\right)$;
as before, it will be the rank of the linear operator behind the linear
system%
\footnote{That this system is linear is not a general fact; instead it is a
property of the EDS which we are handled, and is equivalent to the
linearity of the Maxwell equations.%
}\begin{eqnarray*}
V^{\mu}\lrcorner V^{\nu}\lrcorner\left(F-\dif A\right)=0\\
V^{\mu}\lrcorner V^{\nu}\lrcorner V^{\sigma}\lrcorner\left(\dif\left(\ast F\right)\right)=0\\
V^{\mu}\lrcorner V^{\nu}\lrcorner V^{\sigma}\lrcorner\left(\dif F\right)=0 &  & \forall\mu,\nu,\sigma=0,\cdots,3.\end{eqnarray*}
Using the symmetries of the contraction operator, we see that the
\emph{a priori} independent equations in this set are determined by
the set of pairs of indices\[
\left(0,1\right),\left(0,2\right),\left(0,3\right),\left(1,2\right),\left(1,3\right),\left(2,3\right)\]
and the set of triples\[
\left(0,1,2\right),\left(0,1,3\right),\left(0,2,3\right),\left(1,2,3\right).\]
The linear operator is constant, consequently it has maximal rank
everywhere; so we have that $\text{codim}_{E}V_{4}\left(\mathcal{J}\right)=\#\text{ equations }=14$.
Thus the hypothesis of the Cartan's test are satisfied, and the EDS
$\mathcal{J}$ will be involutive. By using the slicing $x^{0}=\text{ constant}$
we can identify $M\simeq\mR\times L$\[
\Lambda_{x^{0}}\simeq\left(\mR\times L\right)\oplus T^{*}L\oplus\bigwedge^{2}\left(T^{*}L\right)\oplus\bigwedge^{2}\left(T^{*}L\right)\]
with coordinates $\left(a_{0},a,e,b\right)$ induced by the definitions
($\star$ is the $3$-Hodge star induced by the metric)\begin{eqnarray*}
F: & = & \left(\star e\right)\wedge\dif x^{0}+b\\
A: & = & a_{0}\dif x^{0}+a.\end{eqnarray*}
From here we build the hamiltonian version of the equations of motion;
by using the rules\begin{eqnarray*}
\ast\left(\star\left(\dif x^{i}\wedge\dif x^{j}\right)\wedge\dif x^{0}\right) & = & -\dif x^{i}\wedge\dif x^{j}\\
\ast\left(\dif x^{i}\wedge\dif x^{j}\right) & = & \star\left(\dif x^{i}\wedge\dif x^{j}\right)\wedge\dif x^{0}\end{eqnarray*}
the constraint submanifold will be given by the $3$-integral sections
of the EDS\begin{eqnarray*}
\mathcal{J}_{x^{0}}: & = & i_{x^{0}}^{*}\left(\left.\mathcal{J}\right|\Lambda_{x^{0}}\right)\\
 & = & \left\langle \dif a-b,\dif e=0,\dif b=0\right\rangle _{\text{alg}}\end{eqnarray*}
which are the usual constraints in the Hamiltonian version of electromagnetic
field equations.

\subsection{Gotay-Nester algorithm}\label{GotayNesterAlgorithm}
Before starting with the next examples, it is necessary to summarize the
Gotay-Nester algorithm. The initial data is a triple
$\left(M_0,\omega_0,H_0\right)$ where $\left(M_0,\omega_0\right)$ is a
presymplectic manifold and $H_0\in C^\infty\left(M_0\right)$. Then the
algorithm proceed with the following steps:
\begin{enumerate}
\item Calculate $K_0:=\text{ker}\,\omega_0$.
\item For each $l\in\mathbb{N}$, let $\mathcal{C}_{l+1}$ be the ideal in $C^\infty\left(M_l\right)$ generated by $C_{l+1}:=\left\{X\cdot H_l:X\in K_l\right\}$.
\item Define $M_{l+1}$ as the zero locus for functions in $\mathcal{C}_{l+1}$. Admitting that $M_{l+1}$ is a submanifold, define $\omega_{l+1}:=\left.\omega_l\right|_{M_{l+1}},H_{l+1}:=\left.H_l\right|_{M_{l+1}}$, and $K_{l+1}:=\left(TM_{l+1}\right)^\perp$.
\item The algorithm stops whenever $M_{l+1}=M_l$.
\item The dynamics on the final constraint manifold $M_l$ is governed by the solutions of $X_{H_0}\lrcorner\omega_0=\dif H_0$ tangent to $M_l$.
\end{enumerate}
Some shortcuts are in order here: Because $M_{l+1}\subset M_{l}$, we have that $\left(TM_{l}\right)^\perp\subset\left(TM_{l+1}\right)^\perp$, so in each step there are $\text{dim}\left(TM_{l+1}\right)^\perp-\text{dim}\left(TM_{l}\right)^\perp$ complementary vectors to look for. In order to find them out, it is useful the following remark: If $F\in\mathcal{C}_{l+1}$, and it admits a hamiltonian vector, then $X_F\in\left(TM_{l+1}\right)^\perp$.

\subsection{Poisson sigma models}\label{PoissonSigmaExampleSection}

Here we will work in the setting of example \ref{PoissonSigmaExample}. This example has the following feature, which is not shared by the rest of the examples we will deal with: Its Hamilton-Cartan EDS has torsion. In fact, it will be shown that the torsion is equal to the jacobiator of the Poisson structure $\pi$.

\subsubsection{EDS analysis of Poisson sigma models}
 According to our previous discusion, we have that the Hamilton-Cartan EDS $\cI$ is locally generated by
\[
\begin{cases}
\theta^\mu:=\dif x^{\mu}-\pi^{\mu\nu}\eta_{\alpha\nu}\dif\xi^{\alpha},\\
\Gamma_{\sigma}:=\dif\eta_{\alpha\sigma}\wedge\dif\xi^\alpha-\frac{1}{2}\left(\partial_{\sigma}\pi^{\mu\nu}\right)\eta_{\alpha\mu}\eta_{\beta\nu}\dif\xi^{\beta}\wedge\dif\xi^\alpha.\end{cases}
\]
It can be shown (see for example \cite{Calvo07}) that if $J\left(\pi\right)$ is the jacobiator of the bivector $\pi$, then 
\[
\dif\theta^\mu\equiv\left[J\left(\pi\right)\right]^{\mu\nu\rho}\eta_{\alpha\nu}\eta_{\beta\rho}\dif\xi^\alpha\wedge\dif\xi^\beta\mod{\cI}.
\]
So the jacobiator in these systems is related to the torsion of the associated EDS. Because we will study the Cartan characters of a $2$-dimensional integral element, it is not necessary to find an expression for $\dif\Gamma_\mu$.
\\
Let us consider the flag $E_0=0\subset E_1:=\left<v_1\right>\subset E:=\left<v_1,v_0\right>$, where
\[
v_i:=\partial_{\xi^i}+X_i^\mu\partial_{x^\mu}+\Xi_{\alpha\nu}^i\partial_{\eta_{\alpha\nu}}\in T_{\left(\xi,x,\eta\right)}\mathcal{S},i=0,1.
\]
Then $E$ will belong to $V_2\left(\cI\right)$ iff
\[
\begin{cases}
X_0^\rho-\pi^{\rho\nu}\eta_{0\nu}=0,\cr
X_1^\rho-\pi^{\rho\nu}\eta_{1\nu}=0,\cr
\Xi^0_{1\sigma}-\Xi^1_{0\sigma}-\left(\partial_\sigma\pi^{\mu\nu}\right)\eta_{0\mu}\eta_{1\nu}=0
\end{cases}
\]
for $\rho,\sigma=1,\cdots,\text{dim}M$; then
$\text{codim}_EV_2\left(\cI\right)=3n$. On the other side, for the polar
spaces we have that
\begin{align*}
H\left(E_0\right)&=\left\{v\in
  T_{\left(\xi,x,\eta\right)}\mathcal{S}:v\lrcorner\left(\dif
    x^{\mu}-\pi^{\mu\nu}\eta_{\alpha\nu}\dif\xi^{\alpha}\right)=0,\mu=1,\cdots
,\text{dim}M\right\}\cr
H\left(E_1\right)&=\left\{v\in
  H\left(E_0\right):v\lrcorner v_1\lrcorner\left(\dif\eta_{\alpha\sigma}\wedge\dif\xi^\alpha-\frac{1}{2}\left(\partial_{\sigma}\pi^{\mu\nu}\right)\eta_{\alpha\mu}\eta_{\beta\nu}\dif\xi^{\beta}\wedge\dif\xi^\alpha\right)=0\right\}\cr
&=\Big\{v\in
  H\left(E_0\right):\dif\eta_{1\sigma}+\text{terms in }\dif\xi^\beta\Big\},
\end{align*}
so the reduced characters for $E$ will be
$c_0\left(E\right)=\text{dim}M,c_1\left(E\right)=2n$. Then the Cartan test is
satisfied, and the system is involutive at $E$.

\subsubsection{Investigation through the Gotay-Nester algorithm}
It is interesting to see how the Jacobi identity for $\pi$ can be obtained from the Gotay-Nester algorithm associated to the non standard variational problem describing a Poisson sigma model. So let us fix on $\Sigma$ a $1$-submanifold $\Sigma_0$, which is locally described as $\xi^0=0$, and define $\mathcal{S}_0:=\left.\mathcal{S}\right|\Sigma_0$. According to our recipe, on $\Gamma\left(\mathcal{S}_0\right)$ there exists a presymplectic form and a hamiltonian; in order to get them, let us calculate
\begin{multline*}
\dif\tilde\lambda=\dif\eta_{\alpha\mu}\dif\xi^\alpha\wedge\dif x^\mu-\left(\partial_\sigma\pi^{\mu\nu}\right)\eta_{\alpha\mu}\eta_{\beta\nu}\dif x^\sigma\wedge\dif\xi^\alpha\wedge\dif\xi^\beta-2\pi^{\mu\nu}\eta_{\beta\nu}\dif\eta_{\alpha\mu}\wedge\dif\xi^\alpha\wedge\dif\xi^\beta\\
\Rightarrow\qquad\left.\dif\tilde\lambda\right|_{\mathcal{S}_0}=-\dif\eta_{1\mu}\wedge\dif x^\mu\wedge\dif\xi^1
\end{multline*}
and
\begin{multline*}
\partial_{\xi^0}\lrcorner\tilde\lambda=\eta_{0\mu}\dif x^\mu-\pi^{\mu\nu}\eta_{0\mu}\eta_{\beta\nu}\dif\xi^\beta\Rightarrow\left.\partial_{\xi^0}\lrcorner\tilde\lambda\right|_{\mathcal{S}_0}=\eta_{0\mu}\dif x^\mu-\pi^{\mu\nu}\eta_{0\mu}\eta_{1\nu}\dif\xi^1.
\end{multline*}
In order to simplify notation, let us introduce the definitions $\tau:=\xi^0,\sigma:=\xi^1,\eta_\mu:=\eta_{0\mu}$ and $\gamma_\mu:=\eta_{1\mu}$; then if $X_i:=\left(\delta x_i^\mu,\delta\eta^i_\nu,\delta\gamma^i_\rho\right),i=1,2$ indicates arbitrary vectors in $T_{\left(x,\eta,\gamma\right)}\Gamma\left(\mathcal{S}_0\right)$ the presymplectic structure reads
\[
\omega_0\left(X_1,X_2\right)=\int_{\Sigma_0}\left(\delta x_1^\mu\delta\gamma_{\mu}^2-\delta x_2^\mu\delta\gamma_{\mu}^1\right)\dif\sigma
\]
and the hamiltonian will be
\[
H_0:=\int_{\Sigma_0}\eta_\mu\left[\left(x^\mu\right)'-\pi^{\mu\nu}\gamma_\nu\right]\dif\sigma.
\]
Here we denote with a prime the derivative respect to the $\sigma$-variable. So we have that
\[
\text{ker}\,\omega_0=\left\{\left(0,\delta\eta_\mu,0\right)\right\},
\]
and the primary constraints will arise from
\begin{align*}
0&=\dif H_0\left(\delta\eta_\nu\right)\cr
&=\int_{\Sigma_0}\delta\eta_\mu\left[\left(x^\mu\right)'-\pi^{\mu\nu}\gamma_\nu\right]\dif\sigma\qquad\Rightarrow\qquad\left(x^\mu\right)'-\pi^{\mu\nu}\gamma_\nu=0.
\end{align*}
\paragraph*{\textit{Termination of the algorithm}.} Let $C_1$ be the set
$C_1:=\left\{\left(x^\mu\right)'-\pi^{\mu\nu}\gamma_\nu\right\}$ and
$M_1\subset\Gamma\left(\mathcal{S}_0\right)$ its zero locus. 
Now let us define
\[
F:=\int_{\Sigma_0}f_\mu\left[\left(x^\mu\right)'-\pi^{\mu\nu}\gamma_\nu\right]\dif\sigma,
\]
where $f_\mu,\mu=1,\cdots,n$ are arbitrary functions on $\Sigma_0$ and $n=\text{dim}\,M$; then
\begin{align*}
\dif F\left(X_2\right)&=\int_{\Sigma_0}f_\mu\left[\left(\delta x_2^\mu\right)'-\left(\partial_\rho\pi^{\mu\nu}\right)\gamma_\nu\delta x_2^\rho-\pi^{\mu\nu}\delta\gamma^2_\nu\right]\dif\sigma\cr
&=\int_{\Sigma_0}\left\{-f_\mu\pi^{\mu\nu}\delta\gamma_\nu^2-\left[\left(f_\rho\right)'+f_\mu\left(\partial_\rho\pi^{\mu\nu}\right)\gamma_\nu\right]\delta x^\rho_2\right\}\dif\sigma,
\end{align*}
and this function will have hamiltonian vector field, namely
\[
X_F:=\left(-f_\mu\pi^{\mu\nu},0,\left(f_\rho\right)'+f_\mu\left(\partial_\rho\pi^{\mu\nu}\right)\gamma_\nu\right).
\]
The different choices for the arbitrary functions $f_\mu$ will gives the vector fields complementary to $\text{ker}\,\omega_0$. The secondary constraints are obtained from
\begin{align*}
0&=\dif H_0\left(X_F\right)\cr
&=\int_{\Sigma_0}\eta_\mu\left[-\left(f_\nu\pi^{\nu\mu}\right)'+\left(\partial_\rho\pi^{\mu\nu}\right)\gamma_\nu f_\omega\pi^{\omega\rho}-\pi^{\mu\nu}\left(\left(f_\nu\right)'+f_\omega\left(\partial_\nu\pi^{\omega\rho}\right)\gamma_\rho\right)\right]\dif\sigma\cr
&=\int_{\Sigma_0}\eta_\mu\left[f_\nu\left(\partial_\rho\pi^{\nu\mu}\right)\left(x^\rho\right)'+\left(\partial_\rho\pi^{\mu\nu}\right)\gamma_\nu f_\omega\pi^{\omega\rho}-\pi^{\mu\nu}f_\omega\left(\partial_\nu\pi^{\omega\rho}\right)\gamma_\rho\right]\dif\sigma\cr
&=\int_{\Sigma_0}\eta_\mu\left[f_\nu\left(\partial_\rho\pi^{\nu\mu}\right)\pi^{\rho\omega}\gamma_\omega+\left(\partial_\rho\pi^{\mu\nu}\right)\gamma_\nu f_\omega\pi^{\omega\rho}-\pi^{\mu\nu}f_\omega\left(\partial_\nu\pi^{\omega\rho}\right)\gamma_\rho\right]\dif\sigma\cr
&=\int_{\Sigma_0}\eta_\mu f_\nu\gamma_\omega\left[\left(\partial_\rho\pi^{\nu\mu}\right)\pi^{\rho\omega}+\left(\partial_\rho\pi^{\mu\omega}\right)\pi^{\nu\rho}-\pi^{\mu\rho}\left(\partial_\rho\pi^{\nu\omega}\right)\right]\dif\sigma
\end{align*}
where was used that we are restricting on $M_1$, so that $\left(x^\rho\right)'=\pi^{\rho\omega}\gamma_\omega$. Because $\pi$ is a Poisson structure, these constraints are identically satisfied, and $M_1$ is an invariant manifold.

\subsection{Dirac method and integrability conditions}\label{ExampleOfProlongation}

From \cite{Seiler:1995ne} we know that the following PDE system
\[
\begin{cases}
\phi_{zz}+y\phi_{xx}=0\cr
\phi_{yy}=0
\end{cases}
\]
has the equations $\phi_{xxy}=\phi_{xxxx}=0$ as integrability conditions. This system
can be written as the EDS $\mathcal{I}$ generated by
\begin{multline*}
\Big\{\theta:=\dif\phi-p\dif x-q\dif y-r\dif z,\cr
\Gamma_1:=\dif r\wedge\dif x\wedge\dif y+y\dif p\wedge\dif y\wedge\dif z,\Gamma_2:=\dif q\wedge\dif x\wedge\dif z\Big\}
\end{multline*}
on $\mathbb{R}^7$ with coordinates $\left(x,y,z,\phi,p,q,r\right)$. It is
noted also that these integrability conditions can be obtained applying the
prolongation procedure twice. In the following sections we will use the
correspondence settled by Theorem \ref{TeoremaDiracEDS} in order to calculate the
aforementioned integrability conditions through the
Gotay-Nester algorithm. The idea is that, although the system initially considered is highly non regular, there exists (via Cartan-Kuranishi) a prolongation which is regular\footnote{There remains the question of regularity of the foliation introduced in order to define the presymplectic space; it is supposed that this condition is fulfilled.}, and the Dirac constraints can be determined by our theorem \ref{TeoremaDiracEDS}: \emph{Therefore if we know the Dirac constraints, we know something about the prolonged EDS}. In particular, the integrability conditions must arise as these kind of constraints.
\\
Therefore we need to formulate the EDS as a non
standard variational problem; thus let us define the double fibration
\[
\begin{array}{rcccl}
\Lambda:=\mathbb{R}^7&\rightarrow&\Lambda_1:=\mathbb{R}^4&\rightarrow&M:=\mathbb{R}^3\\
\left(x,y,z,\phi,p,q,r\right)&\mapsto&\left(x,y,z,\phi\right)&\mapsto&\left(x,y,z\right),
\end{array}
\]
taking as prolongations structure the given EDS $\cI$, and as lagrangian the
trivial one $\lambda=0$. In order to build an associated Lepagean equivalent
problem, we consider
\[
\tilde{\Lambda}:=\Lambda\oplus\bigwedge^2\left(T^*M\right)\oplus\mathbb{R}^2
\]
with coordinates $(x,y,z,\phi,p,q,r;\alpha,\lambda,\mu)$, $\alpha=A\dif
x\wedge\dif y+B\dif x\wedge\dif z+C\dif y\wedge\dif z,\mu,\lambda\in\mathbb{R}$; the functional to be
extremized will be
\[
S:=\int_M\left(\alpha\wedge\theta+\lambda\Gamma_1+\mu\Gamma_2\right).
\]
We know that it is a covariant Lepage equivalent problem; the following lemma
shows that it is also contravariant.
\begin{lem}
The projection
\[
\nu:\tilde{\Lambda}\rightarrow\Lambda:\left(x,y,z,\phi,p,q,r;\alpha,\lambda,\mu\right)\mapsto\left(x,y,z,\phi,p,q,r\right)
\]
maps extremals of $S$ into integral sections for $\mathcal{I}$; conversely, for each
integral section of $\cI$ given by
\[
\left(x,y,z\right)\mapsto\left(x,y,z,\phi,p,q,r\right)
\]
there exists $\alpha\in\Omega^2\left(M\right),\lambda,\mu\in
C^\infty\left(M\right)$ such that joined to the integral section gives rise to
an extremal of $S$.
\end{lem}
\begin{proof}
The variations respect to $\alpha,\mu$ and $\lambda$ gives us the generators of
$\cI$, so its projection will be an integral section of $\cI$. The variations
respect to $\phi,p,q$ and $r$ leads to
\[
\begin{cases}
\dif\alpha=0,\cr
y\dif\lambda\wedge\dif y\wedge\dif z+\alpha\wedge\dif x=0,\cr
\dif\mu\wedge\dif x\wedge\dif z+\alpha\wedge\dif y=0,\cr
\dif\lambda\wedge\dif x\wedge\dif y+\alpha\wedge\dif z=0.\cr
\end{cases}
\]
This system is equivalent to
\[
\begin{cases}
A_z-B_y+C_x=0,\cr
y\lambda_x+C=0,\cr
\mu_y+B=0,\cr
\lambda_z+A=0;
\end{cases}
\]
in order to find a solution to it, we take $\lambda$ arbitrary, and define $C$ and $A$ by using the second and fourth equation respectively. From third equation we can determine $B$ once $\mu$ is known, and the first of them gives us a differential equation for $\mu$, namely
\[
\mu_{yy}=\lambda_{zz}+y\lambda_{xx}.
\]
So the system has solutions, and our new variational problem is contravariant.
\end{proof}

\subsubsection{Associated dynamical system}
Let $\tilde{\lambda}$ be the form below the integral sign in $S$, and
let us define $N_0\subset\tilde{\Lambda}$ as the submanifold consisting of the points in $\tilde\Lambda$ such that $z=0$. Then
\begin{multline*}
\dif\tilde{\lambda}=\dif\alpha\wedge\left(\dif\phi-p\dif x-q\dif y-r\dif z\right)-\alpha\wedge\left(\dif p\wedge \dif x+\dif q\wedge\dif y+\dif r\wedge\dif z\right)+\cr
+\dif\lambda\wedge\left(\dif r\wedge\dif x\wedge\dif y+y\dif p\wedge\dif y\wedge\dif z\right)+\dif\mu\wedge\dif q\wedge\dif x\wedge\dif z
\end{multline*}
and thus
\[
\left.\dif\tilde{\lambda}\right|_{N_0}=\dif A\wedge\dif x\wedge\dif y\wedge\dif\phi+\dif\lambda\wedge\dif r\wedge\dif x\wedge\dif y;
\]
this allows us to define on $N_0$ the presymplectic form
\[
\omega^z\left(X_1,X_2\right):=\int_{\mathbb{R}^2}\left[\left(\delta A_1\delta \phi_2-\delta A_2\delta \phi_1\right)+\left(\delta\lambda_1\delta r_2-\delta\lambda_2\delta r_1\right)\right]\dif x\wedge\dif y,
\]
where we put
\[
X_i=\left(\delta\phi_i,\delta p_i,\delta q_i,\delta r_i;\delta\alpha_i,\delta\lambda_i,\delta\mu_i\right),\qquad i=1,2
\]
in order to denote generic vectors in the tangent space of
$\Gamma\left(N_0\right)$. It is important to know the kernel of this form.
\begin{lem}
The kernel of $\omega^z$ is given by
\[
\text{ker}\,\omega^z=\left<\left(0,\delta p,\delta q,0;\delta B\dif x\wedge\dif z+\delta C\dif y\wedge\dif z,0,\delta\mu\right)\right>.
\]
\end{lem}
Let us now build the hamiltonian; contracting $\tilde{\lambda}$ in the $z$-direction
\[
\partial_z\lrcorner\tilde{\lambda}=-\left(B\dif x+C\dif z\right)\wedge\left(\dif\phi-p\dif x-q\dif y-r\dif z\right)-r\alpha+y\lambda\dif p\wedge\dif y+\mu\dif q\wedge\dif x,
\]
so that
\begin{multline*}
\left.\partial_z\lrcorner\tilde{\lambda}\right|_{N_0}=\cr
=-\left(B\dif x+C\dif y\right)\wedge\dif\phi+\left(Bq-Cp\right)\dif x\wedge\dif y-rA\dif x\wedge\dif y+y\lambda\dif p\wedge\dif y+\mu\dif q\wedge\dif x
\end{multline*}
we obtain the following expression for the hamiltonian
\[
H^z=\int_{\mathbb{R}^2}\left[\left(C\phi_x-B\phi_y\right)+\left(Bq-Cp-rA\right)+\left(y\lambda p_x-\mu q_y\right)\right]\dif x\wedge\dif y.
\]
\paragraph*{\textit{Primary constraints}}
The first order constraints can be obtained from
\[
\dif H^z\left(\text{ker}\,\omega^z\right)=0.
\]
Taking into account the previous calculations, it results that
\begin{align*}
0&=\dif H^z\left(\delta p\right)&&\\
&=\int_{\mathbb{R}^2}\left[y\lambda\left(\delta p\right)_x-C\delta p\right]\dif x\wedge\dif y&&\\
&=-\int_{\mathbb{R}^2}\left(y\lambda_x+C\right)\delta p\dif x\wedge\dif y&&\Rightarrow\qquad y\lambda_x+C=0\\
0&=\dif H^z\left(\delta q\right)&&\\
&=\int_{\mathbb{R}^2}\left[B\delta q-\mu\left(\delta q\right)_y\right]\dif x\wedge\dif y&&\\
&=\int_{\mathbb{R}^2}\left(B+\mu_y\right)\delta q\dif x\wedge\dif y&&\Rightarrow\qquad B+\mu_y=0\\
0&=\dif H^z\left(\delta B\right)&&\\
&=\int_{\mathbb{R}^2}\left(q-\phi_y\right)\delta B\dif x\wedge\dif y&&\Rightarrow\qquad \phi_y-q=0\\
0&=\dif H^z\left(\delta C\right)&&\\
&=\int_{\mathbb{R}^2}\left(\phi_x-p\right)\delta C\dif x\wedge\dif y&&\Rightarrow\qquad \phi_x-p=0\\
0&=\dif H^z\left(\delta\mu\right)&&\\
&=\int_{\mathbb{R}^2}q_y\delta\mu\dif x\wedge\dif y&&\Rightarrow\qquad q_y=0.
\end{align*}
The set
\[
C_1:=\left\{y\lambda_x+C,B+\mu_y,\phi_y-q,\phi_x-p,q_y\right\}
\]
gives rise to the primary constraints submanifold
$M_1=\left\{C_1=0\right\}\subset\Gamma\left(N_0\right)$. In
order to find out the elements in $\left(TM_1\right)^\perp$, let us use the remark made in the subsection \ref{GotayNesterAlgorithm}: If $F$ is a constraint,
and it admits a hamiltonian vector, then $X_F\in\left(TM_1\right)^\perp$. Let
us take the constraints $q_y=0$ y $\phi_y-q=0$; none of them has a hamiltonian
vector field, although the consequence $\phi_{yy}=0$ admits such a vector. Therefore the function
\[
F_1:=\int_{\mathbb{R}^2}\phi_{yy}f_1\dif x\wedge\dif y
\]
(where $f_1\in C^\infty\left(\mathbb{R}^2\right)$ is arbitrary) has a
hamiltonian vector field. In fact, from the equation
\[
\omega^z\left(X_{F_1},X_2\right)=\dif F_1\left(X_2\right)
\]
and taking into account that
\begin{align*}
\dif F_1\left(X_2\right)&=\int_{\mathbb{R}^2}\left(\delta\phi_2\right)_{yy}f_1\dif x\wedge\dif y\cr
&=\int_{\mathbb{R}^2}\delta\phi_2\left(f_1\right)_{yy}\dif x\wedge\dif y
\end{align*}
we conclude that
\[
X_{F_1}=\left(0,0,0,0;\left(f_1\right)_{yy}\dif x\wedge\dif y,0,0\right).
\]
Therefore we see that $\left(TM_1\right)^\perp=\text{ker}\,\omega^z\oplus\left<X_{F_1}\right>$, and we are ready to calculate the secondary constraints.

\paragraph*{\textit{Secondary constraints}}
These constraints are obtained from $\dif H^z\left(\left(TM_1\right)^\perp\right)=0$ and we know that on $M_1$, $\dif H^z\left(\text{ker}\omega^z\right)$ is identically zero; then
\begin{align*}
0&=\dif H^z\left(X_{F_1}\right)\cr
&=-\int_{\mathbb{R}^2}r\left(f_1\right)_{yy}\dif x\wedge\dif y\cr
&=-\int_{\mathbb{R}^2}r_{yy}f_1\dif x\wedge\dif y\qquad\Rightarrow\qquad r_{yy}=0.
\end{align*}

Let us now define $C_2:=C_1\cup\left\{r_{yy}\right\}$ and $M_2:=\left\{C_2=0\right\}$.
By using the additional constraint $r_{yy}$ we can define
\[
F_2:=\int_{\mathbb{R}^2}r_{yy}f_2\dif x\wedge\dif y
\]
and so
\[
\dif F_2\left(X_2\right)=\int_{\mathbb{R}^2}\left(\delta r_2\right)_{yy}f_2\dif x\wedge\dif y=\int_{\mathbb{R}^2}\delta r_2\left(f_2\right)_{yy}\dif x\wedge\dif y;
\]
the associated hamiltonian vector field is
\[
X_{F_2}=\left(0,0,0,0;0,\left(f_2\right)_{yy},0\right).
\]
It is the complementary vector we were looking for, and so $\left(TM_2\right)^\perp=\left(TM_1\right)^\perp\oplus\left<X_{F_2}\right>$. Thus we can calculate the new constraint, namely
\begin{align*}
0&=\dif H^z\left(X_{F_2}\right)\cr
&=\int_{\mathbb{R}^2}\left(f_2\right)_{yy}yp_x\dif x\wedge\dif y\cr
&=\int_{\mathbb{R}^2}f_2\left(yp_x\right)_{yy}\dif x\wedge\dif y\qquad\Rightarrow\qquad\left(yp_x\right)_{yy}=0.
\end{align*}
But on $M_2$, $p=\phi_x$, and therefore
\begin{align*}
\left(yp_x\right)_{yy}&=\left(y\phi_{xx}\right)_{yy}\cr
&=\left(y\phi_{xxy}+\phi_{xx}\right)_y\cr
&=y\phi_{xxyy}+2\phi_{xxy}\cr
&=2\phi_{xxy}
\end{align*}
on $M_2$.

\paragraph*{\textit{Tertiary constraints.}} Now $C_3:=C_2\cup\left\{\phi_{xxy}\right\}$ and let $M_3$ be its zero locus. 
Continuing with the algorithm, we need to find the vectors in the symplectic orthogonal to $TM_3$ associated to the (new) constraints functions. So we define
\[
F_3:=\int_{\mathbb{R}^2}f_3\phi_{xxy}\dif x\wedge\dif y,
\]
and its derivative along $X_2$ will read
\begin{align*}
\dif F_3\left(X_2\right)&=\int_{\mathbb{R}^2}f_3\left(\delta\phi_2\right)_{xxy}\dif x\wedge\dif y\cr
&=-\int_{\mathbb{R}^2}\left(f_3\right)_{xxy}\delta\phi_2\dif x\wedge\dif y.
\end{align*}
Then
\[
X_{F_3}=\left(0,0,0,0;-\left(f_3\right)_{xxy}\dif x\wedge\dif y,0,0\right),
\]
therefore $\left(TM_3\right)^\perp=\left(TM_2\right)^\perp\oplus\left<X_{F_3}\right>$, and from here we obtain the new constraint
\begin{align*}
0&=\dif H^z\left(X_{F_3}\right)\cr
&=\int_{\mathbb{R}^2}r\left(f_3\right)_{xxy}\dif x\wedge\dif y\cr
&=-\int_{\mathbb{R}^2}r_{xxy}f_3\dif x\wedge\dif y\qquad\Rightarrow\qquad r_{xxy}=0.
\end{align*}

\paragraph*{\textit{Quaternary constraints.}} We define $C_4:=C_3\cup\left\{r_{xxy}\right\}$ and $M_4:=\left\{C_4=0\right\}$.
Then if
\[
F_4:=\int_{\mathbb{R}^2}f_4r_{xxy}\dif x\wedge\dif y
\]
we have that $\dif F_4\left(X_2\right)=-\int_{\mathbb{R}^2}\left(f_4\right)_{xxy}\delta r_2\dif x\wedge\dif y$, and the new vector in $\left(TM_4\right)^\perp$ will be
\[
X_{F_4}=\left(0,0,0,0;0,-\left(f_4\right)_{xxy},0\right).
\]
Then we have that $\left(TM_4\right)^\perp=\left(TM_3\right)^\perp\oplus\left<X_{F_3}\right>$, and for the new constraint
\begin{align*}
0&=\dif H^z\left(X_{F_2}\right)\cr
&=-\int_{\mathbb{R}^2}yp_x\left(f_4\right)_{xxy}\dif x\wedge\dif y\cr
&=\int_{\mathbb{R}^2}\left(yp_x\right)_{xxy}f_4\dif x\wedge\dif y\qquad\Rightarrow\qquad\left(yp_x\right)_{xxy}=0.
\end{align*}
But on $M_4$
\[
\left(yp_x\right)_{xxy}=\left(y\phi_{xx}\right)_{xxy}=\left(y\phi_{xxxx}\right)_{y}=\phi_{xxxx}+y\phi_{xxy}=\phi_{xxxx},
\]
giving rise to the constraint $\phi_{xxxx}=0$.

\paragraph*{\textit{Fifth order constraints.}} With the definitions $C_5:=C_4\cup\left\{\phi_{xxxx}\right\},M_5:=\left\{C_5=0\right\}$ and 
\[
F_5:=\int_{\mathbb{R}^2}f_5\phi_{xxxx}\dif x\wedge\dif y
\]
we have that $\dif H^z\left(X_2\right)=\int_{\mathbb{R}^2}\left(f_5\right)_{xxxx}\delta\phi_2\dif x\wedge\dif y$, so
\[
X_{F_5}=\left(0,0,0,0;\left(f_5\right)_{xxxx}\dif x\wedge\dif y,0,0\right).
\]
The new symplectic complement is given by $\left(TM_5\right)^\perp=\left(TM_4\right)^\perp\oplus\left<X_{F_5}\right>$. Thus it is obtained a new constraint from
\begin{align*}
0&=\dif H^z\left(X_{F_5}\right)\cr
&=\int_{\mathbb{R}^2}r_{xxxx}f_5\dif x\wedge\dif y\qquad\Rightarrow\qquad r_{xxxx}=0.
\end{align*}

\paragraph*{\textit{Termination of the algorithm}.} Let us define $M_6\subset M_5$ as the zero locus of $r_{xxxx}$ in $M_5$; by adding the hamiltonian vector field of $F_6:=\int_{\mathbb{R}^2}f_6r_{xxxx}\dif x\wedge\dif y$ to $\left(TM_5\right)^\perp$, we obtain the vector space $\left(TM_6\right)^\perp$; it can be proved that the function $\dif H^z\left(X_{F_6}\right)$ is identically zero on $M_6$, and we can conclude that it is an invariant submanifold of $\Gamma\left(N_0\right)$. By performing an elimination of the auxiliar variables on the set of constraints $C_6:=C_5\cup\left\{r_{xxxx}\right\}$ and keeping those involving only the original variables $x,y,z,\phi$, we see that the integrability constraints must be $\phi_{xxy}=0,\phi_{xxxx}=0$. A note of caution about the applicability of the method is that this elimination process can in principle be as involved as the problem of finding the original integrability conditions.

\subsection{Non regular EDS and Gotay-Nester algorithm failure}\label{GotayNesterFailure}
We want to construct an example where the Dirac method fails in some sense; in order to do that, we will use the
approach to variational problems developed here. The rationale behind this is
the following: We know from the theorem \ref{TeoremaDiracEDS} that whenever the EDS
associated to the equations of motion admits regular elements, there exists a
one-to-one correspondence between the solutions for this EDS and the solutions
for the dynamical system on the sections of the selected spatial
slice. Therefore we need to find a variational problem whose extremals are
described for an EDS with non regular integral elements. Concretely, we will take a system where the non regularity stems from singularities of the set of integral elements: this lack of regularity cannot be circumvected via prolongation, and the Dirac method would fail. So the Gotay algorithm could run into problems in facing them.

\subsubsection{Initial setting}
Let us consider the EDS $\cI$ generated by
\[
\theta_1:=\dif u-q\dif x,\theta_2:=\dif v-q\dif r
\]
on $\Lambda:=\mathbb{R}^6$ with coordinates $\left(x,y,u,v,q,r\right)$; the independence condition is $\dif x\wedge\dif y\not=0$. The idea is the one given in example \ref{ExampleEDSasNonStandard}, that is, by considering that $\cI$ as the ideal defining a prolongation structure on $\Lambda$, which has the double fibration structure via
\[
\begin{array}{rcccl}
\Lambda&\rightarrow&\Lambda_1:=\mathbb{R}^4&\rightarrow&M:=\mathbb{R}^2\\
\left(x,y,u,v,q,r\right)&\mapsto&\left(x,y,u,v\right)&\mapsto&\left(x,y\right),
\end{array}
\]
and by taking the lagrangian $\lambda=0$. Using our prescriptions, the Lepage equivalent will be constructed on $\tilde{\Lambda}:=\left(\mathbb{R}^2\times\mathbb{R}^4\right)\oplus\left(T^*\mathbb{R}^2\oplus T^*\mathbb{R}^2\right)$, with action
\[
S:=\int_{\mathbb{R}^2}\left(\alpha\wedge\theta_1+\beta\wedge\theta_2\right),
\]
where $\alpha=a\dif x+b\dif y,\beta=m\dif x+n\dif y$. We need to prove that it is a bivariant Lepage equivalent problem.

\begin{lem}
The projection
\[
\nu:\left(x,y,u,v,q,r,\alpha,\beta\right)\mapsto\left(x,y,u,v,q,r\right)
\]
makes $\left(\tilde{\Lambda},0,\alpha\wedge\theta_1+\beta\wedge\theta_2\right)$ into a Lepagean equivalent of $\left(\Lambda,\cI,0\right)$. Furthermore, it is bivariant, that is, every extremal of the variational problem posed by $S$ on sections of $\tilde{\Lambda}$ projects on an integral section of $\cI$ (because the lagrangian is trivial, there are no action to work with here) and conversely, every integral section of $\cI$ can be lifted to an extremal of $S$.
\end{lem}
\begin{proof}
The variations along $\alpha$ and $\beta$ will give us the generators for $\cI$; this shows that it is a covariant Lepage equivalent problem. On the other side, variations along $u,v,q$ and $r$ leads to the following system on $\alpha$ and $\beta$:
\[
\begin{cases}
\dif\alpha=0,\cr
\dif\beta=0,\cr
\beta\wedge\dif r+\alpha\wedge\dif x=0,\cr
\dif\left(q\beta\right)=0.
\end{cases}
\]
For every $\left(x,y\right)\mapsto\left(u,v,q,r\right)$ integral section for $\cI$, this systems admits solutions, so it is contravariant also.
\end{proof}

\subsubsection{Gotay-Nester algorithm}
Now we start with the analysis of the dynamical equations associated to the regular slicing $y=0$. In order to do that, we take $N_0\subset\tilde{\Lambda}$ the submanifold composed by the points in $\tilde\Lambda$ such that $y=0$; it is a bundle on $\mathbb{R}$, and our next task is to define on its space of sections a presymplectic structure and a hamiltonian function. The first thing to note is that for $\tilde{\lambda}:=\alpha\wedge\theta_1+\beta\wedge\theta_2$
\[
\dif\tilde{\lambda}=\dif\alpha\wedge\left(\dif u-q\dif x\right)+\alpha\wedge\dif q\wedge\dif x+\dif\beta\wedge\left(\dif v-q\dif r\right)+\beta\wedge\dif q\wedge\dif r,
\]
so
\[
\left.\dif\tilde{\lambda}\right|_{N_0}=\dif a\wedge\dif x\wedge\dif u+\dif m\wedge\dif x\wedge\left(\dif v-q\dif r\right)+m\dif x\wedge\dif q\wedge\dif r.
\]
Therefore if
\[
X_i=\left(\delta u_i,\delta v_i,\delta q_i,\delta r_i;\delta\alpha_i,\delta\beta_i\right),\qquad i=1,2
\]
denotes a pair of arbitrary elements in $T\Gamma\left(N_0\right)$, the presymplectic structure will be
\begin{multline*}
\omega^y\left(X_1,X_2\right)=\int_{\mathbb{R}}\Big[-\delta a_1\delta u_2-\delta m_1\delta v_2-m\delta r_1\delta q_2+\cr
+\left(q\delta m_1+m\delta q_1\right)\delta r_2+\delta u_1\delta a_2+\left(\delta v_1-q\delta r_1\right)\delta m_2\Big]\dif x
\end{multline*}
and
\[
H^y=\int_{\mathbb{R}}\left[b\left(u_x-q\right)+n\left(v_x-qr_x\right)\right]\dif x.
\]
\paragraph*{\textit{Primary constraints.}} As above
\[
\text{ker}\,\omega^y=\left<\underbrace{\left(0,0,0,0;\delta b\dif y,0\right)}_{Z_1},\underbrace{\left(0,0,0,0;0,\delta n\dif y\right)}_{Z_2}\right>
\]
and therefore the primary constraints will arise from $\dif H^y\left(\text{ker}\,\omega^y\right)=0$. Then
\begin{align*}
0&=\dif H^y\left(Z_1\right)\cr
&=\int_{\mathbb{R}}\delta b\left(u_x-q\right)\dif x\qquad\Rightarrow\qquad u_x-q=0,\cr
0&=\dif H^y\left(Z_2\right)\cr
&=\int_{\mathbb{R}}\delta n\left(v_x-qr_x\right)\dif x\qquad\Rightarrow\qquad v_x-qr_x=0.
\end{align*}
Let us define $C_1:=\left\{u_x-q,v_x-qr_x\right\}$ and $M_1:=\left\{C_1=0\right\}$;
in order to find the two vectors complementary to $\text{ker}\,\omega^y$, we define
\[
F_1:=\int_{\mathbb{R}}f_1\left(u_x-q\right)\dif x,\qquad F_2:=\int_{\mathbb{R}}f_2\left(v_x-qr_x\right)\dif x;
\]
then
\begin{align*}
X_2\cdot F_1&=\int_{\mathbb{R}}f_1\left[\left(\delta u_2\right)_x-\delta q_2\right]\dif x\cr
&=-\int_{\mathbb{R}}\left[\left(f_1\right)_x\delta u_2+f_1\delta q_2\right]\dif x,\cr
X_2\cdot F_2&=\int_{\mathbb{R}}f_2\left[\left(\delta v_2\right)_x-r_x\delta q_2-q\left(\delta r_2\right)_x\right]\dif x\cr
&=\int_{\mathbb{R}}\left[\left(qf_2\right)_x\delta r_2-\left(f_2\right)_x\delta v_2-r_xf_2\delta q_2\right]\dif x.
\end{align*}
Then, under the assumption $m\not=0$, we obtain that
\begin{align*}
X_{F_1}&=\left(0,\frac{qf_1}{m},0,\frac{f_1}{m};\left(f_1\right)_x\dif x,0\right)\cr
X_{F_2}&=\left(0,\frac{qr_xf_2}{m},\frac{q_xf_2}{m},\frac{r_xf_2}{m};0,\left(f_2\right)_x\dif x\right),
\end{align*}
yielding to the secondary constraints
\begin{align*}
0&=\dif H^y\left(X_{F_1}\right)\cr
&=\int_{\mathbb{R}}n\left[\left(\frac{qf_1}{m}\right)_x-q\left(\frac{f_1}{m}\right)_x\right]\dif x\cr
&=\int_{\mathbb{R}}n\frac{q_xf_1}{m}\dif x\qquad\Rightarrow\qquad n\frac{q_x}{m}=0\cr
0&=\dif H^y\left(X_{F_2}\right)\cr
&=\int_{\mathbb{R}}\Big\{-b\frac{q_xf_2}{m}+n\left[\left(\frac{qr_xf_2}{m}\right)_x-r_x\frac{q_xf_2}{m}-q\left(\frac{r_xf_2}{m}\right)_x\right]\Big\}\dif x\cr
&=-\int_{\mathbb{R}}b\frac{q_xf_2}{m}\dif x\qquad\Rightarrow\qquad b\frac{q_x}{m}=0.
\end{align*}
We are now in a very uncomfortable situation: Let us suppose that we define $C_2:=C_1\cup\left\{n\frac{q_x}{m},b\frac{q_x}{m}\right\}$ and $M_2:=\left\{C_2=0\right\}$, and we want to calculate $\text{ker}\,\omega^y\cap TM_2$. This set is composed by those vectors in $\text{ker}\,\omega^y$ that annihilates the differentials of the new constraints; therefore $Z=\left(0,0,0,0;\delta b\dif y,\delta n\dif y\right)\in TM_2$ iff
\[
\begin{cases}
\delta n\frac{q_x}{m}=0\cr
\delta b\frac{q_x}{m}=0.
\end{cases}
\]
Thus if $q_x\not=0$ the algorithm must stops, because in this case $TM_2^\perp=TM_1^\perp$, and it will continue in those points where $q_x=0$. This is a very bad behaviour of the Gotay-Nester algorithm, and it has to do with the singular nature of the new constraints.

\subsubsection{EDS analysis}
The analysis of this PDE system is interesting, because its set of integral elements consists of several parts, each of them with different regularity behaviour. In fact, we will see that these differences introduces the singularities in the Gotay-Nester algorithm.
\\
The integral elements for $\cI$ divides into two sets: the set $W$ consisting of the elements with $q_x=0$, and $U_3$ is the set of integral elements such that $n=b=0$. None of them has regular elements; in particular, the elements in $U_3$ has $0$-forms in its definition, so this set needs to be redefined. The procedure is to pull back the original EDS $\cI$ to the submanifold defined by the collection of problematic $0$-forms. It gives rise to a new EDS $\cI_3$ whose set of integral elements admits a new decomposition into a pair of subsets $V_1$ and $V_3$. The set $V_3$ is characterized by the $0$-form constraint $m=0$, and will not be studied; the set $V_1$ is composed by regular elements. Additionally the prolongation of the portion of the EDS corresponding to the elements in $W$ gives rise to a new EDS $\cI_W$ on $G_2\left(T\mathbb{R}^{10}\right)$ which is regular too.
\\
Therefore it is necessary to know the main characteristics of our EDS $\cI$; first of all, a set of
algebraic generators will be
\[\left\{\theta_1,\theta_2,\dif q\wedge\dif x,\dif
  q\wedge\dif r,\dif\alpha,\dif\beta,\beta\wedge\dif r+\alpha\wedge\dif x,\beta\wedge\dif q\right\}.
\]
Now we proceed to calculate the (reduced) characters
for the flag $E_0:=0\subset E_1:=\left<v_1\right>\subset
E:=\left<v_1,v_2\right>$, where
\begin{align*}
v_1&:=\partial_x+U_1\partial_u+V_1\partial_v+Q_1\partial_q+R_1\partial_r+A_1\partial_a+B_1\partial_b+M_1\partial_m+N_1\partial_n,\cr
v_2&:=\partial_y+U_2\partial_u+V_2\partial_v+Q_2\partial_q+R_2\partial_r+A_2\partial_a+B_2\partial_b+M_2\partial_m+N_2\partial_n
\end{align*}
are in $T_{\left(x,y,u,v,q,r\right)}\mathbb{R}^{10}$. Then these components must
satisfy
\[
\begin{cases}
v_1\lrcorner\left(\dif u-q\dif
    x\right)=0&\Leftrightarrow\quad U_1-q=0,\cr
v_1\lrcorner\left(\dif v-q\dif r\right)=0&\Leftrightarrow\quad V_1-qR_1=0,\cr
v_2\lrcorner\left(\dif u-q\dif
    x\right)=0&\Leftrightarrow\quad U_2=0,\cr
v_2\lrcorner\left(\dif v-q\dif r\right)=0&\Leftrightarrow\quad V_2-qR_2=0,\cr
\left(v_1\wedge v_2\right)\lrcorner\dif q\wedge\dif x=0&\Leftrightarrow\quad Q_2=0,\cr
\left(v_1\wedge v_2\right)\lrcorner\dif q\wedge\dif r=0&\Leftrightarrow\quad Q_2R_1-Q_1R_2=0,\cr
\left(v_1\wedge v_2\right)\lrcorner\dif\alpha=0&\Leftrightarrow\quad A_2-B_1=0,\cr
\left(v_1\wedge v_2\right)\lrcorner\dif\beta=0&\Leftrightarrow\quad M_2-N_1=0,\cr
\left(v_1\wedge v_2\right)\lrcorner\left(\beta\wedge\dif r+\alpha\wedge\dif x\right)=0&\Leftrightarrow\quad nR_1-mR_2-b=0,\cr
\left(v_1\wedge v_2\right)\lrcorner\beta\wedge\dif q=0&\Leftrightarrow\quad nQ_1-mQ_2=0.
\end{cases}
\]
Thus $V_2\left(\cI\right)$ is described by the equations
\[
\begin{cases}
U_1=q,U_2=0,\cr
V_1=qR_1,V_2=qR_2,\cr
Q_2=0,Q_1R_2=0,\cr
A_2-B_1=0,\cr
M_2-N_1=0,\cr
nR_1-mR_2-b=0,\cr
nQ_1=0.
\end{cases}
\]
It is necessary to analyze carefully this system. It results that $V_2\left(\cI\right)=W\cup U_3$, where
\begin{align*}
W:\begin{cases}
U_1=q,U_2=0,\cr
V_1=qR_1,V_2=qR_2,\cr
Q_1=0,Q_2=0,\cr
A_2-B_1=0,\cr
M_2-N_1=0,\cr
nR_1-mR_2-b=0,
\end{cases}
U_3:\begin{cases}
U_1=q,U_2=0,\cr
V_1=qR_1,V_2=0,\cr
Q_2=0,R_2=0,\cr
A_2-B_1=0,\cr
M_2-N_1=0,\cr
b=0,\cr
n=0.
\end{cases}
\end{align*}
Let us calculate the polar spaces for the integral element $E$; we have that
\begin{align*}
H\left(E_0\right)&=\left\{v\in
  T_{\left(x,y,u,v,q,r,a,b,m,n\right)}\mathbb{R}^{10}:v\lrcorner\left(\dif u-q\dif
    x\right)=0,v\lrcorner\left(\dif v-q\dif r\right)=0\right\},\cr
H\left(E_1\right)&=\Big\{v\in
H\left(E_0\right):v\lrcorner\left(v_1\lrcorner\dif q\wedge\dif
    x\right)=0,v\lrcorner\left(v_1\lrcorner\dif q\wedge\dif
    r\right)=0,v\lrcorner\left(v_1\lrcorner\dif\alpha\right)=0,\cr
&v\lrcorner\left(v_1\lrcorner\dif\beta\right)=0,v\lrcorner\left(v_1\lrcorner\left(\beta\wedge\dif r+\alpha\wedge\dif x\right)\right)=0,v\lrcorner\left(v_1\lrcorner\beta\wedge\dif q\right)=0
\Big\}\cr
&=\Big\{v\in
H\left(E_0\right):v\lrcorner\left(Q_1\dif x-\dif q\right)=0,v\lrcorner\left(Q_1\dif r-R_1\dif q\right)=0,\cr
&v\lrcorner\left(A_1\dif x+B_1\dif y-\dif a\right)=0,v\lrcorner\left(M_1\dif x+N_1\dif y-\dif m\right)=0,\cr
&\qquad v\lrcorner\left(m\dif r-R_1\beta-b\dif y\right)=0,v\lrcorner\left(m\dif q-Q_1\beta\right)=0
\Big\}.
\end{align*}
We can consider the reduced Cartan characters for $E$ in each of the sets
$W$ and $U_3$; in every case we have that $c_0\left(E\right)=2$ and
$
c_1\left(E\right)=6
$
on $W$, by supposing that\footnote{In points where $m=0$, some of
the characters decrease by one.} $m\not=0$. Moreover we have that
$
\text{codim}_EW=9,
$
so there are no regular elements in $W$.
\\
On $U_3$ we must take care of the
fact that it is defined through $0$-forms, namely, the constraints $n=0$ and
$b=0$. The regularity of $\cI$ must be studied through the regularity of the EDS $\cI_3$ obtained from $\cI$ by
pulling back to the submanifold
\[
\tilde\Lambda_3:=\left\{\left(x,y,u,v,q,r,a,0,m,0\right):x,y,u,v,q,r,a,m\in\mathbb{R}\right\};
\]
a set of generators for $\cI_3$ will be
\[\left\{\theta_1,\theta_2,\dif q\wedge\dif x,\dif
  q\wedge\dif r,\dif a\wedge\dif x,\dif m\wedge\dif x,m\dif x\wedge\dif r,m\dif
  x\wedge\dif q\right\}.
\]
Then if 
\begin{align*}
w_1&:=\partial_x+U_1\partial_u+V_1\partial_v+Q_1\partial_q+R_1\partial_r+A_1\partial_a+M_1\partial_m,\cr
w_2&:=\partial_y+U_2\partial_u+V_2\partial_v+Q_2\partial_q+R_2\partial_r+A_2\partial_a+M_2\partial_m
\end{align*}
the set of $2$-integral elements $V_2\left(\cI_3\right)$ is described by the equations
\[
U_3':\begin{cases}
U_1=q,U_2=0,\cr
V_1=qR_1,V_2=0,\cr
Q_2=0,Q_1R_2=0,\cr
A_2=0,M_2=0,\cr
mR_2=0.
\end{cases}
\]
Thus we will have that $U_3'=V_1\cup V_3$, where
\begin{equation}\label{SistemaU3V1}
V_1:\begin{cases}
U_1=q,U_2=0,\cr
V_1=qR_1,V_2=0,\cr
Q_2=0,R_2=0,\cr
A_2=0,M_2=0,
\end{cases}
V_3:\begin{cases}
U_1=q,U_2=0,\cr
V_1=qR_1,V_2=0,\cr
Q_2=0,Q_1=0,\cr
A_2=0,M_2=0,\cr
m=0
\end{cases}
\end{equation}
and $\text{codim}_EV_1=8$. On the other side, the polar spaces for the flag $E_0':=0\subset E_1':=\left<w_1\right>\subset
E':=\left<w_1,w_2\right>$ will be
\begin{align*}
H\left(E_0\right)&=\left\{v\in
  T_{\left(x,y,u,v,q,r,a,m\right)}\mathbb{R}^8:v\lrcorner\left(\dif u-q\dif
    x\right)=0,v\lrcorner\left(\dif v-q\dif r\right)=0\right\},\cr
H\left(E_1\right)&=\Big\{v\in
H\left(E_0\right):v\lrcorner\left(w_1\lrcorner\dif q\wedge\dif
    x\right)=0,v\lrcorner\left(w_1\lrcorner\dif q\wedge\dif
    r\right)=0,\cr
&v\lrcorner\left(w_1\lrcorner\dif a\wedge\dif
  x\right)=0,v\lrcorner\left(w_1\lrcorner\dif m\wedge\dif x\right)=0,\cr
&\qquad v\lrcorner\left(w_1\lrcorner\left(m\dif x\wedge\dif r\right)\right)=0,v\lrcorner\left(w_1\lrcorner\left(m\dif x\wedge\dif q\right)\right)=0
\Big\}\cr
&=\Big\{v\in
H\left(E_0\right):v\lrcorner\left(Q_1\dif x-\dif
  q\right)=0,v\lrcorner\left(Q_1\dif r-R_1\dif q\right)=0,\cr
&v\lrcorner\left(A_1\dif x-\dif a\right)=0,v\lrcorner\left(M_1\dif x-\dif
  m\right)=0,\cr
&\qquad v\lrcorner\left(m\dif r-mR_1\dif x\right)=0,v\lrcorner\left(m\dif
  q-mQ_1\dif x\right)=0
\Big\},
\end{align*}
and for $E\in V_1$ we have that
$c_0\left(E\right)=2,c_1\left(E\right)=6$. Therefore, the integral elements in
$V_1$ are regular.
\\
Finally it is necessary to consider the prolongation of the set $W$, in order
to see if some of its elements can be considered as the tangent space of some
solution. The prolongation of an EDS $\cI\subset\Omega^\bullet\left(M\right)$
is the pullback of the contact EDS on $G_2\left(TM\right)$ to the set
$V_2\left(\cI\right)$. Let us introduce the coordinates
\[
\left(x,y,u,v,q,r,a,b,m,n,u_x,v_x,q_x,r_x,a_x,b_x,m_x,n_x,u_y,v_y,q_y,r_y,a_y,b_y,m_y,n_y\right)
\]
on $G_2\left(T\mathbb{R}^{10}\right)$; then $W$ is the subset described by
\[
W:\begin{cases}
u_x=q,u_y=0,\cr
v_x=qr_x,v_y=qr_y,\cr
q_x=0,q_y=0,\cr
a_y-b_x=0,\cr
n_x-m_y=0,\cr
nr_x-mr_y-b=0,
\end{cases}
\]
and the pullback $\cI_W$ of the contact system to $W$ will have the generators
\[
\begin{cases}
\theta_u:=\dif u-q\dif x\cr
\theta_v:=\dif v-q\left(r_x\dif x+r_y\dif y\right)\cr
\theta_q:=\dif q\cr
\theta_r:=\dif r-r_x\dif x-r_y\dif y\cr
\theta_a:=\dif a-a_x\dif x-a_y\dif y\cr
\theta_b:=\dif b-a_y\dif x-b_y\dif y\cr
\theta_m:=\dif m-m_x\dif x-m_y\dif y\cr
\theta_n:=\dif n-m_y\dif x-n_y\dif y
\end{cases}
\Rightarrow\quad
\begin{cases}
\dif\theta_u\equiv0\cr
\dif\theta_v=-q\left(\dif r_x\wedge\dif x+\dif r_y\wedge\dif y\right)\cr
\dif\theta_q=0\cr
\dif\theta_r=-\dif r_x\wedge\dif x-\dif r_y\wedge\dif y\cr
\dif\theta_a=-\dif a_x\wedge\dif x-\dif a_y\wedge\dif y\cr
\dif\theta_b=-\dif a_y\wedge\dif x-\dif b_y\wedge\dif y\cr
\dif\theta_m=-\dif m_x\wedge\dif x-\dif m_y\wedge\dif y\cr
\dif\theta_n=-\dif m_y\wedge\dif x-\dif n_y\wedge\dif y.
\end{cases}
\]
Let $E$ be the flag $E_0:=0\subset E_1:=\left<v_1\right>\subset
E:=\left<v_1,v_2\right>$, with basis
\begin{align*}
v_1&:=\partial_x+U_1\partial_u+V_1\partial_v+Q_1\partial_q+R_1\partial_r+A_1\partial_a+B_1\partial_b+M_1\partial_m+N_1\partial_n+\cr
&+U_x^1\partial_{u_x}+V_x^1\partial_{v_x}+Q_x^1\partial_{q_x}+R_x^1\partial_{r_x}+A_x^1\partial_{a_x}+B_x^1\partial_{b_x}+M_x^1\partial_{m_x}+N_x^1\partial_{n_x}+\cr
&+U_y^1\partial_{u_y}+V_y^1\partial_{v_y}+Q_y^1\partial_{q_y}+R_y^1\partial_{r_y}+A_y^1\partial_{a_y}+B_y^1\partial_{b_y}+M_y^1\partial_{m_y}+N_y^1\partial_{n_y},\cr
v_2&:=\partial_y+U_2\partial_u+V_2\partial_v+Q_2\partial_q+R_2\partial_r+A_2\partial_a+B_2\partial_b+M_2\partial_m+N_2\partial_n+\cr
&+U_x^2\partial_{u_x}+V_x^2\partial_{v_x}+Q_x^2\partial_{q_x}+R_x^2\partial_{r_x}+A_x^2\partial_{a_x}+B_x^2\partial_{b_x}+M_x^2\partial_{m_x}+N_x^2\partial_{n_x}+\cr
&+U_y^2\partial_{u_y}+V_y^2\partial_{v_y}+Q_y^2\partial_{q_y}+R_y^2\partial_{r_y}+A_y^2\partial_{a_y}+B_y^2\partial_{b_y}+M_y^2\partial_{m_y}+N_y^2\partial_{n_y};
\end{align*}
then $V_2\left(\cI_W\right)$ is given by
\[
\begin{cases}
U_1=q,U_2=0,V_1=qr_x,V_2=qr_y,Q_1=0,Q_2=0,R_1=r_x,R_2=r_y,\cr
A_1=a_x,A_2=a_y,B_1=a_y,B_2=b_y,M_1=m_x,M_2=m_y,N_1=m_y,N_2=n_y,\cr
R_x^2-R_y^1=0,A_x^2-A_y^1=0,B_x^2-B_y^1=0,
M_x^2-M_y^1=0,N_x^2-R_y^1=0.
\end{cases}
\]
This means that $\text{codim}_EV_2\left(\cI_W\right)=21$. For $E$ integral element the polar spaces are
\begin{align*}
H\left(E_0\right)&=\Big\{v\in
  TW:v\lrcorner\theta_u=0,v\lrcorner\theta_v=0,v\lrcorner\theta_q=0,v\lrcorner\theta_r=0,\cr
&v\lrcorner\theta_a=0,v\lrcorner\theta_b=0,v\lrcorner\theta_m=0,v\lrcorner\theta_n=0\Big\}\cr
H\left(E_1\right)&=\Big\{v\in H\left(E_0\right):v\lrcorner\left(\dif
    r_x+\cdots\right)=0,v\lrcorner\left(\dif a_x+\cdots\right)=0,\cr
&v\lrcorner\left(\dif a_y+\cdots\right)=0,v\lrcorner\left(\dif m_x+\cdots\right)=0,v\lrcorner\left(\dif m_y+\cdots\right)=0\Big\},
\end{align*}
and the reduced Cartan characters results
$c_0\left(E\right)=8,c_1\left(E\right)=13$; therefore $E$ is a regular
integral element.

\subsubsection{Return to Gotay-Nester algorithm}

Nevertheless, it can be instructive to desingularize these constraints by mimicking the finite dimensional procedure: That is, by considering that the sections fulfilling the problematic constraints are the union of the sections such that $n=b=0$ with the set of sections satisfying the single constraint $q_x=0$. Despite the validity of such a procedure, we consider the continuation of the algorithm in each of these sets.
\begin{itemize}
\item\textbf{Case $n=b=0$.} In this case $C_2':=C_1\cup\left\{n,b\right\}$ and if $M_2'$ is the zero locus for $C_2'$
. Therefore the algorithm must stops, namely, $M_2'$ is an invariant manifold. In order to find the dynamical equations, let us obtain the hamiltonian vector field for $H^y$ on $\Gamma\left(N_0\right)$, restricting it to the submanifold $M_2'$. We will have that
\begin{multline}\label{FlaHamiltonianVectorField}
X_2\cdot H^y=\int_{\mathbb{R}}\Big\{\delta b_2\left(u_x-q\right)+\delta n_2\left(v_x-qr_x\right)+\cr
+b\left[\left(\delta u_2\right)_x-\delta q_2\right]+n\left[\left(\delta v_2\right)_x-r_x\delta q_2-q\left(\delta r_2\right)_x\right]\Big\}\dif x
\end{multline}
and so $\left.X_2\cdot H^y\right|_{M_2'}=0$; if the hamiltonian vector field for $H^y$ reads $X_{H^y}=\left(u_y,v_y,q_y,r_y,a_y\dif x+b_y\dif y,m_y\dif x+n_y\dif y\right)$, the tangent solution to $M_2'$ of the Hamilton eqs must be\footnote{All the hamiltonian vector fields for $H^y$ has the form $X_{H^y}=\mu_1Z_1+\mu_2Z_2$, and the tangent condition requires $\mu_1=\mu_2=0$.} $X_{H^y}=0$ and the dynamical equations reads
\[
\begin{cases}
u_y=0\cr
v_y-qr_y=0\cr
q_y=0\cr
a_y=0\cr
m_y=0.
\end{cases}
\]
It can be compared with the regular EDS $V_1$ described by Eq. \ref{SistemaU3V1}: This branch of the regularization procedure finds the solutions associated to regular elements in $V_1$.
\item\textbf{Case $q_x=0$.} Now $C_2'':=C_1\cup\left\{q_x\right\}$, and $M_2''=\left\{C_2''=0\right\}$. 
If we define
\[
G:=\int_{\mathbb{R}}gq_x\dif x
\]
its derivative along $X_2$ results
\begin{align*}
X_2\cdot G&=\int_{\mathbb{R}}g\left(\delta q_2\right)_x\dif x=-\int_{\mathbb{R}}g_x\delta q_2\dif x.
\end{align*}
Therefore
\[
X_G=\left(0,q\frac{g_x}{m},0,\frac{g_x}{m},0;0,0\right)\in\left(TM_2'\right)^\perp
\]
and the invariance condition will reads
\begin{align*}
0&=X_G\cdot H^y\cr
&=\int_{\mathbb{R}}n\left[\left(q\frac{g_x}{m}\right)_x-q\left(\frac{g_x}{m}\right)_x\right]\dif x\cr
&=\int_{\mathbb{R}}g_x\frac{nq_x}{m}\dif x\qquad\Rightarrow\qquad\left(n\frac{q_x}{m}\right)_x=0.
\end{align*}
This condition is fulfilled on $M_2''$, so it is an invariant manifold for the dynamics defined by $\omega^y$ and $H^y$. Thus let us calculate the hamiltonian vector field $X_{H^y}$, and then restricts it to $M_2''$; by taking into account Eq. \ref{FlaHamiltonianVectorField} we have that
\[
\left.X_2\cdot H^y\right|_{M_2''}=\int_{\mathbb{R}}\Big[-b_x\delta u_2-\left(b-nr_x\right)\delta q_2-n_x\delta v_2\left(nq\right)_x\delta r_2\Big]\dif x.
\]
Therefore the equations of motion are
\[
\begin{cases}
u_y=0\cr
v_y-qr_y=0\cr
q_y=0\cr
a_y=b_x\cr
b_y=\delta b\cr
m_y=n_x\cr
n_y=\delta n.
\end{cases}
\]
In this case the solutions of the dynamical equations are in a one-to-one correspondence to solutions associated to the regular elements in the prolongation of $W$; moreover, the additional step in the Gotay algorithm has to do with the extra prolongation performed.
\end{itemize}

\subsection{Some remarks}\label{RemarkFieldTheory}

It must be noted that there exists a field theory whose Dirac constraints has
similar behaviour to the constraints found in the last example: It is the system
composed of a massless spin $1$ charged particle and the electromagnetic field
\cite{GotayNotas}. A detailed analysis of such a system from the point of
view adopted in this article is being performed; the true nature of the problems encountered there is under investigation yet \cite{HernanSanti}. Nevertheless, it is interesting to note here that the non
standard perspective permits us to analyse any EDS (in particular, any PDE
system) using techniques of (pre)symplectic geometry. The example
\ref{ExampleOfProlongation} is a sample in this direction.

\section{Conclusions}

In this paper it was built an entire dynamical theory for non standard
variational problems, at least for those admitting a Cartan form.
We were able also to describe the
constraint submanifold in every case where such a description is possible.
It was proved that this submanifold is the space of integral sections
of an EDS which is associated to the EDS generated by the Hamilton-Cartan
equations and the slicing of the space-time. We saw that the Dirac
constraints has two sources: They can arise from the differential
closure of the Hamilton-Cartan EDS, although some others can appears
because of the involutiveness requeriment, which is a consequence
of the stability condition defining the constraint submanifold. Four relevant examples showing these aspects were discussed in some extent.

\begin{acknowledgement*}
The author would like to thank to his advisors, Hugo Montani
and Hern\'{a}n Cendra, for many ideas, comments and corrections.
He wants to thanks warmly to Sergio Grillo for useful comments on the work, and CONICET for finantial support. Also, the author offers special thanks to the referees for their useful comments which have improved the first version of this paper.
\end{acknowledgement*}

\appendix

\section{On (formal) variational calculus}\label{Section:1}

The purpose of this appendix is to set some common facts about geometrical concepts involved in the description of spaces of sections and the variational calculus used throughout the work \cite{Gotay:1997eg,Gotay:2004ib}.

\subsection{Variational derivative}

Given the bundle $F\rightarrow B$, let us consider on $\Gamma\left(F\right)$
a monoparametric family $\sigma_{t}$ of sections such that $\sigma_{0}=\sigma$;
for every $b\in B$ we can calculate the following vector\[
V\left(b\right):=\vec{\frac{\text{d}}{\text{d}t}}\left[\sigma_{t}\left(b\right)\right]\Big|_{t=0}\in T_{\sigma\left(b\right)}F\]
which has the following properties:

\begin{itemize}
\item It is a vertical vector, that is, it belongs to $\text{ker}\left(\tau_{*}\right)$.
\item The map $b\mapsto V\left(b\right)$ covers $\sigma$, that is, it
makes the following diagram commutative\[
\begin{diagram}
\node[2]{VF}\arrow{s,r}{\tau_F}\\
\node{B}\arrow{ne,t}{V}\arrow{e,b}{\sigma}\node{F}
\end{diagram}
\] 
\end{itemize}
Therefore $V$ is a section of the pullback bundle $\sigma^{*}\left(VF\right)$.
So we make the following definition.

\begin{defn}
The \emph{tangent space at $\sigma$} of the space of sections $\Gamma\left(F\right)$
of a bundle $F\stackrel{\tau}{\longrightarrow}B$ is the space of
sections $\Gamma\left(\sigma^{*}\left(VF\right)\right)$.
\end{defn}
\begin{note*}
We want to remark that the previous one is just a definition, and
does not implies the existence of differential structure in particular
cases.
\end{note*}
Now we proceed in the opposite direction: Let us suppose that
we have an element $V\in\Gamma\left(\sigma^{*}\left(VF\right)\right)$,
and we need to build a monoparametric family of sections $\sigma_{t}\in\Gamma\left(F\right)$
such that

\begin{itemize}
\item They satisfies the initial condition $\sigma_{0}=\sigma$.
\item The section $V$ can be reconstructed through the formula \[
V\left(b\right):=\vec{\frac{\text{d}}{\text{d}t}}\left[\sigma_{t}\left(b\right)\right]\Big|_{t=0}.\]

\end{itemize}
This is the case when $V$ admits an extension%
\footnote{This means that there exist on $F$ a vector field $\hat{V}$ whose
restriction to the image of $\sigma$ is equal to $V$%
} $\hat{V}\in\Gamma\left(VF\right)$; in general it is a problem with no solution.
Nevertheless there exists a particular case for which there always exists
such an extension, namely whenever $F$ is a vector bundle.

\begin{prop}
\label{pro:Existe-extension}Let $F\stackrel{\tau}{\longrightarrow}B$
a vector bundle, $\sigma\in\Gamma\left(F\right)$ a section and $V\in\Gamma\left(\sigma^{*}\left(VF\right)\right)$
a tangent vector at $\sigma$. Then there exists $\hat{V}\in\Gamma\left(VF\right)$
which extends $V$.
\end{prop}
\begin{proof}
Because $F$ is a vector bundle, we have the identification $V_{\sigma\left(b\right)}\left(F\right)\simeq\tau^{-1}\left(b\right)$,
and then we can consider $V$ as a section for $F$. Then defining
the family of sections $\sigma_{t}:b\mapsto\sigma\left(b\right)+tV\left(b\right)$
we obtain a vector field on $F$ (the tangent vector field) extending
the given vector.
\end{proof}
A \emph{functional}  on $F$ is a function on $\Gamma\left(F\right)$ (from
a set-theoretical point of view). We say that a functional $S$ on $F$ is \emph{differentiable} at
$\sigma\in\Gamma\left(F\right)$ iff for every extensible $V\in\Gamma\left(\sigma^{*}\left(VF\right)\right)$
the number\[
V|_{\sigma}\cdot S:=\frac{\text{d}}{\text{d}t}\left(S\left[\phi_{t}\left(\sigma\right)\right]\right)\Big|_{t=0}\]
exists and it is independent of the extension. Here $\phi_{t}:\Gamma\left(F\right)\rightarrow\Gamma\left(F\right)$
is the flux in $\Gamma\left(F\right)$ defined through $\psi_{t}:F\rightarrow F,\psi_{0}=\text{id}_{\Lambda_{1}}$
associated to some extension for $V$ (that is, $\phi_{t}\left(\sigma\right):b\mapsto\left(\psi_{t}\circ\sigma\right)\left(b\right)$).
With respect to this subclass of functionals we define the \emph{variational
derivative.}

\begin{defn}
The variational derivative of the differentiable functional $S$ in
$\sigma\in\Gamma\left(F\right)$ is the form $\delta S\left[\sigma\right]:T_{\sigma}\Gamma\left(F\right)\rightarrow\mR$,
given by\[
\delta S\left[\sigma\right]\left(V\right)=V|_{\sigma}\cdot S.\]

\end{defn}
\begin{rem}
The uniqueness of the derivative of a functional depends strongly
on each case; it has to do with the choice of the extension of the vector $V$.
\end{rem}

\subsection{Variational derivative for special functionals}

We want to study the variational derivative of some class of differentiable
functionals on $F$, that is, those that can be given by the formula\[
S_{\lambda}[\sigma]:=\int_{B}\sigma^{*}\left(\lambda\right)\]
for some $\lambda\in\Omega^{n}\left(F\right),n=\dime B$.

\begin{prop}
\label{pro:The-variational-derivative}\cite{GotayCartan} The variational
derivative for $S_{\lambda}$ at $\sigma$ is given by\[
\delta S_{\lambda}\left[\sigma\right]\left(V\right)=\int_{B}\sigma^{*}\left(\mathcal{L}_{\hat{V}}\lambda\right),\]
where $\hat{V}\in\Gamma\left(VF\right)$ is some extension for $V$. In particular, on these kind of functionals the variational derivative is unique.
\end{prop}

\section{Some facts about exterior differential systems}\label{Section:2}

In this section we introduce the basic concepts of the theory of EDS; the sources of the material
presented here are \cite{BCG,CartanBeginners,nkamran2000,OlvEquiv}.
The problem is to find out conditions that ensures us the existence
of submanifolds on which a certain set of forms vanish. Since
any of such submanifold annihilates the algebraic ideal generated
by these forms and its differentials, let us introduce the following definition.

\begin{defn}
An \emph{exterior differential system} on the manifold $M$ is an
ideal $\cI\subset\Omega^{\bullet}\left(M\right)$ closed with respect to
the exterior differential operator. We say then that $\cI$ is \emph{differentially
closed.} A manifold which annihilates all the elements in $\cI$
is called \emph{integral submanifold} of the EDS.
\end{defn}
It will be supposed that the EDS does not contain $0$-forms; it is not a true
restriction in cases where the zero locus of these $0$-forms is a submanifold
of $M$. In order to build up the integral submanifolds of a given EDS, we
find out the planes (of the corresponding dimension) that annihilates
$\cI$, asking about the conditions under which these planes
are tangent to some submanifold. The following definition is motivated
by these ideas.

\begin{defn}
An \emph{integral element} of dimension $k$ for an EDS $\cI$ is
a subspace $E\subset T_{x}M$ of dimension $k$ such that $\alpha|E=0$
for all $\alpha\in\cI$. The set of $k$-integral elements associated
to a given $\cI$ will be denoted by $V_{k}\left(\cI\right)$, and
it is naturally included in the manifold $G_{k}\left(TM\right)$,
the $k$-grassmannian on $M$.
\end{defn}
The verification that a given subspace is integral for some EDS involves
the resolution of a (mostly large) linear system, as the following
fact shows.

\begin{lem}
\label{lem:For-integral-elements}Let us define $\cI^{k}:=\cI\cap\Omega^{k}\left(M\right)$;
moreover, let us fix \[
\left(\alpha_{1}^{1},\cdots,\alpha_{k_{1}}^{1},\alpha_{1}^{2},\cdots,\alpha_{k_{2}}^{2},\cdots,\alpha_{1}^{p},\cdots,\alpha_{k_{p}}^{p}\right)\subset\Omega^{\bullet}\left(M\right)\]
a set of generators (in the algebraic sense) for the EDS $\cI$
such that $\left\{ \alpha_{1}^{l},\cdots,\alpha_{k_{l}}^{l}\right\} \subset\cI^{l}$
for all $l$. Then 
\begin{itemize}
\item $V_{k}\left(\cI\right)|_{x}=\left\{ E\in G_{k}\left(T_{x}M\right):\alpha|E=0\quad\forall\alpha\in\cI^{k}\right\} .$
\item The subspace $E\subset T_{x}M$ is $r$-integral for $\cI$ if and
only if $\alpha_{l}^{m}|E=0$ for all $1\leq l\leq k_{m}$ and $1\leq m\leq r$.
\end{itemize}
\end{lem}
It is important to note that in order to verify integrability in terms
of algebraic generators of an EDS, it is necessary to prove that the
given subspace annihilates all the generators of degree less than
or equal to its dimension.

If the EDS $\cI$ is defined on a bundle $F\stackrel{\tau}{\longrightarrow}B$,
we ask about integral submanifolds which are sections of the given
bundle, called \emph{integral sections}\footnote{The set of integral sections for an EDS on a fiber bundle will be denoted by the symbol $\Gamma\left(\cI\right)$.}; if $n=\text{dim}B$, the
procedure that allows to find them works by fixing a nonzero element
$\Omega\in\Omega^{n}\left(B\right)$ and looking for $n$-integral manifolds
for $\cI$ on which $\tau^{*}\left(\Omega\right)\not=0$. This discussion
motivates the following definition.

\begin{defn}
An \emph{EDS with independence condition} on a manifold $M$ is a
pair $\left(\cI,\Omega\right)$ composed by an EDS $\cI$ and a differential
$n$-form $\Omega$. The integral elements (resp. submanifolds) of
such an object are the integral elements (resp. submanifolds) of $\cI$
such that $\Omega\not=0$ on them.
\end{defn}
The independence condition, although apparently innocuous, changes
dramatically the analysis of an EDS, as we will see in the discussion
on the Cartan-K\"{a}hler theorem.

\subsection{K\"{a}hler-ordinary integral elements}

First we need to speak about the notions of regularity which are necessary
in the hypothesis of the Cartan-K\"{a}hler theorem. The first thing that
we have to impose is the regularity condition for the integral elements in the differential geometric
sense.

\begin{defn}
Let $\cI\subset\Omega^{\bullet}\left(M\right)$ be a EDS. An $n$-integral
element $E\in V_{n}\left(\cI\right)$ is \emph{K\"{a}hler-ordinary} iff
there exists a neighborhood $E\in U\subset G_{n}\left(TM\right)$
such that $U\cap V_{n}\left(\cI\right)$ is a submanifold of $G_{n}\left(TM\right)$.
\end{defn}
This is a typical requeriment which ensures (at least locally) a description
of the set of $n$-integral elements as the level set of some differentiable
functions. In general the sets $V_{n}\left(\cI\right)$ are algebraic
subvarieties of $G_{n}\left(TM\right)$ and so there exists singular
points.

\subsection{Polar spaces and K\"{a}hler-regularity}

In order to build up integral submanifolds of dimension $n$ from
an integral submanifold of $\left(n-1\right)$-dimensional, we need
to consider the possible directions in which it is possible.

\begin{defn}
Given an element $E\in V_{k}\left(\cI\right)$, we will define its
\emph{polar space} as the subspace generated for all the integral
elements of dimension $k+1$ associated to $\cI$ which contains $E$. The polar space for an integral element $E$ will be denoted by $H\left(E\right)$.
\end{defn}
The following lemma gives us a way in which we can calculate a polar
space.

\begin{lem}
If $E\subset T_{x}X$ and $\left\{ \bv_{1},\cdots,\bv_{k}\right\} $
is a basis of $E$, then\[
H\left(E\right)=\left\{ w\in T_{x}X:\phi\left(\bv_{1},\cdots,\bv_{k},w\right)=0,\quad\forall\phi\in\cI^{k+1}\right\} .\]

\end{lem}
From the previous lemma we get some intuition about how the polar
spaces are determined. That is, supposing that $\cI$ has generators
like in lemma \ref{lem:For-integral-elements} and that $E=\left\langle \bv_{1},\cdots,\bv_{n}\right\rangle $;
the vector $\bx\in T_{x}X$ will belongs to $H\left(E\right)$ iff\begin{equation}
\begin{cases}
\alpha_{\nu}^{2}\left(\bv_{i_{1}},\bx\right)=0, & 1\leq\nu\leq k_{2},1\leq i_{1}\leq n\\
\alpha_{\mu}^{3}\left(\bv_{j_{1}},\bv_{j_{2}},\bx\right)=0, & 1\leq\mu\leq k_{3},1\leq j_{1}<j_{2}\leq n\\
\qquad\vdots & \qquad\vdots\\
\alpha_{\sigma}^{n+1}\left(\bv_{1},\cdots,\bv_{n},\bx\right)=0, & 1\leq\sigma\leq k_{n+1}.\end{cases}\label{eq:Linear-System-for-polar}\end{equation}
So, instead of having a set of algebraic equations, we have a linear
system determining $\bx$, which depends on a flag in $E$; the next
notion to be introduced assures the regularity of this system, in
the sense that it deals with the invariance of its rank.

\begin{defn}
A K\"{a}hler-ordinary element $E\in V_{n}\left(\cI\right)$ is \emph{K\"{a}hler-regular}
if \[
\text{codim}H\left(E\right)=\text{codim}H\left(E'\right)\]
 for all $E'$ belonging to a neighborhood of $E$ in $V_{n}\left(\cI\right)\subset G_{n}\left(TX\right)$.
\end{defn}
We need to develop some strategies in order to deal with this condition,
because it is difficult to verify in practice. The main theoretical
resource in this direction is the \emph{Cartan's Test for Involutivity}.

\begin{thm}
[Cartan's Test for involutivity]Let $E_{k},0\leq k\leq n$ a flag
of integral elements for $\cI$ at $x$, and let us define\[
c_{k}:=\text{codim}H\left(E_{k}\right),0\leq k\leq n-1.\]
Then\[
\text{codim}_{E_{n}}V_{n}\left(\cI\right)\geq c_{0}+\cdots+c_{n-1},\]
where the codimension of $V_{n}\left(\cI\right)$ at $E_{n}$ is measured
in the algebraic sense: It is defined as the maximum number of smooth
functions on $G_{n}\left(TX\right)$ which vanish on $V_{n}\left(\cI\right)$
and has independent differentials at $E_{n}$. 

Moreover, $V_{n}\left(\cI\right)$ is smooth of codimension exactly
$c_{0}+\cdots+c_{n-1}$ at $E_{n}$ if and only if the $E_{k}$ are
all K\"{a}hler-regular for $0\leq k\leq n-1$.
\end{thm}

\subsection{Cartan-K\"{a}hler Theorem}

Once we have introduced all the necessary ingredients, we can formulate
a version of the Cartan-K\"{a}hler Theorem \cite{CartanBeginners,nkamran2000}.

\begin{thm}
[Cartan-K\"{a}hler Theorem]Assume $\cI$ is an analytic EDS on $M$
and $P^{n}\subset M$ is an analytic submanifold which is K\"{a}hler-regular
and such that, at each $p\in P$, $H\left(T_{p}P\right)$ has dimension
$n+r+1$. Moreover, let us assume that $R\subset M$ is an analytic
submanifold of codimension $r$ such that $P\subset R$ and $T_{p}R$
intersects transversely with $H\left(T_{p}P\right)$. Then for each
$p\in P$ there is a neighborhood $U\subset R$ of $p$ and a unique
analytic $\left(n+1\right)$-dimensional integral manifold $N\subset U$
containing $P\cap U$.
\end{thm}
The main drawback of this theorem is that we need to work in the analytic
setting; for the systems which we consider here this hypothesis is
fulfilled, although it is important to remark that it is a strong
condition to be required.

\subsection{Cartan-Kuranishi Theorem}

According to the Cartan's test for involutivity, we can apply the
Cartan-K\"{a}hler theorem only in those cases in which the codimension
of the set of integral elements has the appropiate value. It can happen
that the integral manifolds we are looking for cannot be obtained
by using this existence theorem. For example, to ensure that the
set of $3$-integral sections for an EDS can be obtained by restriction
of $4$-integral sections to some particular submanifold, the hypothesis
of the Cartan-K\"{a}hler theorem needs to be verified for each $3$-integral
section. These ideas motivate the following concept, central in the
theory of PDEs.

\begin{defn}
Let $\cI\subset\Omega^\bullet\left(M\right)$ be an EDS on $M$, and $E\in V_n\left(\cI\right)$. A flag $0\subset E_1\subset\cdots\subset E_n=E$ is a \emph{regular flag} iff $E_k$ is a regular $k$-integral element of $\cI$ for $1\leq k\leq n-1$. 
\end{defn}

The regular flags allows us to solve the PDE behind an EDS as a sucession of
Cauchy-Kowalevky problems; we can introduce a fundamental concept related to
the integrability of the underlying PDE.

\begin{defn}
An EDS is \emph{$n$-involutive} iff every $n$-integral element is the terminus of a regular flag.
\end{defn}

The main consequence of this definition and the Cartan-K\"{a}hler theorem is the following proposition (see for example \cite{BCG}).

\begin{prop}
If an EDS is $n$-involutive, through any $n$-integral element passes an integral manifold of dimension $n$.
\end{prop}

One may prove there exists an operation, called \emph{prolongation},
which allows us to associate to every EDS $\cI$ on a manifold $M$ and every $n\in\mathbb{N}$ a new EDS $\cI'$ on a manifold $M'$ such that
\begin{itemize}
\item There exists a submersion $p:M'\rightarrow M$, and
\item the $n$-integrals elements of $\cI$ and $\cI'$ are in \emph{one to one correspondence} via the map $p$.
\end{itemize}
The key point is that, under certain
technical assumptions \cite{BCG}, after a finite number of prolongations
one can obtain an involutive EDS: This fact is known as \emph{Cartan-Kuranishi
theorem}. In this work we are assuming that these technical requeriments are fulfilled elsewhere.

\bibliographystyle{hplain}

\medskip
Received xxxx 20xx; revised xxxx 20xx.
\medskip

\end{document}